\documentclass[12pt]{article}
\pdfoutput=1

\usepackage{booktabs}
\usepackage{tabulary}
\usepackage{multirow}
\usepackage{rotating}
\usepackage{epsfig}
\usepackage{psfrag}
\usepackage{latexsym}
\usepackage{indentfirst}
\usepackage{fancyhdr}
\usepackage{dsfont}
\usepackage{adjustbox}
\usepackage{amsmath}
\usepackage{amssymb}
\usepackage{amsfonts}
\usepackage{mathrsfs}
\usepackage{amsthm}
\usepackage{pifont}
\usepackage{dsfont}
\usepackage{multirow}
\usepackage{array}
\usepackage{chngpage}
\usepackage{longtable}
\usepackage{cite}
\usepackage{bbold}
\usepackage{color}
\usepackage{braket}
\usepackage{colordvi}
\usepackage{fancybox}
\usepackage[footnotesize]{caption2}
\usepackage{graphicx}
\usepackage[center,footnotesize,hang]{subfigure}
\usepackage{bbm}
\usepackage{url}
\usepackage[table]{xcolor}
\usepackage[colorlinks, linkcolor=red, anchorcolor=black, citecolor=violet]{hyperref}
\usepackage{multirow}
\usepackage{harpoon}
\usepackage{textcomp}  
\usepackage{enumitem}
\usepackage{mathrsfs}  

\usepackage{tikz}
\usepackage{diagbox}

\newcommand{\PreserveBackslash}[1]{\let\E^c=\\#1\let\\=\E^c}
\newcolumntype{C}[1]{>{\PreserveBackslash\centering}p{#1}}
\newcolumntype{R}[1]{>{\PreserveBackslash\raggedleft}p{#1}}
\newcolumntype{L}[1]{>{\PreserveBackslash\raggedright}p{#1}}
\allowdisplaybreaks  

\newcommand{\bq}{\begin{eqnarray}}
\newcommand{\nq}{\end{eqnarray}}

\newcommand{\ignore}[1]{}

\newcommand{\MeV}{\mathrm{MeV}}
\newcommand{\meV}{\mathrm{meV}}
\newcommand{\GeV}{\mathrm{GeV}}

\begin{document}
\title{
\begin{flushright}
\hfill\mbox{\small USTC-ICTS/PCFT-21-20} \\[5mm]
\begin{minipage}{0.2\linewidth}
\normalsize
\end{minipage}
\end{flushright}
{\Large \bf
Flavor mixing and CP violation from the interplay of $S_4$ modular group and gCP
\\[2mm]}
}
\date{}

\author{
Bu-Yao Qu\footnote{E-mail: {\tt
qubuyao@mail.ustc.edu.cn}},  \
Xiang-Gan Liu\footnote{E-mail: {\tt
hepliuxg@mail.ustc.edu.cn}},  \
Ping-Tao Chen\footnote{E-mail: {\tt
chenpt@mail.ustc.edu.cn}},  \
Gui-Jun~Ding\footnote{E-mail: {\tt
dinggj@ustc.edu.cn}}
\\*[20pt]
\centerline{
\begin{minipage}{\linewidth}
\begin{center}
{\it \small
Interdisciplinary Center for Theoretical Study and  Department of Modern Physics,\\
University of Science and Technology of China, Hefei, Anhui 230026, China}\\
\end{center}
\end{minipage}}
\\[10mm]}
\maketitle
\thispagestyle{empty}

\begin{abstract}
We have performed a systematical analysis of lepton and quark masses models based on $\Gamma_4\cong S_4$ modular symmetry with gCP symmetry. We have considered both cases that neutrinos are Majorana particles and Dirac particles. All possible nontrivial representation assignments of matter fields are considered, and the most general form of fermion mass matrices are given. The phenomenologically viable models with the lowest number of free parameters together with the results of fit are presented. We find out nine lepton models with seven real free parameters including the real and imaginary parts of modulus for Majorana neutrinos, which can accommodate the lepton masses and neutrino oscillation data. The prediction for leptogenesis is studied in an example lepton model. The observed baryon asymmetry as well as lepton masses and mixing angles can be explained. For Dirac neutrinos, four lepton models with five real free couplings are compatible with experimental data. Ten quark models containing seven couplings are found to be able to accommodate the hierarchical quark masses and mixing angles and CP violation phase. Furthermore, the $S_4$ modular symmetry can provide a unified description of lepton and quark flavor structure, and a benchmark model is presented.
\end{abstract}

\section{Introduction}
Understanding the hierarchical fermion masses and the flavor mixing structure of quarks and leptons from the first principle is a longstanding challenge in particle physics. The measurement of neutrino mixing parameters provides new clews to the above mentioned flavor puzzle.
There is still lack of a guiding principle to explain this flavour puzzle, and one of the most extensively studied schemes is the flavor symmetry, which is traditionally based on continuous Lie groups or discrete finite groups which relate the three generations of fermions. See Ref.~\cite{Feruglio:2019ktm} for the latest review.
In the traditional flavor symmetry, the flavor groups are broken along certain direction in the flavor space by the vacuum expectation value (VEV) of some scalar fields called flavons. In order to realize the desired symmetry breaking pattern, certain shaping symmetry and additional dynamics and fields are generally necessary. As a consequence, the resulting model look quite complex. In order to overcome this drawback, the modular invariance from a bottom-up perspective has been proposed~\cite{Feruglio:2017spp}. The modular symmetry plays the role of flavor symmetry, the flavons are replaced by the so-called modulus $\tau$ and the Yukawa couplings are modular forms which are holomorphic functions of $\tau$. This framework has been extended to invariance under more general discrete groups and the modular forms become more general automorphic forms~\cite{Ding:2020zxw}.

In this scheme, modular symmetry is governed by the infinite discrete group $\Gamma=\mathrm{SL}(2,\mathbb{Z})$. The modular invariant models are classified by the level $N$ which is a positive integer, and the matter fields are assumed to transform in irreducible representations of the finite modular group $\Gamma_N\equiv\bar{\Gamma}/\bar{\Gamma}(N)$ or its double covering group $\Gamma'_N\equiv\Gamma/\Gamma(N)$. For a small number of finite modular groups, some fermion masses models based on the modular invariance and their phenomenology have been studied, such as $\Gamma_2\cong S_3$~\cite{Kobayashi:2018vbk,Kobayashi:2018wkl,Kobayashi:2019rzp,Okada:2019xqk}, $\Gamma_3\cong A_4$~\cite{Feruglio:2017spp,Criado:2018thu,Kobayashi:2018vbk,Kobayashi:2018scp,deAnda:2018ecu,Okada:2018yrn,Kobayashi:2018wkl,Novichkov:2018yse,Nomura:2019jxj,Okada:2019uoy,Nomura:2019yft,Ding:2019zxk,Okada:2019mjf,Nomura:2019lnr,Kobayashi:2019xvz,Asaka:2019vev,Gui-JunDing:2019wap,Zhang:2019ngf,Nomura:2019xsb,Wang:2019xbo,Kobayashi:2019gtp,King:2020qaj,Ding:2020yen,Okada:2020rjb,Nomura:2020opk,Okada:2020brs,Yao:2020qyy,Feruglio:2021dte}, $\Gamma_4\cong S_4$~\cite{Penedo:2018nmg,Novichkov:2018ovf,deMedeirosVarzielas:2019cyj,Kobayashi:2019mna,King:2019vhv,Criado:2019tzk,Wang:2019ovr,Gui-JunDing:2019wap,Wang:2020dbp}, $\Gamma_5\cong A_5$~\cite{Novichkov:2018nkm,Ding:2019xna,Criado:2019tzk}, $\Gamma_7\cong \mathrm{PSL}(2,\mathbb{Z}_7)$~\cite{Ding:2020msi}, $\Gamma'_3\cong T'$~\cite{Liu:2019khw,Lu:2019vgm}, $\Gamma'_4\cong S'_4$~\cite{Novichkov:2020eep,Liu:2020akv} and $\Gamma'_5\cong A'_5$~\cite{Wang:2020lxk,Yao:2020zml}. There have also been attempts to implement modular symmetry in the Grand Unified Theories (GUTs) to address both the lepton and quark flavor problems~\cite{deAnda:2018ecu,Kobayashi:2019rzp,Du:2020ylx,Zhao:2021jxg,Chen:2021zty,King:2021fhl,Ding:2021zbg}.
In the modular invariance approach, the crucial elements are the modular forms of level $N$. For the even modular weights, the modular forms can be arranged into multiplets of the finite modular group $\Gamma_{N}$~\cite{Feruglio:2017spp}. If the modular weights are general integers, the modular forms can be organized into irreducible multiplets of the double covering group $\Gamma'_{N}$~\cite{Liu:2019khw}.
In addition to the integer weight modular forms, there are fractional weight modular forms for some  particular level $N$. Accordingly the modular group $\mathrm{SL}(2,\mathbb{Z})$ should be extended to its metaplectic covering group and the finite metaplectic group acts as flavor symmetry~\cite{Liu:2020msy}.
The modular forms of weights $k/2$ with finite metaplectic modular group $\tilde{\Gamma}_{4}\cong \tilde{S}_4$~\cite{Liu:2020msy} and the weight $k/5$ modular forms with finite metaplectic modular group $\tilde{\Gamma}_5\cong A'_5\times Z_5$~\cite{Yao:2020zml} have been studied in the bottom-up modular invariance approach. It has been shown that metaplectic flavor symmetries can be derived from compactifications on tori with magnetic background fluxes~\cite{Kikuchi:2020frp,Almumin:2021fbk}.
The VEV of $\tau$ is usually treated as a free parameter in modular invariant models in order to match the experimental data, it is remarkable that the hierarchical fermion mass matrices may arise due to the proximity of the modulus to the residual symmetry preserved points $\tau=i, -1/2+i\sqrt{3}/2, i\infty$~\cite{Okada:2020ukr,Feruglio:2021dte,Novichkov:2021evw}. Furthermore, the generalized CP symmetry can be consistently imposed in the context of symplectic modular symmetry for single modulus with $g=1$~\cite{Novichkov:2019sqv,Baur:2019kwi,Baur:2019iai} and for multi-moduli with $g\geq 2$~\cite{Ding:2021iqp}. Notice that the symplectic group coincides with the modular group $\mathrm{SL}(2,\mathbb{Z})$ when $g=1$.
In the symmetric basis for the modular generators $S$ and $T$ with $\rho_{\mathbf{r}}(S)=\rho^T_{\mathbf{r}}(S)$ and $\rho_{\mathbf{r}}(T)=\rho^T_{\mathbf{r}}(T)$, the gCP transformation would coincide with the canonical CP. The gCP invariance enforces all coupling constants to be real if the Clebsch-Gordan (CG) coefficients are also real in the symmetric basis. Thus the gCP symmetry could further reduce the number of free parameters of the modular invariant models and leads to a higher predictive power.

The $S_4$ modular group has five irreducible representations: two singlets $\mathbf{1}$ and $\mathbf{1}'$, a doublet $\mathbf{2}$, and two triplets $\mathbf{3}$ and $\mathbf{3}'$. Similar to the $A_4$ modular symmetry, the three generations of right-handed lepton fields and right-handed charged leptons are usually assumed to transform as triplet and singlet respectively under $S_4$ modular symmetry in the known $S_4$ modular invariant models, and the doublet assignment for the lepton fields has not been considered although it provides new features and possibilities unavailable in the $A_4$ modular symmetry. Moreover, $S_4$ modular symmetry has been used to explain the flavor structure of lepton so far, but it is not clear whether the $S_4$ modular symmetry can help to address the quark flavor problem except few GUT models~\cite{Zhao:2021jxg,King:2021fhl,Ding:2021zbg}. In this paper, we intend to perform a systematical analysis of lepton and quark models based on $\Gamma_4\cong S_4$ modular symmetry and gCP, and we aim at the viable models involving the lowest number of free parameters. For lepton models, we find that thirteen viable models can successfully describe the experimental data of lepton masses and mixing parameters in terms of seven real parameters including $\text{Re}\tau$ and $\text{Im}\tau$. In the quark sector,  at least seven real couplings are necessary in order to accommodate the measured values of quark masses and CKM mixing matrix, Furthermore, we find that agreement with the experimental data of quarks and lepton sectors can be achieved for a common value of $\tau$, and a benchmark model is presented. In modular invariant models which also fulfill gCP invariance, the VEV of the modulus $\tau$ is the unique source of both modular symmetry breaking and CP violation. Thus imposing gCP symmetry would lead to strong correlations between the low energy CP violation phases and CP asymmetry in leptogenesis. We shall discuss the baryon asymmetry generated via unflavored thermal leptogenesis in an example model of lepton.

This paper is organized as follows. In section~\ref{sec:ModularSymmetry}, we briefly review the modular symmetry and modular forms of level 4.
In section~\ref{sec:models}, we give the most general forms of the Yukawa superpotential and the Majorana mass term for different possible assignments of matter fields, the corresponding mass matrices are presented. Moreover, we show that different assignments can lead to the same fermion mass matrices. In section~\ref{sec:numerical results}, we find out the phenomenologically viable lepton and quark models with the smallest number of free parameters, and the results of fit are presented. The prediction for leptogenesis is studied in a minimal lepton model. Finally, we draw the conclusions in section~\ref{sec:conclusion}. The finite modular group $\Gamma_4\cong S_4$ and the compact expression of CG coefficients are listed in Appendix~\ref{sec:S4_group}.
The concrete forms of modular multiplets at weight 4, 6, 8 are presented in Appendix~\ref{sec:modularform_of_N=4}. We give the general modular invariant superpotentials and mass matrices for two right-handed neutrino case in Appendix~\ref{sec:two right-handed neutrino}.

\section{\label{sec:ModularSymmetry}Modular symmetry and modular forms of level $N=4$}
In the modular invariant framework with single modulus, the modular symmetry is described by the modular group which is the special linear group $\mathrm{SL}(2,\mathbb{Z})$ of degree two over integers:
\begin{equation}
\mathrm{SL}(2,\mathbb{Z})=\left\{\begin{pmatrix}
a & b \\ c & d
\end{pmatrix}  \Big| ad- bc=1\,,\quad  a,b,c,d \in \mathbb{Z}\,\right\}\,.
\end{equation}
$\mathrm{SL}(2,\mathbb{Z})$ is often denoted $\Gamma$ as well. It have two generators $S$ and $T$ with
\begin{equation}
S=\begin{pmatrix}
0 & 1 \\ -1 & 0
\end{pmatrix}\,, \qquad T=\begin{pmatrix}
1 & 1 \\ 0 & 1
\end{pmatrix}\,,
\end{equation}
which satisfy the relations
\begin{equation}
S^4=(ST)^3=\mathbb{1}\,, \quad S^2 T=T S^2\,.
\end{equation}
The modular group $\Gamma$ has an important class of normal subgroups called principal congruence subgroup of level $N$ which is defined as
\begin{equation}
\Gamma(N)=\left\{ \gamma \in \mathrm{SL}(2,\mathbb{Z}) ~~\Big |~~ \gamma \equiv \begin{pmatrix}
1 & 0 \\ 0 & 1
\end{pmatrix} \mod N \right\}\,.
\end{equation}
Note that $\Gamma(1)=\mathrm{SL}(2,\mathbb{Z})$ and $T^N\in \Gamma(N)$. We can obtain the finite modular group from the quotient group:
\begin{equation}
\label{eq:mul-rules-GammaN}\Gamma_N\equiv \bar{\Gamma}/\bar{\Gamma}(N):~~~~ S^2=(ST)^3=T^N=1\,, ~\quad~ N<6\,,
\end{equation}
where $\bar{\Gamma}$ and $\bar{\Gamma}(N)$ are the projective groups $\bar{\Gamma}=\Gamma/\{\pm 1\}$ and $\bar{\Gamma}(N)=\Gamma(N)/\{\pm 1\}$. Notice that $\Gamma(N)\cong \bar{\Gamma}(N)$ for $N>2$. The group $\Gamma_N$ is usually called inhomogeneous finite modular group of level $N$, and similarly the homogeneous finite modular group $\Gamma'_N$ can be defined as $\Gamma'_N\equiv \Gamma/\Gamma(N)$ which can also be generated by $S$ and $T$ obeying the multiplication rules $S^4=(ST)^3=T^N=1$~\cite{Liu:2019khw}, and additional relations are need to render the group finite for $N\geq6$~\cite{deAdelhartToorop:2011re}. The group $\Gamma_N$ is isomorphic to quotient group of $\Gamma'_N$ over its center $\{1, S^2=-1\}$, consequently $\Gamma'_N$ is the double covering group of $\Gamma_N$, and $\Gamma'_N$ has twice as many elements as $\Gamma_N$. Top-down constructions in string theory generally lead to homogeneous finite modular group $\Gamma'_N$ rather than the inhomogeneous finite modular group $\Gamma_N$~\cite{Baur:2019kwi,Baur:2019iai}.

The modular group $\mathrm{SL}(2,\mathbb{Z})$ acts on the upper half plane $\mathcal{H}=\{\tau \in \mathbb{C} ~|~ \text{Im}(\tau)>0\}$ by the linear fractional transformation:
\begin{equation}
\gamma \tau = \dfrac{a\tau+b}{c\tau+d}\,,\qquad \gamma=\begin{pmatrix}
a & b \\c & d
\end{pmatrix}\in \Gamma\,.
\end{equation}
Obviously $\gamma$ and $-\gamma$ give the same action on the modulus $\tau$, thus each linear fractional transformation corresponds to an element of the projective special linear group $\bar{\Gamma}$. If all the points in the orbit of a modulus $\tau$ is identified, we obtain the coset space $\mathcal{H}/\Gamma$ which is the so-called fundamental domain $\mathcal{D}$ of $\mathrm{SL}(2,\mathbb{Z})$:
\begin{equation}
\label{eq:fundamental-domain}\mathcal{D}=\left\{\tau\in \mathcal{H} ~\Big|~ |\tau|\geq 1\,,~ -1/2 \leq \text{Re}(\tau) \leq 1/2\right\}\,,
\end{equation}
which is a hyperbolic triangle bounded by the vertical lines $\text{Re}(\tau)=\frac{1}{2}$, $\text{Re}(\tau)=-\frac{1}{2}$
and the circle $|\tau|=1$. Every point $\tau\in\mathcal{H}$ is equivalent to a point of $\mathcal{D}$ via the action of $\mathrm{SL}(2,\mathbb{Z})$, and no two distinct points inside $\mathcal{D}$ are equivalent under the action of $\mathrm{SL}(2,\mathbb{Z})$ and two points of $\mathcal{D}$ are in the same orbit only if they lie on the boundary of $\mathcal{D}$.

The modular form of integral weight $k$ and level $N$ is a holomorphic function of $\tau$, and it transforms under $\Gamma(N)$ as follows
\begin{equation}
f(\gamma\tau)=(c\tau+d)^{k}f(\tau)\,, \qquad \forall \gamma=\begin{pmatrix}
a & b \\ c & d
\end{pmatrix}\in \Gamma(N)\,.
\end{equation}
Therefore the weight $k$ ``differential form'' $f(\tau)(d\tau)^{k/2}$ is invariant under the action of every element of $\Gamma(N)$.
The modular forms of weight $k$ and level $N$ span a finite dimensional linear space $\mathcal{M}_{k}(\Gamma(N))$. The product of a modular form of weight $k_1$ with a modular form of weight $k_2$
is a modular form of weight $k_1+k_2$. Thus the set $\mathcal{M}(\Gamma(N))=\bigoplus^{\infty}_{k=0}\mathcal{M}_{k}(\Gamma(N))$ of all modular forms of level $N$ forms a graded ring. Furthermore, it has been proved that the finite dimensional space $\mathcal{M}_{2k}(\Gamma(N))$ can be decomposed into irreducible representations of the finite modular groups $\Gamma_{N}$~\cite{Feruglio:2017spp,Liu:2019khw} up to the automorphy factor $(c\tau+d)^{2k}$.
That is to say, it is always possible to choose a basis in $\mathcal{M}_{2k}(\Gamma(N))$ so that $Y^{(2k)}_\mathbf{r}=(f_1(\tau),f_2(\tau),\dots)^T$ transform under the full modular group $\Gamma$ as
\begin{equation}
Y^{(2k)}_\mathbf{r}(\gamma\tau)=(c\tau+d)^{2k}\rho_\mathbf{r}(\gamma)Y^{(2k)}_\mathbf{r}(\tau)\,, \qquad \forall \gamma=\begin{pmatrix}
		a & b \\ c & d
	\end{pmatrix}\in \Gamma\,,
\end{equation}
where $\rho_\mathbf{r}(\gamma)$ is the irreducible representation of quotient group $\Gamma_N$.

In the present work, we are interested in the level $N=4$, the linear space of the modular forms of level 4 is well established, and it can be constructed by making use of Dedekind eta function or the theta constants ~\cite{Gui-JunDing:2019wap,Liu:2020msy,Novichkov:2020eep}:
\begin{align}
\label{eq:Mk_Gamma4}
\nonumber
\mathcal{M}_{k}(\Gamma(4))&=\bigoplus_{a+b=2k,\,a,b\ge0} \mathbb{C} \frac{\eta^{2b-2a}(4\tau)\eta^{5a-b}(2\tau)}{\eta^{2a}(\tau)}\,\\
&=\bigoplus_{a+b=2k,\, a,b\ge 0} \mathbb{C} \theta^{a}_2(\tau)\theta^{b}_3(\tau)\,,
\end{align}
where Dedekind eta function $\eta(\tau)$ is defined by
\begin{equation}	
\label{eq:eta_function}
\eta(\tau)=q^{1/24}\prod_{n=1}^\infty \left(1-q^n \right),\qquad q\equiv e^{i 2 \pi\tau}\,,
\end{equation}
and the theta constants are defined as
\begin{equation}
\theta_2(\tau)= \sum_{m\in\mathbb{Z}} e^{2\pi i \tau (m+1/2)^2}\,,\quad \theta_3(\tau)= \sum_{m\in\mathbb{Z}} e^{2\pi i \tau m^2}\,.
\end{equation}
Thus the dimension of the modular space $\mathcal{M}_{2k}(\Gamma(N))$ is equal to $4k+1$. In the working basis given in Appendix~\ref{sec:S4_group}, all the modular forms of weight $k$ and level 4 can be expressed as the homogeneous polynomials of degree $2k$ in the modular functions $\vartheta_{1}$ and $\vartheta_{2}$ which are linear combinations of $\theta_2(\tau)$ and $\theta_3(\tau)$ as follows~\cite{Liu:2020msy},
\begin{eqnarray}
\nonumber &&\vartheta_1(\tau)=\omega^2\theta_3(\tau)+(i+\omega)\theta_2(\tau)\,, \\
&&\vartheta_2(\tau)=\frac{\sqrt{2}+\sqrt{6}}{2}\theta_3(\tau)+e^{i\pi/4}\theta_2(\tau)\,,
\end{eqnarray}
with $\omega=e^{2\pi i/3}$. In particular, the weight 2 modular multiplets $Y_\mathbf{2}^{(2)}$ and $Y_\mathbf{3}^{(2)}$ can be written as~\cite{Liu:2020msy}
\begin{align}
\label{eq:wt2MF_express}
\nonumber
&Y^{(2)}_\mathbf{2}\equiv \begin{pmatrix}
Y_1 \\ Y_2
\end{pmatrix}=\frac{e^{i\pi/3}}{9+6\sqrt{3}}\begin{pmatrix}
2\sqrt{2}\vartheta_1\vartheta_2^3 - \vartheta_1^4 \\ \vartheta_2^4 + 2\sqrt{2}\vartheta_1^3 \vartheta_2
\end{pmatrix}\,,\\
&Y^{(2)}_\mathbf{3}\equiv \begin{pmatrix}
Y_3 \\ Y_4 \\ Y_5
\end{pmatrix}=\frac{e^{i\pi/3}}{3(2+\sqrt{3})}\begin{pmatrix}
3\vartheta_1^2\vartheta_2^2 \\ \vartheta_2^4 - \sqrt{2}\vartheta_1^3 \vartheta_2 \\ \vartheta_1^4 + \sqrt{2}\vartheta_1 \vartheta_2^3
\end{pmatrix}\,.
\end{align}
The weight 2 modular forms of level 4 can also be constructed from the derivative of eta function~\cite{Penedo:2018nmg,Novichkov:2018ovf} or the products of the Dedekind eta function~\cite{Gui-JunDing:2019wap}, the resulting $q-$expansions of the modular forms would be identical when going to the same representation basis of the modular generators $S$ and $T$. The higher weight modular forms can be generated by the tensor product of the weight 2 modular forms, and their specific forms can be found in Appendix~\ref{sec:modularform_of_N=4}. We summarize the modular multiplets of level 4 up to weight 8 in table~\ref{Tab:LeveL4_MM}.
\begin{table}[th!]
\centering
\begin{tabular}{|c|c|}\hline \hline
Modular weight $2k$ & Modular forms $Y^{(2k)}_{\mathbf{r}}$ \\ \hline
			
$2k=2$ & $Y^{(2)}_{\mathbf{2}}, ~Y^{(2)}_{\mathbf{3}}$\\ \hline
			
$2k=4$ & $Y^{(4)}_{\mathbf{1}},~Y^{(4)}_{\mathbf{2}},~ Y^{(4)}_{\mathbf{3}}, ~Y^{(4)}_{\mathbf{3}'}$\\ \hline
		
$2k=6$ & $Y^{(6)}_{\mathbf{1}},~Y^{(6)}_{\mathbf{1}'},~Y^{(6)}_{\mathbf{2}},~Y^{(6)}_{\mathbf{3}I},~Y^{(6)}_{\mathbf{3}II},~Y^{(6)}_{\mathbf{3}'}$\\ \hline

$2k=8$ & $Y^{(8)}_{\mathbf{1}},~Y^{(8)}_{\mathbf{2}I},~Y^{(8)}_{\mathbf{2}II},~Y^{(8)}_{\mathbf{3}I},~Y^{(8)}_{\mathbf{3}II},~Y^{(8)}_{\mathbf{3'}I},~Y^{(8)}_{\mathbf{3}'II}$\\ \hline \hline
\end{tabular}
\caption{\label{Tab:LeveL4_MM}Summary of the even weight modular forms at level $N=4$, the subscript $\mathbf{r}$ denotes the irreducible representations of the inhomogeneous finite modular group $\Gamma_4\cong S_4$. Here $Y^{(6)}_{\mathbf{3}I}$ and $Y^{(6)}_{\mathbf{3}II}$ denote the two linearly independent weight 8 modular forms in the doublet representation $\mathbf{2}$, and we adopt a similar notation for $Y^{(8)}_{\mathbf{2}I}$, $Y^{(8)}_{\mathbf{2}II}$ and $Y^{(8)}_{\mathbf{3}I},~Y^{(8)}_{\mathbf{3}II}$ and $Y^{(8)}_{\mathbf{3'}I},~Y^{(8)}_{\mathbf{3}'II}$.   }
\end{table}
\section{\label{sec:models}Fermion mass models based on $S_4$ modular symmetry with generalized CP}

We shall briefly review the modular invariance approach in the following, then recapitulate on the consistency condition which should be fulfilled to consistently combine modular symmetry with generalized CP symmetry (gCP).
Furthermore, we perform a systematical classification of quark and lepton mass models based on the modular symmetry $\Gamma_4\cong S_4$ and gCP.

\subsection{\label{subsec:framework} The framework}
We formulate our models in the framework of modular invariant approach with $\mathcal{N}=1$ global supersymmetry~\cite{Feruglio:2017spp}. The field content consists of a set of chiral matter superfields $\Phi_I$ and a modulus superfield $\tau$, their modular transforms under $\mathrm{SL}(2,\mathbb{Z})$ are given by:
\begin{equation}
\label{eq:modularTrs_Phi}
	\tau\to \gamma\tau=\frac{a\tau+b}{c\tau+d}\,,\qquad
	\Phi_I\to (c\tau+d)^{-k_I}\rho_I(\gamma)\Phi_I\,,
\end{equation}
where $-k_I$ is called the modular weight of the matter field $\Phi_I$, and $\rho_I(\gamma)$ is the unitary representation of $\Gamma_N$.
The K\"ahler potential is taken to be the minimal form following the convention of~\cite{Feruglio:2017spp}
\begin{equation}
\label{eq:kahler}\mathcal{K}(\Phi_I,\bar{\Phi}_I; \tau,\bar{\tau}) =-h\Lambda^2 \log(-i\tau+i\bar\tau)+ \sum_I (-i\tau+i\bar\tau)^{-k_I} |\Phi_I|^2~~~,
\end{equation}
which gives rise to the kinetic terms of the matter fields and the modulus field after the modular symmetry breaking caused by vacuum expectation value (VEV) of $\tau$.
Notice that the modular invariance doesn't fix the K\"ahler potential in the bottom-up approach~\cite{Chen:2019ewa}, and the K\"ahler potential could receive unsuppressed contributions from modular forms. However, generally both traditional flavor symmetry and modular symmetry are present in top-down approach such as the string derived standard-like models~\cite{Nilles:2020nnc,Nilles:2020kgo,Ohki:2020bpo}, the off-diagonal contributions to the K\"ahler metric are forbidden by the traditional flavor group and the minimal K\"ahler potential in Eq.~\eqref{eq:kahler} appears as the leading order term. Even including the whole modular dependence in the K\"ahler potential in these top-down models, the resulting phenomenological predictions don't differ from those which have been obtained by using just the standard K\"ahler potential Eq.~\eqref{eq:kahler}.
The superpotential $\mathcal{W}(\Phi_I,\tau)$ can be expanded in power series of the involved supermultiplets $\Phi_I$,
\begin{equation}
\mathcal{W}(\Phi_I,\tau) =\sum_n Y_{I_1...I_n}(\tau)~ \Phi_{I_1}... \Phi_{I_n}\,,
\end{equation}
where $Y_{I_1...I_n}$ is a modular multiplet of weight $k_Y$ as introduced in previous section. Modular invariance requires that each term of the $\mathcal{W}(\Phi_I,\tau)$ satisfies the following conditions:
\begin{equation}
	k_Y=k_{I_1}+...+k_{I_n},~\quad~ \rho_Y\otimes\rho_{I_1}\otimes\ldots\otimes\rho_{I_n}\ni\mathbf{1}\,.
\end{equation}

In order to improve the prediction power of the modular invariance approach, we include the generalized CP symmetry further. It is known that the complex modulus $\tau$ transforms under the action of gCP as~\cite{Novichkov:2019sqv,Baur:2019kwi,Acharya:1995ag,Dent:2001cc,Giedt:2002ns}
\begin{equation}
\tau\stackrel{\mathcal{CP}}{\longmapsto}-\tau^{*}
\end{equation}
up to modular transformations. The gCP transforms a generic chiral superfield $\Phi$ into the hermitian conjugate superfield
\begin{equation}
\Phi(x)\stackrel{\mathcal{CP}}{\longmapsto} X_{\mathbf{r}}\overline{\Phi}(x_{\mathcal{P}})\,,
\end{equation}
with $x=(t,\vec{x})$ and $x_{\mathcal{P}}=(t,-\vec{x})$, the CP transformation matrix $X_{\mathbf{r}}$ is a unitary matrix acting on flavor space.
The chiral superfield $\Phi(x)$ is assigned to an irreducible unitary representation $\rho_{\mathbf{r}}$ of the finite modular group, then the form of the matrix $X_{\mathbf{r}}$ is strongly constrained due to the presence of modular symmetry. Firstly applying a gCP transformation followed by a modular transformation and subsequently an inverse CP transformation, the complex modulus $\tau$ and the matter field $\Phi$ transform as follows
\begin{eqnarray}
\nonumber&\hskip-0.4in \tau\stackrel{\mathcal{CP}}{\longrightarrow}-\tau^{*}\stackrel{\gamma}{\longrightarrow} -\frac{a\tau^{*}+b}{c\tau^{*}+d}\stackrel{\mathcal{CP}^{-1}}{\longrightarrow}\frac{a\tau-b}{-c\tau+d}\,,\\
\label{eq:consistency-chain}&\hskip-0.4in \Phi(x)\stackrel{\mathcal{CP}}{\longrightarrow} X_{\mathbf{r}}\overline{\Phi}(x_{\mathcal{P}})
\stackrel{\gamma}{\longrightarrow}(c\tau^{*}+d)^{-k} X_{\mathbf{r}}\rho^{*}_{\mathbf{r}}(\gamma)\overline{\Phi}(x_{\mathcal{P}})\stackrel{\mathcal{CP}^{-1}}{\longrightarrow}
(-c\tau+d)^{-k} X_{\mathbf{r}}\rho^{*}_{\mathbf{r}}(\gamma)X^{-1}_{\mathbf{r}}\Phi(x)\,,
\end{eqnarray}
where $-k$ denotes the modular weight of $\Phi$. The closure of the modular transformations and gCP transformations entails that the following consistency condition has to be satisfied~\cite{Novichkov:2019sqv,Ding:2021iqp},
\begin{equation}
\label{eq:cc}X_{\mathbf{r}}\rho^{*}_{\mathbf{r}}(\gamma)X^{-1}_{\mathbf{r}}=\chi^{-k}(\gamma)\rho_{\mathbf{r}}(u(\gamma))\,,
\end{equation}
where $u(\gamma)$ is an outer automorphism of the modular group,
\begin{equation}
\gamma=\begin{pmatrix}
a  &  b \\
c  &  d
\end{pmatrix}\longmapsto u(\gamma)=\chi(\gamma)\begin{pmatrix}
a  &  -b \\
-c  &  d
\end{pmatrix}\,.
\end{equation}
Here $\chi(\gamma)$ is called character and it is a homomorphism of $\mathrm{SL}(2,\mathbb{Z})$ into $\left\{+1, -1\right\}$. From the relations $S^4=(ST)^3=1$ satisfied by the modular generators $S$ and $T$, it is easy to know that only two possible values of the character are allowed~\cite{automorphism-sl2z}
\begin{equation}
\chi(S)=\chi(T)=1\,,~~\text{or}~~\chi(S)=\chi(T)=-1\,.
\end{equation}
Consequently two possible gCP symmetries can be defined in the context of modular invariance. We see that the CP transformation $X_{\mathbf{r}}$ maps the modular group element $\gamma$ onto another element $u(\gamma)$ and the group structure of the modular symmetry is preserved, i.e. $u(\gamma_1\gamma_2)=u(\gamma_1)u(\gamma_2)$. Hence, it is sufficient to impose the consistency condition of Eq.~\eqref{eq:cc} on the generators $S$ and $T$. For the first kind of gCP associated with the trivial character $\chi(S)=\chi(T)=1$, the consistency condition becomes
\begin{equation}
\label{eq:consc_gen}X_{\mathbf{r}}\rho^{*}_{\mathbf{r}}(S)X^{-1}_{\mathbf{r}}=\rho_{\mathbf{r}}(S^{-1}),\qquad X_{\mathbf{r}}\rho^{*}_{\mathbf{r}}(T)X^{-1}_{\mathbf{r}}=\rho_{\mathbf{r}}(T^{-1})\,,
\end{equation}
The second kind of gCP associated with the nontrivial character $\chi(S)=\chi(T)=-1$ has been studied in~\cite{Novichkov:2020eep,Ding:2021iqp}, the explicit form of $X_{\mathbf{r}}$ depends on both the modular weight $-k$ and representation assignment $\mathbf{r}$ of the matter field, and obviously it would be reduced the first gCP for $(-1)^{-k}\rho_{\mathbf{r}}(S^2)=1$. Notice that $\rho_{\mathbf{r}}(S^2)=\pm1$ because of $S^4=1$. As a result, it is only relevant for the case of $(-1)^{-k}\rho_{\mathbf{r}}(S^2)=-1$ which implies odd $k$ for the inhomogeneous finite modular group $\Gamma_N$, the gCP transformation $X_{\mathbf{r}}$ is determined by~\cite{Novichkov:2020eep,Ding:2021iqp}
\begin{equation}
\label{eq:consc-2}X_{\mathbf{r}}\rho^{*}_{\mathbf{r}}(S)X^{-1}_{\mathbf{r}}=-\rho_{\mathbf{r}}(S^{-1}),\qquad X_{\mathbf{r}}\rho^{*}_{\mathbf{r}}(T)X^{-1}_{\mathbf{r}}=-\rho_{\mathbf{r}}(T^{-1})\,,
\end{equation}
which can be satisfied if and only if the level $N$ is even, the dimension of the representation $\rho_{\mathbf{r}}$ is even together with the vanishing trace of $\rho_{\mathbf{r}}(S)$ and $\rho_{\mathbf{r}}(T)$. If one intends to impose the second gCP in a model, the three generations of matter fields should be assigned to the direct sum of one-dimensional and two-dimensional representations of the finite modular group $\Gamma'_N$ or $\Gamma_N$, and second gCP acts nontrivially on the two matter fields in the doublet representation while the gCP transformation of the other matter field in the singlet representation can only be the first one. Moreover, the minus sign in Eq.~\eqref{eq:consc-2} implies that the fermion mass matrix would be block diagonal and consequently some mixing angles would be constrained to be vanishing if the second gCP is implemented. Because none of quark or lepton mixing angles are vanishing in spite of some very small quark mixing angles, we shall not consider the second gCP symmetry and focus on the first gCP in the present work. As regards the CP transformation of the modular forms, it has been shown that the integral weight modular forms are in the irreducible representations of $\Gamma'_N$ fulfilling $(-1)^{-k}\rho_{\mathbf{r}}(S^2)=1$~\cite{Liu:2019khw} so that only the first gCP acts on the modular forms and they transform in the same way as the matter fields under gCP, i.e., $Y_{\mathbf{r}}(\tau)\stackrel{\mathcal{CP}}{\longmapsto} Y_{\mathbf{r}}(-\tau^{*})= X_{\mathbf{r}}Y^{*}(\tau)$ if the basis of the modular space is properly chosen~\cite{Ding:2021iqp}.

The explicit form of the gCP transformation $X_{\mathbf{r}}$ is determined by the consistency condition in Eq.~\eqref{eq:cc} up to an overall phase for any given irreducible representation $\mathbf{r}$. For the concerned first gCP and the finite modular group $\Gamma_4\cong S_4$ with the basis listed in table~\ref{tab:S4_rep}, solving the consistency conditions of Eq.~\eqref{eq:consc_gen}, we find that the generalized CP transformation $X_{\mathbf{r}}$ is in common with the representation matrix of $S$,
\begin{equation}
\label{eq:Xr_S4p}X_{\mathbf{r}}=\rho_{\mathbf{r}}(S)\,,
\end{equation}
which is a combination of the modular symmetry transformation $S$ and the  canonical CP transformation. Modular invariance requires that the action is invariant under the modular transformation $S$, thus the gCP transformation in Eq.~\eqref{eq:Xr_S4p} is essentially the canonical CP transformation. Furthermore, for the level 4 modular forms built from $Y^{(2)}_\mathbf{2}(\tau)$ and $Y^{(2)}_\mathbf{3}(\tau)$ up to weight 8, it is straightforward to check that they transform under gCP as follows
\begin{equation}
Y^{(k)}_{\mathbf{r}}(\tau)\stackrel{\mathcal{CP}}{\longmapsto} Y^{(k)}_{\mathbf{r}}(-\tau^{*})=X_{\mathbf{r}}Y^{(k)*}_{\mathbf{r}}(\tau)\,,
\end{equation}
which is consistent with the general results of~\cite{Ding:2021iqp}. As given in Appendix~\ref{sec:S4_group}, all the Clebsch-Gordan (CG) coefficients in our working basis are real, thus the gCP symmetry would constrain all the coupling constants to be real.

In the modular invariant theory with gCP symmetry, both modular and CP symmetries are uniquely broken by the VEV of the modulus $\tau$. In particular, all CP violation phases arises from non-vanishing real part of $\tau$. In the following, we shall perform a systematical classification of
the Yukawa superpotential according to the transformation properties of the matter fields under the $\Gamma_4\cong S_4$ modular symmetry. We assume the Higgs doublets $H_u$ and $H_d$ are $S_4$ trivial singlet $\mathbf{1}$ and their modular weights $k_{H_u, H_d}$ are vanishing. Notice that $k_{H_u, H_d}$ can always be set to zero by redefining the modular weights of matter fields.

\subsection{\label{sec:Dirac mass} Classifying the Yukawa couplings }
\paragraph{}

The modular invariance approach is formulated in the framework of supersymmetry, and we adopt the gauge symmetry, lepton superfields, quark superfields and Higgs multiplets of the minimal supersymmetric standard model. We consider both scenarios that neutrinos are Dirac or Majorana particles, and the neutrino masses are generated by the type I seesaw mechanisms if neutrinos are Majorana particles. It is known that at least two right-handed neutrinos are necessary to generate the non-vanishing solar and atmospheric neutrino mass squared differences. We have considered both cases with two and three right-handed neutrinos. For simplicity, we denote the left-handed lepton and quark doublets as $F$ and the right-handed lepton and quark singlets as $F^c$,  i.e. $F^c \in \{u^c, d^c , E^c, N^c\}$ and $F \in \{Q, L\}$. The three generations of matter fields can be assigned to transform as a triplet $F^{(c)} \sim \mathbf{3}^i$ under $S_4$ modular group, the direct sum of doublet and singlet $F^{(c)}\sim\mathbf{2}\oplus \mathbf{1}^i$, or the direct sum of the three singlets $F^{(c)}\sim\mathbf{1}^{i_1}\oplus \mathbf{1}^{i_2}\oplus \mathbf{1}^{i_3}$. If only two right-handed neutrinos are introduced, they can transform as a doublet or two singlets under $S_4$, as discussed in Appendix~\ref{sec:two right-handed neutrino}. Therefore, there will be many possible $S_4$ modular invariant models for quarks and leptons. The main purpose of this paper is to classify all of these possible fermion mass superpotentials. In the following, we will consider modular forms of weight less than 10, and the analytical results reached can be easily extended to much higher weight modular forms analogously.

Before going into the concrete discussion below, let us make an explanation of the notations. We denote the $S_4$ singlet and triplet representations as $\mathbf{1}\equiv\mathbf{1}^0$, $\mathbf{1}'\equiv\mathbf{1}^1$, $\mathbf{3}\equiv\mathbf{3}^0$, $\mathbf{3}'\equiv\mathbf{3}^1$. We use $i,j,k,l$ to represent the indices of the singlet or the triplet representations, and they can only take the values $0$ or $1$, i.e., $i,j,k,l\in\{0,1\}$. The lowercase letter $a,b$ are used to label the component of the modular multiplets, and they can only take the value $1$, $2$ and $3$, i.e. $a,b\in\{1,2,3\}$. For simplicity of the formula, we introduce the superfluous notations $Y_{\mathbf{1},2}^{(k)}$, $Y_{\mathbf{1},3}^{(k)}$ and $Y_{\mathbf{2},3}^{(k)}$ which are set to zero.
Moreover, we use the capital letters $A$, $B$, $C$ to describe the degeneracy of the modular multiplets. For instance, there are two weight 6 modular forms $Y^{(6)}_{\mathbf{3}I}$ and $Y^{(6)}_{\mathbf{3}II}$ in the triplet representation $\mathbf{3}$. Furthermore, we introduce the operations $<>$ and $\prec\succ$ and they are defined as $<i>=i~(\mathrm{mod}~2)$ and $\prec i\succ=i~(\mathrm{mod}~3)$ which take values in the range of $\{0,1\}$ and $\{1,2,3\}$ respectively. Notice that we define $\prec i\succ=3$ if $i$ is divisible by 3.
It is remarkable that the general analytical expression of the fermion mass matrix can be read out for each possible representation assignment of the matter fields.

In this section, we will investigate the Yukawa superpotential for the fermion masses, which can be generally written as
\begin{equation}
\mathcal{W}_F=\alpha(F^cFH_{u/d}f(Y))_{\bf{1}}\,,
\end{equation}
where all independent $S_4$ contractions should be considered and different singlet combinations are associated with different coefficients. The function $f(Y)$ is some modular form multiplet fixed by the weight and representation assignments of the matter fields $F$ and $F^c$. In the following, we give the concrete form of the Yukawa superpotential and the corresponding fermion mass matrix for different $S_4$ transformation properties of $F$ and $F^c$.

\begin{itemize}[labelindent=-0.8em, leftmargin=1.5em]

\item{$F^c\sim\mathbf{3}^i\,,~~F\sim\mathbf{3}^j\,.$}

Let us first consider the case that both left-handed and right-handed fermions transform as triplets under $S_4$. The modular weights of $F^c$ and $F$ are denoted as $k_{F^c}$ and $k_{F}$ respectively. The general Yukawa supepotential for this assignment is given by
\begin{equation}
\begin{aligned}
\mathcal{W}_F=&
\alpha\left((F^cF)_{\mathbf{1}^{<i+j>}}Y_{\mathbf{1}^{<i+j>}}^{(k_{F^c}+k_{F})}\right)_{\mathbf{1}}H_{u/d}+\sum_A \beta_{A}\left((F^cF)_{\mathbf{2}}Y_{\mathbf{2}A}^{(k_{F^c}+k_{F})}\right)_{\mathbf{1}}H_{u/d}\\
&+\sum_{l=0}^{1}\sum_{B} \gamma^l_B\left((F^cF)_{\mathbf{3}^{l}}Y_{\mathbf{3}^{l}B}^{(k_{F^c}+k_{F})}\right)_{\mathbf{1}}H_{u/d}\\
=&\sum_{a=1}^3\sum_{b=1}^3 F^c_aF_b
\bigg\{\alpha Y_{\mathbf{1}^{<i+j>},\prec a+b-1\succ}^{(k_{F^c}+k_{F})}+(-1)^{\prec a+b+1\succ(i+j)}\sum_A \beta_{A}Y_{\mathbf{2}A,\prec a+b-2\succ}^{(k_{F^c}+k_{F})}\\
&+\sum_{l=0}^1\sum_B \gamma^l_BY_{\mathbf{3}^lB,\prec 3-a-b\succ}^{(k_{F^c}+k_{F})}\left[\delta_{ab}\left(1-(-1)^{(i+j+l)}\right)-(\epsilon_{ba\prec-b-a\succ})^{(i+j+l+1)}\right]\bigg\}H_{u/d}\,.
\end{aligned}
\end{equation}
Here we have assumed that the modular form multiplets $Y^{(k_{F^c}+k_{F})}_{\mathbf{1}}$, $Y^{(k_{F^c}+k_{F})}_{\mathbf{1}'}$, $Y^{(k_{F^c}+k_{F})}_{\mathbf{2}}$, $Y^{(k_{F^c}+k_{F})}_{\mathbf{3}}$ and $Y^{(k_{F^c}+k_{F})}_{\mathbf{3}'}$ in all $S_4$ irreducible representations are present. As shown in table~\ref{Tab:LeveL4_MM}, certain modular multiplets at some specific modular weights are not allowed and thus the corresponding terms should be dropped.
The fermion mass matrix can be read out from this superpotential:
\begin{equation}
\begin{aligned}
M_F=&{}
\alpha\left(\begin{matrix}
Y_{\mathbf{1}^{<i+j>}}^{(k_{F^c}+k_{F})}&0 &0\\
0 &0 &Y_{\mathbf{1}^{<i+j>}}^{(k_{F^c}+k_{F})} \\
0 &Y_{\mathbf{1}^{<i+j>}}^{(k_{F^c}+k_{F})} &0
\end{matrix}\right)v_{u/d}\\
+&
\beta_{A}\left(\begin{matrix}
0&(-1)^{i+j}Y_{\mathbf{2}A,1}^{(k_{F^c}+k_{F})} &Y_{\mathbf{2}A,2}^{(k_{F^c}+k_{F})} \\
(-1)^{i+j}Y_{\mathbf{2}A,1}^{(k_{F^c}+k_{F})} &Y_{\mathbf{2}A,2}^{(k_{F^c}+k_{F})} &0 \\
Y_{\mathbf{2}A,2}^{(k_{F^c}+k_{F})} &0 &(-1)^{i+j}Y_{\mathbf{2}A,1}^{(k_{F^c}+k_{F})}
\end{matrix}\right)v_{u/d}\\
+&
\gamma^l_B\left(\begin{matrix}
[1-(-1)^{i+j+l}]Y_{\mathbf{3}^lB,1}^{(k_{F^c}+k_{F})}&(-1)^{i+j+l}Y_{\mathbf{3}^lB,3}^{(k_{F^c}+k_{F})} &-Y_{\mathbf{3}^lB,2}^{(k_{F^c}+k_{F})}\\
-Y_{\mathbf{3}^lB,3}^{(k_{F^c}+k_{F})} &[1-(-1)^{i+j+l}]Y_{\mathbf{3}^lB,2}^{(k_{F^c}+k_{F})} &(-1)^{i+j+l}Y_{\mathbf{3}^lB,1}^{(k_{F^c}+k_{F})}\\
(-1)^{i+j+l}Y_{\mathbf{3}^lB,2}^{(k_{F^c}+k_{F})} &-Y_{\mathbf{3}^lB,1}^{(k_{F^c}+k_{F})} &[1-(-1)^{i+j+l}]Y_{\mathbf{3}^lB,3}^{(k_{F^c}+k_{F})}
\end{matrix}\right)v_{u/d}\,,
\end{aligned}\label{eq:3 and 3}
\end{equation}
where repeated indices are implicitly summed over. If $F$ and $F^c$ are quark and charged lepton fields, the case of  $k_{F^c}+k_{F}=0$ is not viable, since it gives rise to three degenerate mass eigenvalues. On the other hand, if $\mathcal{W}_F$ describe the neutrino Dirac coupling under the assumption of Majorana neutrinos, the vanishing modular weight $k_{F^c}+k_{F}=0$ is allowed.

\item{$F^c \sim\mathbf{1}^{i_1}\oplus\mathbf{1}^{i_2}\oplus\mathbf{1}^{i_3} \,,~~ F\sim\mathbf{3}^j\,.$}

In this case, the three generations of left-handed fermions $F$ transform as a triplet of $S_4$ and the right-handed fields $F^c$ are assigned to be singlets of $S_4$. The modular weight of $F$ and $F^c$ are denoted by $k_F$ and $k_{F^c_{1,2,3}}$ respectively.
Notice that permutating the assignments of the three right-handed fermions $F^c$ amount to multiplying certain permutation matrix from the right side of the mass matrix, consequently the results for the charged fermion masses and mixing matrix are left invariant. The superpotential for this assignment can be written as
\begin{equation}
\begin{aligned}
\mathcal{W}_F=&{}[\alpha(F^c_1Ff_{1}(Y))_{\bf{1}}+\beta(F^c_2Ff_{2}(Y))_{\bf{1}}+\gamma(F^c_3Ff_{3}(Y))_{\bf{1}}]H_{u/d}\\
=&\sum_{b=1}^3\sum_A\alpha_{A}F^c_1F_bY_{\mathbf{3}^{< i_1+j >}A,\prec 2-b\succ}^{(k_{F^c_1}+k_{F})}H_{u/d}
+\sum_{b=1}^3\sum_B\beta_{B}F^c_2F_bY_{\mathbf{3}^{< i_2+j >}B,\prec 2-b\succ}^{(k_{F^c_2}+k_{F})}H_{u/d} \\
+&\sum_{b=1}^3\sum_C\gamma_{C}F^c_3F_bY_{\mathbf{3}^{< i_3+j >}C,\prec 2-b\succ}^{(k_{F^c_3}+k_{F})}H_{u/d}\,.
\end{aligned}
\end{equation}
which leads to the following fermion mass matrix
\begin{equation}
M_F=\left(\begin{matrix}
\alpha_AY_{\mathbf{3}^{< i_1+j >}A,1}^{(k_{F^c_1}+k_{F})}&\alpha_AY_{\mathbf{3}^{< i_1+j >}A,3}^{(k_{F^c_1}+k_{F})}&\alpha_AY_{\mathbf{3}^{< i_1+j >}A,2}^{(k_{F^c_1}+k_{F})}\\
\beta_BY_{\mathbf{3}^{< i_2+j >}B,1}^{(k_{F^c_2}+k_{F})}&\beta_BY_{\mathbf{3}^{< i_2+j >}B,3}^{(k_{F^c_2}+k_{F})}&\beta_BY_{\mathbf{3}^{< i_2+j >}B,2}^{(k_{F^c_2}+k_{F})}\\
\gamma_CY_{\mathbf{3}^{< i_3+j >}C,1}^{(k_{F^c_3}+k_{F})}&\gamma_CY_{\mathbf{3}^{< i_3+j >}C,3}^{(k_{F^c_3}+k_{F})}&\gamma_CY_{\mathbf{3}^{< i_3+j >}C,2}^{(k_{F^c_3}+k_{F})}
\end{matrix}\right)v_{u/d}\label{eq:1+1+1 and 3}\,.
\end{equation}
If two right-handed fields are assigned to have the same modular weight and representation assignment and they couple with a unique modular multiplet, two rows of the mass matrix would be proportional such that one mass eigenvalue would be vanishing.
In some specific cases, the mass matrix can also gives zero eigenvalue. For example, from Appendix \ref{sec:modularform_of_N=4} we can see the modular forms $Y_{\mathbf{3}I}^{(8)}$ and $Y_{\mathbf{3}II}^{(8)}$ are parallel to $Y_{\mathbf{3}}^{(2)}$ and $Y_{\mathbf{3}}^{(4)}$ respectively. As a consequence, the fermion mass matrix for the assignment $(k_{F^c_1}+k_{F},k_{F^c_2}+k_{F},k_{F^c_3}+k_{F})=(2,4,8)$ and $(<i_1+j>,<i_2+j>,<i_3+j>)=(0,0,0)$ will have zero mass eigenvalue as well.

\item{$F^c \sim\mathbf{2}\oplus\mathbf{1}^i\,,~~F\sim\mathbf{3}^j\,.$}

Without loss of generality, we assign the first two right-handed fermions $F^c_D=(F^c_1,F^c_2)^T$ to transform as a doublet under $S_4$ and the third one $F^c_3$ is the singlet. The modular weights of these fields are $k_{F^c_D}$, $k_{F^c_3}$ and $k_{F}$. The general superpotential is of the following form:
\begin{eqnarray}
\nonumber\mathcal{W}_F&=&{}[\alpha(F^c_DFf_{1}(Y))_{\bf{1}}+\beta(F^c_3Ff_{2}(Y))_{\bf{1}}]H_{u/d}\\
\nonumber&=&\sum_{a=1}^2\sum_{b=1}^3\sum_{l=0}^1\sum_A(-1)^{(a+1)(j+l)}\alpha^l_{A}F^c_aF_bY_{\mathbf{3}^lA,\prec 2+a-b \succ}^{(k_{F^c_D}+k_F)}H_{u/d}\\
&&~~+\sum_{b=1}^3\sum_B\beta_{B}F^c_3F_bY_{\mathbf{3}^{< i+j >}B,\prec 2-b \succ}^{(k_{F^c_3}+k_F)}H_{u/d}\,.
\end{eqnarray}
The mass matrix  which can be read out from this superpotential is
\begin{equation}
M_F=\left(\begin{matrix}
\alpha^l_{A}Y_{\mathbf{3}^lA,2}^{(k_{F^c_D}+k_F)}& \alpha^l_{A}Y_{\mathbf{3}^lA,1}^{(k_{F^c_D}+k_F)}& \alpha^l_{A}Y_{\mathbf{3}^lA,3}^{(k_{F^c_D}+k_F)}\\
(-1)^{j+l}\alpha^l_{A}Y_{\mathbf{3}^lA,3}^{(k_{F^c_D}+k_F)}& (-1)^{j+l}\alpha^l_{A}Y_{\mathbf{3}^lA,2}^{(k_{F^c_D}+k_F)}& (-1)^{j+l}\alpha^l_{A}Y_{\mathbf{3}^lA,1}^{(k_{F^c_D}+k_F)}\\
\beta_BY_{\mathbf{3}^{< i+j >}B,1}^{(k_{F^c_3}+k_F)}&\beta_BY_{\mathbf{3}^{< i+j >}B,3}^{(k_{F^c_3}+k_F)}&\beta_BY_{\mathbf{3}^{< i+j >}B,2}^{(k_{F^c_3}+k_F)}
\end{matrix}\right)v_{u/d}\label{eq:2+1 and 3}\,.
\end{equation}
In some cases, the rank of $M_F$ is less than three due to the structure of the modular forms. For instance, the rank of the mass matrix is two for the assignment $(k_{F^c_D}+k_{F},k_{F^c_3}+k_{F})=(2,4)$.
\item{$F^c\sim\mathbf{3}^i\,,~~F \sim\mathbf{2}\oplus\mathbf{1}^j\,.$}

We interchange the representation assignments of the left-handed and the right-handed fields discussed in above. The left-handed fermions are assigned to the direct sum of a doublet $F_D=(F_1,F_2)\sim\mathbf{2}$ and a singlet $F_3\sim\mathbf{1}^j$, while the right-handed fermions $F^c=(F^c_1,F^c_2,F^c_3)$ transform as a triplet under $S_4$.
The modular weights of these fields are denoted as $k_{F^c}$, $k_{F_D}$ and $k_{F_3}$.
Then we can straightforwardly read out the Yukawa superpotential for this kind of assignment,
\begin{equation}
\begin{aligned}
\mathcal{W}_F=&{}[\alpha(F^cF_Df_{F_D}(Y))_{\bf{1}}+\beta(F^cF_3f_{F_3}(Y))_{\bf{1}}]H_{u/d}\\
=&\sum_{a=1}^3\sum_{b=1}^2\sum_{l=0}^1\sum_A(-1)^{(b+1)(i+l)}\alpha^l_{A}F_bF^c_aY_{\mathbf{3}^{l}A,\prec 2+b-a \succ}^{(k_{F^c}+k_{F_D})}
H_{u/d}\\
&+\sum_{a=1}^3\sum_B\beta_{B}F_3F^c_aY_{\mathbf{3}^{< i+j >}B,\prec 2-a \succ}^{(k_{F^c}+k_{F_3})}H_{u/d}\,.
\end{aligned}
\end{equation}
The resulting mass matrix is given by
\begin{equation}
M_F=\left(\begin{matrix}
\alpha^l_{A}Y_{\mathbf{3}^lA,2}^{(k_{F^c}+k_{F_D})}&(-1)^{i+l}\alpha^l_{A}Y_{\mathbf{3}^lA,3}^{(k_{F^c}+k_{F_D})} &\beta_BY_{\mathbf{3}^{< i+j >}B,1}^{(k_{F^c}+k_{F_3})}\\
\alpha^l_{A}Y_{\mathbf{3}^lA,1}^{(k_{F^c}+k_{F_D})}&(-1)^{i+l}\alpha^l_{A}Y_{\mathbf{3}^lA,2}^{(k_{F^c}+k_{F_D})} &\beta_BY_{\mathbf{3}^{< i+j >}B,3}^{(k_{F^c}+k_{F_3})}\\
\alpha^l_{A}Y_{\mathbf{3}^lA,3}^{(k_{F^c}+k_{F_D})}&(-1)^{i+l}\alpha^l_{A}Y_{\mathbf{3}^lA,1}^{(k_{F^c}+k_{F_D})} &\beta_BY_{\mathbf{3}^{< i+j >}B,2}^{(k_{F^c}+k_{F_3})}
\end{matrix}\right)v_{u/d}\label{eq:3 and 2+1}\,,
\end{equation}
which is the transpose of the mass matrix in Eq.~\eqref{eq:2+1 and 3} with the indices $i$ and $j$ exchanged.

\item{$F^c\sim\mathbf{2}\oplus\mathbf{1}^i\,,~~F\sim\mathbf{2}\oplus\mathbf{1}^j\,.$}

In this case, both left-handed and right-handed fields are assigned to the direct sum of $S_4$ doublet and singlet. We denote $F_D=(F_1,F_2)$, $F^c_D=(F^c_1,F^c_2)$ which transform as doublet under $S_4$ while $F_3$, $F^c_3$ are singlets. The modular weights of these fields are $k_{F^c_D}$, $k_{F^c_3}$, $k_{F_D}$ and $k_{F_3}$.
Then the Yukawa superpotential is given by
\begin{equation}
\begin{aligned}
W_F=&{}[\alpha(F^c_DF_Df_{DD}(Y))_{\bf{1}}+\beta(F^c_DF_3f_{D3}(Y))_{\bf{1}}+\gamma(F^c_3F_Df_{3D}(Y))_{\bf{1}}+\delta (F^c_3F_3f_{33}(Y))_{\mathbf{1}}]H_{u/d}\\
=&\Big[\alpha^l_{1}(F^c_1F_2+(-1)^lF^c_2F_1)Y_{\mathbf{1}^l}^{(k_{F^c_D}+k_{F_D})}+\alpha_{2A}(F^c_1F_1Y_{\mathbf{2}A,1}^{(k_{F^c_D}+k_{F_D})}+F^c_2F_2Y_{\mathbf{2}A,2}^{(k_{F^c_D}+k_{F_D})})\\
&+\beta_B F_3(F^c_1Y_{\mathbf{2}B,2}^{(k_{F^c_D}+k_{F_3})}+(-1)^jF^c_2Y_{\mathbf{2}B,1}^{(k_{F^c_D}+k_{F_3})}) +\gamma_C F^c_3(F_1Y_{\mathbf{2}C,2}^{(k_{F^c_3}+k_{F_D})}\\
&+(-1)^iF_2Y_{\mathbf{2}C,1}^{(k_{F^c_3}+k_{F_D})})
+\delta F^c_3F_3Y_{\mathbf{1}^{< i+j >}}^{(k_{F^c_3}+k_{F_3})}\Big]H_{u/d}\,,
\end{aligned}
\end{equation}
which leads to
\begin{equation}
M_F=\left(\begin{matrix}
\alpha_{2A}Y_{\mathbf{2}A,1}^{(k_{F^c_D}+k_{F_D})}&\alpha^l_{1}Y_{\mathbf{1}^l}^{(k_{F^c_D}+k_{F_D})}&\beta_B Y_{\mathbf{2}B,2}^{(k_{F^c_D}+k_{F_3})}\\
(-1)^l\alpha^l_{1}Y_{\mathbf{1}^l}^{(k_{F^c_D}+k_{F_D})}&\alpha_{2A}Y_{\mathbf{2}A,2}^{(k_{F^c_D}+k_{F_D})}&(-1)^j\beta_B Y_{\mathbf{2}B,1}^{(k_{F^c_D}+k_{F_3})}\\
\gamma_CY_{\mathbf{2}C,2}^{(k_{F^c_3}+k_{F_D})}&(-1)^i\gamma_CY_{\mathbf{2}C,1}^{(k_{F^c_3}+k_{F_D})} &\delta Y_{\mathbf{1}^{< i+j >}}^{(k_{F^c_3}+k_{F_3})}
\end{matrix}\right)v_{u/d}\label{eq:2+1 and 2+1}\,.
\end{equation}

The above mass matrix can be divided into four parts
\begin{equation}
M_F=\left(\begin{matrix}
M_{DD} & M_{D3} \\
M_{3D} & M_{33} \end{matrix}\right)\,.
\end{equation}
Let us firstly consider the (33) entry $M_{33}$ which involves the modular forms in the singlet representations of $S_4$. From table~\ref{Tab:LeveL4_MM}, we see that there are only four singlet modular forms $Y^{(4)}_{\mathbf{1}}$, $Y^{(6)}_{\mathbf{1}}$, $Y^{(6)}_{\mathbf{1}'}$ and $Y^{(8)}_{\mathbf{1}}$ up to weight 8. Hence $M_{33}$ would be vanishing if the following conditions are fulfilled
\begin{equation}
\begin{aligned}
&k_{F_3}+k_{F^c_3}<0\,,\\
\text{ or }~~&{}k_{F_3}+k_{F^c_3}=2\,,~
\begin{cases}
\rho_{F_3}=\mathbf{1}\,,\rho_{F^c_3}=\mathbf{1}\\
\rho_{F_3}=\mathbf{1}'\,,\rho_{F^c_3}=\mathbf{1}'
\end{cases}\\
\text{ or }~~&k_{F_3}+k_{F^c_3}\neq 6~\,,
\begin{cases}
\rho_{F_3}=\mathbf{1}'\,,\rho_{F^c_3}=\mathbf{1}\\
\rho_{F_3}=\mathbf{1}\,,\rho_{F^c_3}=\mathbf{1}'
\end{cases}\,.
\end{aligned}
\end{equation}
Notice that odd weight $k_{F_3}+k_{F^c_3}=1,3,5, \ldots$ can also lead to vanishing $M_{33}$, but the rank of $M_F$ would be less than three so that at least one mass eigenvalue is zero. Then we proceed to consider the $M_{3D}$ block consisted of the $(31)$ and $(32)$ entries, it would be vanishing if the modular weights fulfill $k_{F^c_3}+k_{F_D}\leq0$ or $k_{F^c_3}+k_{F_D}=1,3,5, \ldots$. For the case of odd modular weight $k_{F^c_3}+k_{F_D}=1,3,5, \ldots$, some mixing angles or masses are vanishing\footnote{Non-zero fermion masses requires that $k_{F^c_3}+k_{F_3}$ and $k_{F^c_D}+k_{F_D}$ are even while $k_{F^c_D}+k_{F_3}$ is odd in this case. As a result, both up and down quark (charged lepton and neutrino) mass matrices are block diagonal simultaneously such that some mixing angles are vanishing.}. As regards the $M_{D3}$ block consisted of the $(13)$ and $(23)$ entries, it would be vanishing if the modular weight $k_{F^c_D}+k_{F_3}$ is non-positive or odd. However, odd $k_{F^c_D}+k_{F_3}$ leads to vanishing fermion masses or mixing angles. Although either $M_{3D}$ or $M_{D3}$ can be vanishing, they can not be vanishing simultaneously otherwise some masses or mixing angles are constrained to be zero.

\item{$F^c\sim\mathbf{1}^{i_1}\oplus\mathbf{1}^{i_2}\oplus\mathbf{1}^{i_3}\,,~~F\sim\mathbf{2}\oplus\mathbf{1}^j\,.$}

Analogous to previous cases, we find the Yukawa superpotential takes the following form,
\begin{eqnarray}
\nonumber\mathcal{W}_F&=&[\alpha(F^c_1F_Df_{1D}(Y))_{\bf{1}}+\beta(F^c_2F_Df_{2D}(Y))_{\bf{1}}+\gamma(F^c_3F_Df_{3D}(Y))_{\bf{1}}\\
\nonumber&&+\delta_1 (F^c_1F_3f_{13}(Y))_{\mathbf{1}}+\delta_2 (F^c_2F_3f_{23}(Y))_{\mathbf{1}}+\delta_3 (F^c_3F_3f_{33}(Y))_{\mathbf{1}}]H_{u/d}\\
\nonumber&=&\Big[\alpha_A F^c_1(F_1Y_{\mathbf{2}A,2}^{(k_{F^c_1}+k_{F_D})}+(-1)^{i_1}F_2Y_{\mathbf{2}A,1}^{(k_{F^c_1}+k_{F_D})})+\delta_1 F^c_1F_3Y_{\mathbf{1}^{<i_1+j>}}^{(k_{F^c_1}+k_{F_3})}\\
\nonumber&&+\beta_B F^c_2(F_1Y_{\mathbf{2}B,2}^{(k_{F^c_2}+k_{F_D})}+(-1)^{i_2}F_2Y_{\mathbf{2}B,1}^{(k_{F^c_2}+k_{F_D})})+\delta_2 F^c_2F_3Y_{\mathbf{1}^{<i_2+j>}}^{(k_{F^c_2}+k_{F_3})}\\
&&+\gamma_C F^c_3(F_1Y_{\mathbf{2}C,2}^{(k_{F^c_3}+k_{F_D})}+(-1)^{i_3}F_2Y_{\mathbf{2}C,1}^{(k_{F^c_3}+k_{F_D})})+\delta_3 F^c_3F_3Y_{\mathbf{1}^{<i_3+j>}}^{(k_{F^c_3}+k_{F_3})}\Big]H_{u/d}\,.
\end{eqnarray}
The fermion mass matrix is determined to be
\begin{equation}
M_F=\left(\begin{matrix}
\alpha_A Y_{\mathbf{2}A,2}^{(k_{F^c_1}+k_{F_D})}&(-1)^{i_1}\alpha_A Y_{\mathbf{2}A,1}^{(k_{F^c_1}+k_{F_D})} &\delta_1 Y_{\mathbf{1}^{<i_1+j>}}^{(k_{F^c_1}+k_{F_3})}\\
\beta_B Y_{\mathbf{2}B,2}^{(k_{F^c_2}+k_{F_D})}&(-1)^{i_2}\beta_B Y_{\mathbf{2}B,1}^{(k_{F^c_2}+k_{F_D})} &\delta_2 Y_{\mathbf{1}^{<i_2+j>}}^{(k_{F^c_2}+k_{F_3})}\\
\gamma_CY_{\mathbf{2}C,2}^{(k_{F^c_3}+k_{F_D})}&(-1)^{i_3}\gamma_CY_{\mathbf{2}C,1}^{(k_{F^c_3}+k_{F_D})} &\delta_3 Y_{\mathbf{1}^{<i_3+j>}}^{(k_{F^c_3}+k_{F_3})}
\end{matrix}\right)v_{u/d}\label{eq:1+1+1 and 2+1}\,.
\end{equation}

In above, we don't consider the singlet assignment for the left-handed fields $F$, because generally more free coupling constants would be necessary in the resulting quark and lepton models, and we are mainly concerned with the models with small number of free parameters in the present work.

\end{itemize}

\subsection{\label{sec:Majorana-mass} Classifying the Majorana mass terms }
\paragraph{}
In this subsection, we explore the superpotential for the Majorana mass terms, which can be written as
\begin{equation}
\mathcal{W}_{F^c}=\Lambda(F^cF^c f(Y))_{\bf{1}}\,,
\end{equation}
where $\Lambda$ is the characteristic scale of flavor dynamics, and all independent invariant singlets should be included. The function $f(Y)$ refers to the modular multiplets to ensure modular invariance, and it is fixed by the modular weight and representation of $F^c$.

\begin{itemize}[labelindent=-0.8em, leftmargin=1.5em]

\item{$F^c\sim\mathbf{3}^k\,.$}

As shown in the Appendix~\ref{sec:S4_group}, the contraction $\mathbf{3}^i\times\mathbf{3}^i\rightarrow\mathbf{3}$ is antisymmetric. Thus the triplet modular forms transforming as $\mathbf{3}$ don't contribute to the Majorana mass terms of $F^c$. The superpotential $\mathcal{W}_{F^c}$ reads as
\begin{equation}
\begin{aligned}
\mathcal{W}_{F^c}
=&{}\left[\alpha\left((F^cF^c)_{\mathbf{1}}Y_{\mathbf{1}}^{(2k_{F^c})}\right)_{\mathbf{1}}+\sum_A \beta_{A}\left((F^cF)_{\mathbf{2}}Y_{\mathbf{2}A}^{(2k_{F^c})}\right)_{\mathbf{1}}
+\sum_{B} \gamma'_B\left((F^cF^c)_{\mathbf{3}'}Y_{\mathbf{3}'B}^{(2k_{F^c})}\right)_{\mathbf{1}}\right]\Lambda\\
=&\sum_{a=1}^3\sum_{b=1}^3 \Lambda F^c_aF^c_b\{\alpha Y_{\mathbf{1},\prec a+b-1\succ}^{(2k_{F^c})}+\sum_A \beta_{A}Y_{\mathbf{2}A,\prec a+b-2\succ}^{(2k_{F^c})}+\sum_B \gamma'_{B}Y_{\mathbf{3}' A,\prec 3-a-b\succ}^{(2k_{F^c})}(3\delta_{ab}-1)\}\,.
\end{aligned}
\end{equation}
The Majorana mass matrix of $F^c$ is symmetric
\begin{equation}
M_{F^c}=\left(\begin{matrix}
\alpha Y_{\mathbf{1}}^{(2k_{F^c})}+2\gamma'_{B}Y_{\mathbf{3}'B,1}^{(2k_{F^c})}&\beta_{A}Y_{\mathbf{2}A,1}^{(2k_{F^c})}-\gamma'_{B}Y_{\mathbf{3}'B,3}^{(2k_{F^c})} &\beta_{A}Y_{\mathbf{2}A,2}^{(2k_{F^c})}-\gamma'_{B}Y_{\mathbf{3}'B,2}^{(2k_{F^c})}\\
\beta_{A}Y_{\mathbf{2}A,1}^{(2k_{F^c})}-\gamma'_{B}Y_{\mathbf{3}'B,3}^{(2k_{F^c})} &\beta_{A}Y_{\mathbf{2}A,2}^{(2k_{F^c})}+2\gamma'_{B}Y_{\mathbf{3}'B,2}^{(2k_{F^c})} &\alpha Y_{\mathbf{1}}^{(2k_{F^c})}-\gamma'_{B}Y_{\mathbf{3}'B,1}^{(2k_{F^c})}\\
\beta_{A}Y_{\mathbf{2}A,2}^{(2k_{F^c})}-\gamma'_{B}Y_{\mathbf{3}'B,2}^{(2k_{F^c})} &\alpha Y_{\mathbf{1}}^{(2k_{F^c})}-\gamma'_{B}Y_{\mathbf{3}'B,1}^{(2k_{F^c})} &\beta_{A}Y_{\mathbf{2}A,1}^{(2k_{F^c})}+2\gamma'_{B}Y_{\mathbf{3}'B,3}^{(2k_{F^c})}
\end{matrix}\right)\Lambda
\label{eq:Majorana 3}\,.
\end{equation}

\item{$F^c \sim\mathbf{2}\oplus\mathbf{1}^i\,.$}

The contraction $\mathbf{2}\times\mathbf{2}\rightarrow\mathbf{1}'$ is antisymmetric and consequently it has no contribution to mass matrix. The general superpotential for the Majorana mass is
\begin{equation}
\begin{aligned}
\mathcal{W}_{F^c}=&{}[\alpha(F^c_DF^c_Df_{DD}(Y))_{\bf{1}}+2\beta(F^c_DF^c_3f_{D3}(Y))_{\bf{1}}+\gamma (F^c_3F^c_3f_{33}(Y))_{\mathbf{1}}]\Lambda\\
=&\Big[2\alpha_{1}F^c_1F^c_2Y_{\mathbf{1}^l}^{(2k_{F^c_D})}+\alpha_{2A}(F^c_1F^c_1Y_{\mathbf{2}A,1}^{(2k_{F^c_D})}+F^c_2F^c_2Y_{\mathbf{2}A,2}^{(2k_{F^c_D})})\\
&+2\beta_B F^c_3(F^c_1Y_{\mathbf{2}B,2}^{(k_{F^c_D}+k_{F^c_3})}+(-1)^iF^c_2Y_{\mathbf{2}B,1}^{(k_{F^c_D}+k_{F^c_3})}) +\gamma  F^c_3F^c_3Y_{\mathbf{1}}^{(2k_{F^c_3})}\Big]\Lambda\,,
\end{aligned}
\end{equation}
which gives rise to the following mass matrix,
\begin{equation}
M_{F^c}=\left(\begin{matrix}
\alpha_{2A}Y_{\mathbf{2}A,1}^{(2k_{F^c_D})}&\alpha_{1}Y_{\mathbf{1}}^{(2k_{F^c_D})}&\beta_B Y_{\mathbf{2}B,2}^{(k_{F^c_D}+k_{F^c_3})}\\
\alpha_{1}Y_{\mathbf{1}}^{(2k_{F^c_D})}&\alpha_{2A}Y_{\mathbf{2}A,2}^{(2k_{F^c_D})}&(-1)^i\beta_B Y_{\mathbf{2}B,1}^{(k_{F^c_D}+k_{F^c_3})}\\
\beta_BY_{\mathbf{2}B,2}^{(k_{F^c_3}+k_{F^c_D})}&(-1)^i\beta_BY_{\mathbf{2}B,1}^{(k_{F^c_3}+k_{F^c_D})} &\gamma Y_{\mathbf{1}}^{(2k_{F^c_3})}
\end{matrix}\right)\Lambda\label{eq:Majorana 2+1}\,.
\end{equation}

\item{$F^c \sim\mathbf{1}^{i_1}\oplus\mathbf{1}^{i_2}\oplus\mathbf{1}^{i_3}\,.$}

In the same fashion, we can read out the most general Majorana mass terms for the singlet assignment of $F^c$,
\begin{equation}
\begin{aligned}
\mathcal{W}_{F^c}&{}=\sum_{a=1}^3\sum_{b=1}^3 \Lambda \alpha_{ab}F^c_aF^c_bf_{ab}(Y) \\
&=\sum_{a=1}^3\Lambda \alpha_{aa}F^c_aF^c_aY_{\mathbf{1}}^{(2k_{F^c_a})}+2\sum_{1\leq a<b\leq3}\Lambda \alpha_{ab}F^c_aF^c_bY_{\mathbf{1}^{<i_a+i_b>}}^{(k_{F^c_a}+k_{F^c_b})}\,,
\end{aligned}
\end{equation}
and the mass matrix is
\begin{equation}
M_{F^c}=\left(\begin{matrix}
\alpha_{11}Y_{\mathbf{1}}^{(2k_{F^c_1})} ~&\alpha_{12}Y_{\mathbf{1}^{<i_1+i_2>}}^{(k_{F^c_1}+k_{F^c_2})} ~&\alpha_{13}Y_{\mathbf{1}^{<i_1+i_3>}}^{(k_{F^c_1}+k_{F^c_3})} \\
\alpha_{12}Y_{\mathbf{1}^{<i_1+i_2>}}^{(k_{F^c_1}+k_{F^c_2})} ~& \alpha_{22}Y_{\mathbf{1}}^{(2k_{F^c_2})} ~&\alpha_{23}Y_{\mathbf{1}^{<i_2+i_3>}}^{(k_{F^c_2}+k_{F^c_3})} \\
\alpha_{13}Y_{\mathbf{1}^{<i_1+i_3>}}^{(k_{F^c_1}+k_{F^c_3})} ~&\alpha_{23}Y_{\mathbf{1}^{<i_2+i_3>}}^{(k_{F^c_2}+k_{F^c_3})} ~&\alpha_{33}Y_{\mathbf{1}}^{(2k_{F^c_3})}
\end{matrix}\right)\Lambda\,.
\end{equation}
If the three generations of $F^c$ all transform as singlets under $S_4$, the Lagrangian would be less constrained by modular symmetry and consequently more free parameters would be introduced in the Yukawa coupling and the Majorana mass term.

\end{itemize}

\subsection{Equivalence of different assignments }
\paragraph{}

The possible quark models with $S_4$ modular symmetry can be obtained by combining the possible forms of the up quarks and down quarks Yukawa couplings discussed in previous section~\ref{sec:Majorana-mass}.
Similarly the possible lepton models can be obtained for Dirac neutrinos, and the Majorana mass terms of the right-handed neutrinos should be considered as well if neutrinos are Majorana particles. In the present work, the left-handed quark and lepton fields are assumed to transform as a triplet or the direct sum of doublet and singlet under $S_4$, all the three possible assignments: triplet, doublet plus singlet and three singlets for the right-handed quark and lepton fields would be considered. It is notable that different assignments can lead to the same predictions for fermion masses and mixing matrix. For instance, if both left-handed leptons $F$ and right-handed charged leptons $F^c$ transform as $S_4$ triplets $F^c\sim\mathbf{3}^{i}$ and $F\sim\mathbf{3}^{j}$, the mass matrix can be read from Eq.~\eqref{eq:3 and 3} for any given modular weights. It can be easily seen that the representation assignment $F\sim\mathbf{3}^{<i+1>}\,,~F^c\sim\mathbf{3}^{<j+1>}$ gives the same charged lepton mass matrix.

Two different kinds of representation assignments can also give mass matrices related by phase transformations, as summarized in table~\ref{tab:mass-trans}. As an example, let us consider the case $L_D=(L_1, L_2)^{T}\sim\mathbf{2}$, $L_3\sim\mathbf{1}^{i_3}$, $E^c_a\sim\mathbf{1}^{j_a}$ with $a=1, 2, 3$, the general form of the charged lepton mass matrix is given by Eq.~\eqref{eq:1+1+1 and 2+1}. If we change the representation assignment
\begin{equation}
L_3:\mathbf{1}^{i_3}\rightarrow\mathbf{1}^{<i_3+1>}\,,~E^c_a:\mathbf{1}^{j_a}\rightarrow \mathbf{1}^{<j_a+1>}\,,
\end{equation}
the charged lepton mass matrix would turn into
\begin{equation}
M_e\rightarrow M_e\mathrm{diag}\{1,-1,1\}\,.
\end{equation}
Analogously changing the representation of the right-handed neutrinos, the light neutrino mass matrix would change as
\begin{eqnarray}
\nonumber&&\text{Dirac~neutrinos}:\left\{
\begin{array}{ll}
M_{\nu}\rightarrow M_{\nu}\mathrm{diag}\{1,-1,1\}  &  \text{for}~ N^c\sim\mathbf{3}^k~\text{or}
~N^c_a\sim\mathbf{1}^{k_a}\,,\\[0.1in]
M_{\nu}\rightarrow \mathrm{diag}\{1,-1,1\}M_{\nu}\mathrm{diag}\{1,-1,1\}  &  \text{for}~ N^c\sim\mathbf{2}\oplus\mathbf{1}^k\,,
\end{array}
\right.\\[0.1in]
&&\text{Majorana~neutrinos}: M_{\nu}\rightarrow \mathrm{diag}\{1,-1,1\}M_{\nu}\mathrm{diag}\{1,-1,1\}\,.
\end{eqnarray}
The phase matrix $\mathrm{diag}\{1,-1,1\}$ can be absorbed into the lepton fields, the lepton masses and mixing parameters are left invariant.
As a consequence, without loss of generality, we can take $F\sim\mathbf{3}$ for the triplet assignment of the left-handed fields and $F\sim\mathbf{2}\oplus\mathbf{1}$ for the doublet plus singlet assignment.

Since the signal neutrinoless double beta decay has not been observed, the nature of neutrinos is still unknown. We shall consider both Majorana and Dirac neutrinos in this work. The light neutrino masses are generated by the type-I seesaw mechanism for Majorana neutrinos, and the light neutrino mass matrix is given by the seesaw formula $M_{\nu}=-M_D^TM_{N^c}^{-1}M_D$, where $M_D$ and $M_{N^c}$ are Dirac mass matrix and the Majorana mass matrix of the right-handed neutrinos respectively. For Dirac neutrinos,  additional symmetry is generally necessary to forbid the right-handed neutrino Majorana mass term and it is usually taken to be $U(1)_L$ lepton number. In the context of modular invariance approach, the Majorana mass terms of the right-handed neutrino can be naturally forbidden by taking the modular weights of right-handed neutrinos $N^c$ to be negative integers, because there are no modular forms of negative weight.

\begin{table}
\centering
\resizebox{\textwidth}{!}{
\begin{tabular}{|c|c|c||c|}\hline \hline
\multirow{2}{*}{\diagbox[width=5.82cm,trim=l]{$~~~~~~~~~~~~~~~~F^c$}{$F~~~~~~~~~~~~~~$}} &\multirow{2}{*}{$\mathbf{3}^j\rightarrow\mathbf{3}^{<j+1>}$} &\multirow{2}{*}{$\mathbf{2}\oplus\mathbf{1}^{j}\rightarrow\mathbf{2}\oplus\mathbf{1}^{<j+1>}$} &\multirow{2}{*}{Majorana mass matrix} \\ & & & \\ \hline
\multirow{2}{*}{$\mathbf{3}^i\rightarrow\mathbf{3}^{<i+1>}$} &\multirow{2}{*}{$M_F\rightarrow M_F$} &\multirow{2}{*}{$M_F\rightarrow M_FP$} &\multirow{2}{*}{$M_{F^c}\rightarrow M_{F^c}$ }\\ & & & \\ \hline
\multirow{2}{*}{$\mathbf{2}\oplus\mathbf{1}^i\rightarrow\mathbf{2}\oplus\mathbf{1}^{<i+1>}$} &\multirow{2}{*}{$M_F\rightarrow PM_F$} &\multirow{2}{*}{$M_F\rightarrow PM_FP$} &\multirow{2}{*}{$M_{F^c}\rightarrow PM_{F^c}P$} \\ & & & \\ \hline
$\mathbf{1}^{i_1}\oplus\mathbf{1}^{i_2}\oplus\mathbf{1}^{i_3}$ &\multirow{2}{*}{$M_F\rightarrow M_F$} &\multirow{2}{*}{$M_F\rightarrow M_FP$} &\multirow{2}{*}{$M_{F^c}\rightarrow M_{F^c}$} \\ $\rightarrow\mathbf{1}^{<i_1+1>}\oplus\mathbf{1}^{<i_2+1>}\oplus\mathbf{1}^{<i_3+1>}$ & & & \\ \hline \hline
\end{tabular}}
\caption{Transformation of the fermion mass matrix under changing the representation assignments of matter fields, and $P$ is the diagonal matrix $\mathrm{diag}\{1,-1,1\}$. In the case that both $F$ and $F^c$ are assigned to direct sum of doublet and singlet under $S_4$, the couplings $\alpha_{1}^l$ associated with the operators $(F^c_DF_D)_{\mathbf{1}^l}$ should be transformed into $-\alpha^l_{1}$, and the coupling $\alpha_1$ associated with the operator $(F^c_DF^c_D)_{\mathbf{1}}$ should also change to $-\alpha_1$.
\label{tab:mass-trans}}
\end{table}

\section{\label{sec:numerical results}Phenomenologically viable models and numerical results}

From the general analytical expressions of the mass matrix for different representation of matter fields, we can straightforwardly obtain the possible lepton and quark models based on $S_4$ modular symmetry. In this work, we are interested in the models with small number of free parameters: lepton models with less than 9 free parameters and quark models with less than 11 free parameters. For each model, we perform a conventional $\chi^2$ analysis and we use the well-known package \texttt{TMinuit} to numerically search for minimum of the $\chi^2$ function and determine the best values of the input parameters. Then we evaluate masses and mixing parameters of quarks and leptons at the best fit points, and determine whether they are within the experimentally allowed $3\sigma$ regions.
The overall scale factor of the mass matrix can be adjusted to reproduce any one of the mass eigenvalues. For instance, the overall factors of the charged lepton, up type quark and down type quark mass matrices are fixed by the measured values of the electron, top quark and down quark masses respectively. The overall scale of the neutrino mass matrix is determined by the solar neutrino mass square difference $\Delta m^2_{21}$. We scan over the parameter space of the models, the ratios of coupling coefficients are taken as random numbers whose absolute values freely vary in the range of $[0,10^5]$. Moreover, the vacuum expectation value (VEV) of the complex modulus $\tau$ is also treated as a free parameter to optimize the agreement between predictions and experimental data. Since each point of $\tau$ in the complex upper-half plane can be mapped into the fundamental domain $\mathcal{D}$ given in Eq.~\eqref{eq:fundamental-domain} by a modular transformation, thus it is sufficient to limit the modulus VEV $\langle\tau\rangle$ in the fundamental domain $\mathcal{D}$. Under the CP transformation $\tau\rightarrow-\tau^{*}$, gCP invariance implies that the fermion mass matrix becomes $M_{F}(-\tau^{*})=\rho^{*}_{F^c}(S)M^{*}_{F}(\tau)\rho^{\dagger}_{F}(S)$ for the charged fermion and $M_{F^c}(-\tau^{*})=\rho^{*}_{F^c}(S)M^{*}_{F^c}(\tau)\rho^{\dagger}_{F^c}(S)$ for the Majorana mass matrix of $F^c$~\cite{Ding:2021iqp}. Therefore at the CP dual point $\tau\rightarrow-\tau^{*}$, the predictions for fermion masses and mixing angles are left unchanged while the signs of all CP violation phases are flipped.

We use the fermion mass ratios, mixing angles and CP violation phases to construct the $\chi^2$ function, the experimental data of the leptons and quarks are summarized in table~\ref{tab:lepton_data}.
The data of the lepton mixing parameters are taken from the latest global fit of NuFIT v5.0 including the atmospheric neutrino data from Super-Kamiokande~\cite{Esteban:2018azc} and the neutrino mass spectrum is taken to be normal ordering for illustration. The charged lepton mass ratios are taken from~\cite{Ross:2007az}, and the quark mixing parameters and mass ratios are adopted from~\cite{Antusch:2013jca}, and they are calculated at the GUT scale $M_{\text{GUT}}=2\times 10^{16}$ GeV in a minimal SUSY breaking scenario, with SUSY breaking scale $M_{\text{SUSY}} = 1$ TeV and $\tan \beta = 7.5, \bar{\eta}_b=0.09375$. The leptonic Dirac CP phase $\delta^{l}_{CP}$ has not been accurately measured, therefore we don't include the contribution of $\delta^{l}_{CP}$ in the $\chi^2$ function. If all observables at the best fit point of a model are compatible with the experimental data at $3\sigma$ level, this model would be regarded as phenomenologically viable. In the following, we report the fitting results of the viable models with the minimal number of free parameters, and all numerical results are shown with six significant digits. Notice that $\mathrm{Re}\langle\tau\rangle$ and all CP violation phases flipped their signs while all other observables and free parameters are unchanged at the CP dual point.

\begin{table}[t!]
\centering
\begin{tabular}{| c | c || c | c|} \hline \hline
\multicolumn{2}{|c||}{Leptons} & \multicolumn{2}{c|}{Quarks} \\ \hline
Observables & Central value and $1\sigma$ error &  Observables & Central value and $1\sigma$ error\\ \hline
$m_e/m_\mu $ & $0.0048 \pm 0.0002$ & $m_u/m_c$ & $(1.9286 \pm 0.6017)\times 10^{-3}$ \\
$m_\mu/m_\tau$ & $0.0565 \pm 0.0045$ & $m_c/m_t$ & $(2.7247 \pm 0.1200)\times 10^{-3}$\\
$\Delta m_{21}^2 / 10^{-5}\text{eV}^2$ & $7.42^{+0.21}_{-0.20}$ & $m_d/m_s$ & $(5.0528 \pm 0.6192)\times 10^{-2}$\\
$\Delta m_{31}^2 / 10^{-3}\text{eV}^2$ & $2.517^{+0.026}_{-0.028}$ & $m_s/m_b$ & $(1.7684 \pm 0.0975)\times 10^{-2}$ \\ \hline
$\delta^{l}_{CP}/\pi$ & $1.0944^{+0.1500}_{-0.1333}$ & $\delta^q_{CP}$ & $69.213^\circ \pm 3.115^\circ$\\
$\sin^2\theta^{l}_{12}$ & $0.304^{+0.012}_{-0.012}$ & $\theta^q_{12}$ & $0.22736\pm 0.00073$ \\
$\sin^2\theta^{l}_{13}$ & $0.02219^{+0.00062}_{-0.00063}$ & $\theta^q_ {13}$ & $0.00338\pm 0.00012$ \\
$\sin^2\theta^{l}_{23}$ & $0.573^{+0.016}_{-0.020}$ & $\theta^q_{23}$ & $0.03888\pm 0.00062$ \\ \hline \hline
\end{tabular}
\caption{\label{tab:lepton_data}The central values and the $1\sigma$ errors of the mass ratios and mixing angles and CP violation phases in lepton and quark sectors. The charged lepton mass ratios averaged over $\tan\beta$ are taken from~\cite{Ross:2007az,Feruglio:2017spp}, and we adopt the values of the lepton mixing parameters from NuFIT v5.0 with Super-Kamiokanda atmospheric data for normal ordering~\cite{Esteban:2020cvm}. The data of quark mass ratios and mixing parameters are taken from ~\cite{Antusch:2013jca} with the SUSY breaking scale $M_{\text{SUSY}}=1$ TeV and $\tan\beta=7.5, \bar{\eta}_b=0.09375$.
}
\end{table}

\subsection{Lepton models}

For Majorana neutrinos, the minimal phenomenologically viable models only depend on five free parameters besides the complex modulus $\tau$, and we find nine such models labelled as L1$\sim$L9. Notice that the model L2 was firstly presented in~\cite{Novichkov:2019sqv}.
The $S_4$ representation and modular weights of the lepton fields in each models are listed in table~\ref{tab:lepton models}. The best fit values of the coupling constants and the corresponding predictions for the lepton masses and mixing parameters are summarized in table~\ref{tab:lepton bset-fit}.
Although we have considered the minimal seesaw model with two right-handed neutrinos, three right-handed neutrinos are involved in these minimal models. It turns out that more free parameters are needed to accommodate the experimental data in the modular models with two right-handed neutrinos.
In most modular symmetry models, both left-handed leptons $L$ and right-handed neutrinos $N^c$ are assumed to transform as a triplet under the finite modular group while the right-handed charged leptons $E^c$ are singlets. Our models L2, L3, L4 and L5 belong to this category. It is notable that we find new possible assignments here. All the lepton fields $L$, $E^c$ and $N^c$ are $S_4$ triplet in the model L1. Both $L$ and $N^c$ transform as triplet $\mathbf{3}$ or $\mathbf{3}'$ under $S_4$ while the right-handed charged leptons are in the reducible representation $\mathbf{2}\oplus\mathbf{1}'$ in the models L6 and L7. Furthermore, both $L$ and $E^c$ are assigned to the direct sum of the doublet and singlet of $S_4$ in the models L8 and L9. From table~\ref{tab:lepton bset-fit}, we can see that all these models can accommodate the experimental data very well, the atmospheric mixing angle $\theta_{23}$ is predicted to be in the second octant. The Dirac CP violation phase $\delta^{l}_{CP}$ is determined to be sizable in these models, and it distributes in the range of $[1.27\pi, 1.65\pi]$. The upcoming generation of long-baseline neutrino oscillation experiments such as DUNE~\cite{Acciarri:2016crz,Acciarri:2015uup,Strait:2016mof,Acciarri:2016ooe}
and Hyper-Kamiokande~\cite{Abe:2016ero} can significantly improve the sensitivity to $\theta_{23}$ and $\delta^{l}_{CP}$. It is expected that a $5\sigma$ discovery of CP violation can be reached after ten years of data taking over $50\%$ of the parameter space. Thus our predictions for $\theta_{23}$ and $\delta^{l}_{CP}$ can be tested  in near future.

The neutrino mass scale can be probed from direct kinematic searches, neutrinoless double decay and cosmology. The cosmological observation is sensitive to the sum of light neutrino masses $\sum m_i$, and the most stringent bound is $\sum m_{i}<0.12$ eV at $95\%$ confidence level from the Planck Collaboration~\cite{Aghanim:2018eyx}. All the minimal models satisfy this bound except L8 and L9 which give $\sum m_i\simeq 121$ meV very close to the upper limit. Notice that the cosmological bound on the neutrino masses significantly depend on the data sets that need to be combined in order to break the degeneracies of the many cosmological parameters~\cite{Aghanim:2018eyx}.  Combining the Planck lensing with the baryon acoustic oscillation (BAO) data and the acoustic scale measured by the CMB, the neutrino mass is constrained to be $\sum_i m_i<600$ meV~\cite{Aghanim:2018eyx}. The limits also become weaker when one departs from the framework of $\Lambda$CDM plus neutrino mass to frameworks with more cosmological parameters. The direct kinematic searches provide the most model independent approach to test the neutrino mass, and the neutrino mass extracted from ordinary beta decay is
\begin{equation}
m_{\beta}=\sqrt{\sum_i |U_{ei}|^2m^2_i}
=\sqrt{\cos^2\theta^l_{12}\cos^2\theta^l_{13}m^2_1+\sin^2\theta^l_{12}\cos^2\theta^l_{13}m^2_2+\sin^2\theta^l_{13}m^2_3}~\,,
\end{equation}
where $U$ is the lepton mixing matrix. From the values of lepton mixing angles and neutrino masses, we can determine the effective mass $m_{\beta}$, as shown in table~\ref{tab:lepton bset-fit}. The predictions for $m_{\beta}$
are around 15 meV, and they are much below the upper limit $m_{\beta}<1.1$ eV given by KATRIN~\cite{Aker:2019uuj}.
It is expected that KATRIN can advance the sensitivity on $m_{\beta}$ by one order of magnitude down to 0.2 eV after 5 years, and the next generation experiments such as Project 8 may be able to reach the 50 meV level~\cite{Doe:2013jfe}. Therefore a positive signal of KATRIN or Project 8 in near future could rule out our models.

It is known that the neutrinoless double beta ($0\nu\beta\beta$) decays of even-even nuclei are important to test the Majorana nature of neutrinos, they can provide valuable information on the neutrino mass spectrum and the CP violation phases. The amplitude of the neutrinoless double beta decay is proportional to the effective Majorana mass $m_{\beta\beta}$ which is given by
\begin{eqnarray}
\nonumber m_{\beta\beta}&=&\Big|\sum_i U^2_{ei}m_i\Big|\\
\label{eq:mbb}&=&\Big|\cos^2\theta^l_{12}\cos^2\theta^l_{13}m_1+\sin^2\theta^l_{12}\cos^2\theta^l_{13}e^{i\alpha_{21}}m_2+\sin^2\theta^l_{13}e^{i(\alpha_{31}-2\delta^l_{CP})}m_3\Big|\,.
\end{eqnarray}
The strongest bound on $m_{\beta\beta}$ is set by the KamLAND-Zen experiment $m_{\beta\beta}<(61-165)$ meV~\cite{KamLAND-Zen:2016pfg}, where the largest uncertainty arises from the computation of the associated nuclear matrix element. There are many $0\nu\beta\beta$ decay experiments in the plan and construction, which aim to improve the current bounds on $m_{\beta\beta}$. The future large scale $0\nu\beta\beta$ decay experiments have the potential of measuring the decay half-life exceeding $10^{28}$ years. For instance, the SNO+ Phases II is expected to reach a sensitivity of $19-46$ meV~\cite{Andringa:2015tza}. The LEGEND experiment intends to achieve a sensitivity of 15-50 meV by operating 1000 kg of detectors for 10 years~\cite{Abgrall:2017syy}. The nEXO is the successor of EXO-200, and its projected $m_{\beta\beta}$ sensitivity is $5.7-17.7$ meV after 10 years of data taking~\cite{Albert:2017hjq}. Using the master formula of Eq.~\eqref{eq:mbb},  we can determine the values of the effective Majorana neutrino mass $m_{\beta\beta}$ at the best fitting points, as given in table~\ref{tab:lepton bset-fit}. We see that the latest bound of KamLAND-Zen experiment is well satisfied and the predictions are within the reach of future tonne-scale $0\nu\beta\beta$ experiments except the models L5 and L6 which are experimentally very challenging because of the quite low values of $m_{\beta\beta}$.

\begin{figure}[h]
\centering
\includegraphics[width=0.9\textwidth]{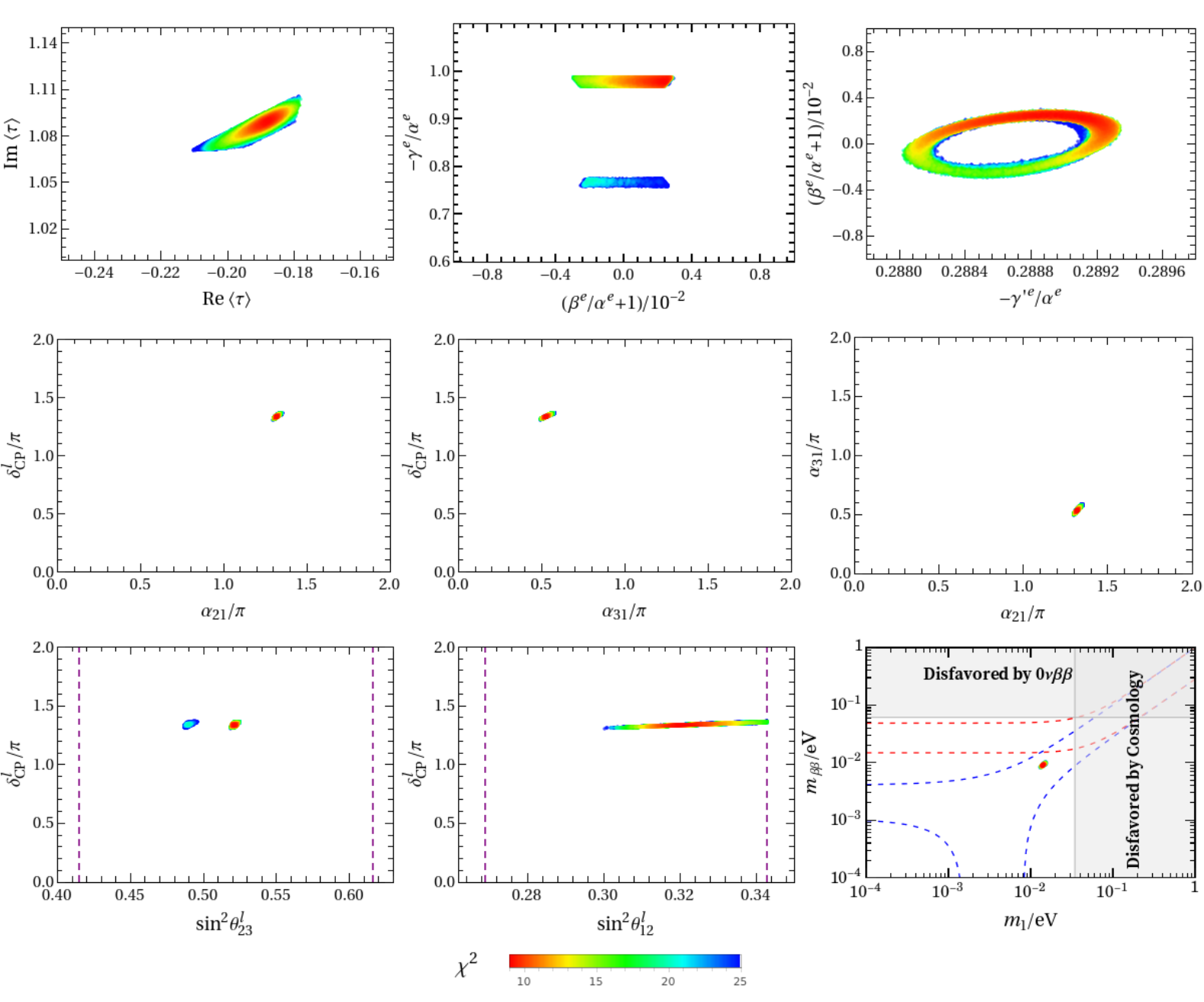}
\caption{\label{fig:model L1}
The correlations among the input parameters, lepton mixing angles, CP
violation phases and neutrino masses in the	model L1. The lepton masses and mixing angles are required to lie in the experimentally preferred $3\sigma$ regions~\cite{Esteban:2020cvm}.
Notice that the transformation $\tau\rightarrow-\tau^{*}$ leaves all observables unchanged except shifting the signs of the CP phases $\delta^{l}_{CP}$, $\alpha_{21}$ and $\alpha_{31}$, consequently we don't show the CP conjugate region. In the last panel, the red (blue) dashed lines indicate the most general  allowed regions for inverted ordering (normal ordering) neutrino mass spectrum respectively, where the neutrino oscillation parameter are varied over their $3\sigma$ ranges. The present upper limit $m_{\beta\beta}<(61-165)$ meV from KamLAND-Zen~\cite{KamLAND-Zen:2016pfg} is shown by horizontal grey band. The vertical grey exclusion band denotes the current bound coming  from the cosmological data of $\sum_im_i<0.120$eV at 95\% confidence level  obtained by the Planck collaboration~\cite{Aghanim:2018eyx}.  }
\end{figure}

\begin{figure}[h]
\centering
\includegraphics{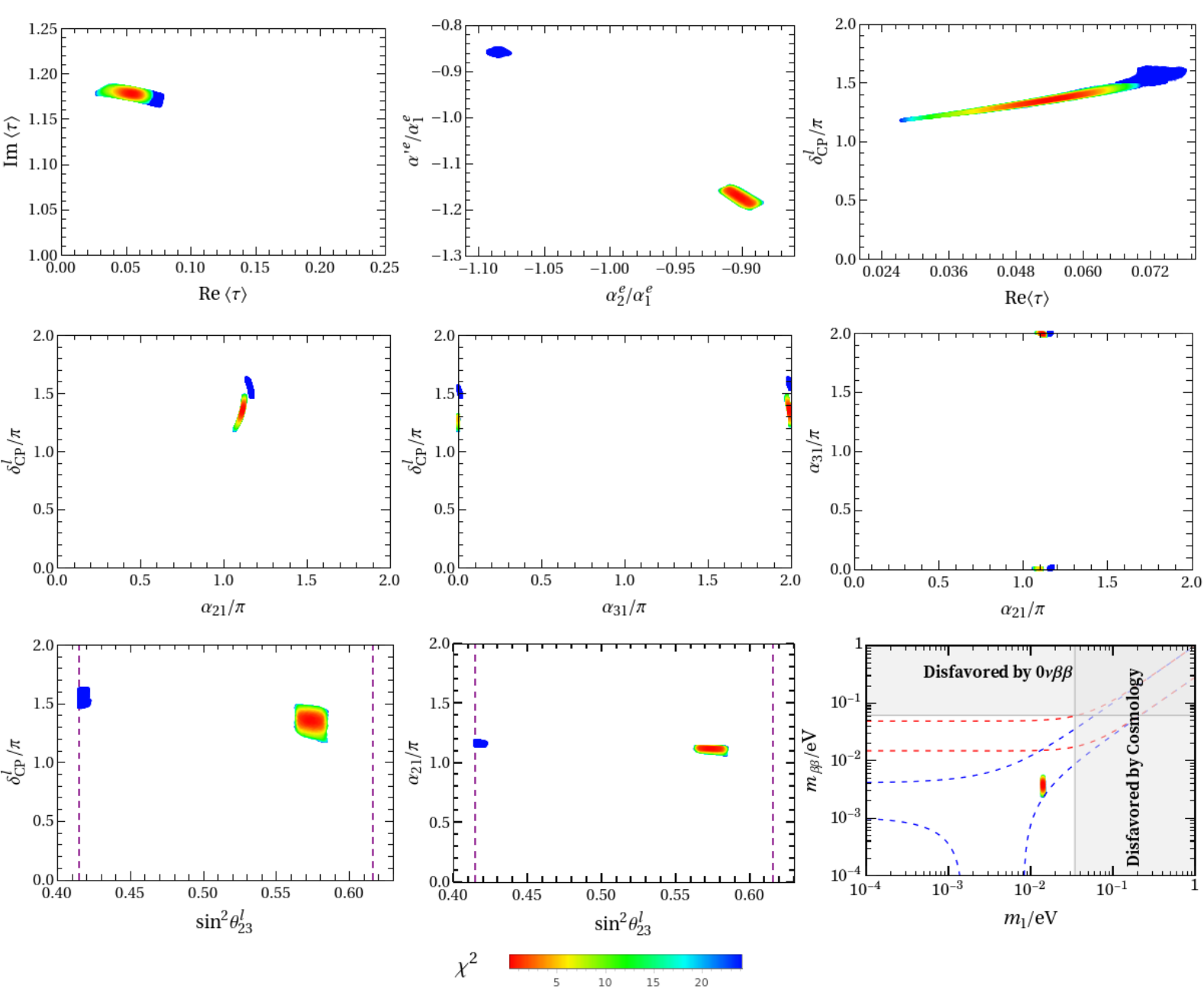}
\caption{\label{fig:model L2}The correlations among the input parameters, lepton mixing angles, CP violation phases and neutrino masses in the model L6, where we adopt the same convention as figure~\ref{fig:model L1}.}
\end{figure}

It is remark that these minimal viable models only use five real parameters together with the complex modulus $\tau$ to describe 12 observables: 3 charged lepton masses, 3 neutrino masses and 3 lepton mixing angles and 3 CP violating phases. Thus the values of the free parameters are strongly constrained by the experimental data and the different observables should be correlated with each other. For example, the light neutrino mass matrix only depends on the modulus $\tau$ up to an overall scale in the model L1 while there are four real couplings in the charged lepton superpotential with $\mathcal{W}_{e}=\alpha^e (E^cL)_{\mathbf{1}}Y^{(4)}_{\mathbf{1}} +\beta^e (E^cLY^{(4)}_{\mathbf{2}})_{\mathbf{1}}+\gamma^e (E^cLY^{(4)}_{\mathbf{3}})_{\mathbf{1}}
+\gamma'^{e} (E^cLY^{(4)}_{\mathbf{3}'})_{\mathbf{1}}$. It is notable that the hierarchical masses of charged leptons can be reproduced although the four coupling constants $\alpha^e$, $\beta^e$, $\gamma^e$ and $\gamma'^{e}$  are of the same order of magnitude, as can be seen from table~\ref{tab:lepton bset-fit}. Thus the charged lepton masses are also dictated by modular symmetry, and the hierarchical mass eigenvalues arise from the departure of $\langle\tau\rangle$ from the self-dual fixed point $\tau=i$~\cite{Okada:2020brs,Feruglio:2021dte,Novichkov:2021evw}. Furthermore, we take the models L1 and L6 as examples, and we comprehensively scan the parameter space of these two models. The lepton masses and mixing angles are required to lie in the experimentally preferred $3\sigma$ regions~\cite{Esteban:2020cvm}, we display the correlations among the free parameters and observables in figure~\ref{fig:model L1} and figure~\ref{fig:model L2}. It is worth mentioning that the experimental data can only be accommodated in small regions of parameter space such that the predictions for the lepton mixing parameters are quite precise and their allowed regions are small as well.

Since the signal of $0\nu\beta\beta$ decay has not been observed, the possibility that neutrinos are Dirac particles can not be excluded at present. Generally additional symmetry such as lepton number conservation is  necessary to forbid the Majorana mass term of the right-handed neutrinos.
Modular invariance can naturally enforce Dirac neutrinos if the modular weights of the right-handed neutrinos are negative. In the same fashion, we can analyze the possible Dirac neutrino mass models with $S_4$ modular symmetry and gCP symmetry. We find that the phenomenologically viable models make use of at least five couplings besides the modulus $\tau$, and four minimal models are found.
The modular transformation properties of the lepton fields and the results of the $\chi^2$ analysis are reported in table~\ref{tab:Dirac lepton models}. Notice that the model D3 was already discussed in~\cite{Wang:2020dbp}. These models can be tested by the measurement of $\theta^{l}_{23}$ and $\delta^{l}_{CP}$ at future long baseline neutrino oscillation experiments~\cite{Acciarri:2016crz,Acciarri:2015uup,Strait:2016mof,Acciarri:2016ooe,Abe:2016ero}.
The effective neutrino mass $m_{\beta}$ in beta decay is predicted to be an order of magnitude below the expected sensitivity of the KATRIN experiment~\cite{Aker:2019uuj}.

\begin{table}[]
\resizebox{1.0\textwidth}{!}{
\begin{tabular}{|c|c|c|c|c|c|c|c|c|c|}
\hline \hline
&L1 &L2 &L3 &L4 &L5 &L6 &L7 &L8 &L9\\
\hline
$\rho_L$ & $\mathbf{3}$ & $\mathbf{3}$ & $\mathbf{3}$ & $\mathbf{3}$ & $\mathbf{3}$ & $\mathbf{3}$  &$\mathbf{3}$ & $\mathbf{2}\oplus\mathbf{1}$ & $\mathbf{2}\oplus\mathbf{1}$ \\
\hline
$\rho_{E^c}$ & $\mathbf{3}$ & $\mathbf{1}\oplus\mathbf{1}'\oplus\mathbf{1}$ & $\mathbf{1}\oplus\mathbf{1}'\oplus\mathbf{1}'$ & $\mathbf{1}\oplus\mathbf{1}'\oplus\mathbf{1}$ & $\mathbf{1}\oplus\mathbf{1}'\oplus\mathbf{1}$ & $\mathbf{2}\oplus\mathbf{1}'$  & $\mathbf{2}\oplus\mathbf{1}'$ & $\mathbf{2}\oplus\mathbf{1}$ & $\mathbf{2}\oplus\mathbf{1}'$ \\
\hline
$\rho_{N^c}$ & $\mathbf{3}$ & $\mathbf{3}'$ & $\mathbf{3}'$ & $\mathbf{3}$ & $\mathbf{3}$ & $\mathbf{3}$  & $\mathbf{3}'$ & $\mathbf{3}$ & $\mathbf{3}$ \\ \hline\hline
$k_{L}$ & $-1$ & $2$ & $2$ & $-1$ & $-1$ & $-1$  & $2$ & $1,1$ & $1,1$ \\
\hline
$k_{E^c}$ & $5$ & $0,2,2$ & $0,2,4$ & $3,5,7$ & $3,5,9$ & $7,5$  & $2,2$ & $1,1$ & $1,1$\\
\hline
$k_{N^c}$ & $1$ & $0$ & 0
& $1$ & $1$ & $1$  & $0$ & $1$ & $1$\\
\hline \hline
\end{tabular}}
\caption{\label{tab:lepton models} Summary of the representation and modular weight assignments of the matter fields in the minimal phenomenologically viable lepton models based on $S_4$ modular symmetry and gCP symmetry, the neutrinos are assumed to be Majorana particles. Notice that the Higgs fields are invariant under $S_4$ with zero modular weight.
}
\end{table}

\begin{table}[]
\resizebox{\textwidth}{!}
{
\begin{tabular}{|c|c||c|c|c||c|c|c|}
\hline\hline
Models &L1  &Models  &L2  &L3  &Models  &L4  &L5  \\\hline
$\mathrm{Re}\langle\tau\rangle$ & $-0.187799$  &$\mathrm{Re}\langle\tau\rangle$ &0.101211  &0.101211  &$\mathrm{Re}\langle\tau\rangle$ & $-0.178216$  &0.0667032    \\\hline
$\mathrm{Im}\langle\tau\rangle$ &1.08926  &$\mathrm{Im}\langle\tau\rangle$ &1.01587  &1.01587  &$\mathrm{Im}\langle\tau\rangle$ &1.09848  &1.17412   \\\hline
$\beta^e/\alpha^e$ & $-0.997574$  & $\beta^e/\alpha^e$ & 11.1874  &11.1873  & $\beta^e/\alpha^e$ & 1441.41  & 1012.34    \\\hline
$\gamma^e/\alpha^e$ & $-0.978209$  & $\gamma^e/\alpha^e$ & 0.00260186 & 0.00209056  & $\gamma^e_1/\alpha^e$ & 48.4687  & 7.25588   \\\hline
$\gamma'{}^e/\alpha^e$ & $-0.288994$  & $\gamma^D/\beta^D$ & 0.0139677  & 0.0139677  & $\gamma^e_2/\alpha^e$ & $-49.8784$  & $-55.3767$   \\\hline
$\alpha^e v_d(\MeV)$ & 228.251  & $\alpha^e v_d(\MeV)$ & 42.0558  & 42.0555  & $\alpha^e v_d(\MeV)$ & 0.384973  & 0.629924  \\\hline
$\frac{(\alpha^{D}v_u)^2}{\beta^N\Lambda}(\meV)$ &28.4038  & $\frac{(\beta^D v_u)^2}{\alpha^N\Lambda}(\meV)$ &9.40363  &9.40363  & $\frac{(\alpha^D v_u)^2}{\beta^N\Lambda}(\meV)$ & 27.5524 &22.6716 \\\hline\hline
$\sin^2\theta^{l}_{12}$ &0.320753 &$\sin^2\theta_{12}$ & 0.305480  & 0.305480  &$\sin^2\theta^{l}_{12}$ &0.307702 &0.304328  \\\hline
$\sin^2\theta^{l}_{13}$ &0.0219057 &$\sin^2\theta_{13}$ &0.0221663  &0.0221663  &$\sin^2\theta^{l}_{13}$ &0.0221664  &0.0221541  \\\hline
$\sin^2\theta^{l}_{23}$ &0.521416 &$\sin^2\theta_{23}$ &0.485929  &0.485929  &$\sin^2\theta^{l}_{23}$ &0.503135  &0.491327  \\\hline
$\delta^l_{CP}/\pi$ &1.33248 &$\delta^l_{CP}/\pi$ &1.64135  &1.64135  &$\delta^l_{CP}/\pi$ &1.32001  &1.46324  \\\hline
$\alpha_{21}/\pi$ &1.32139 &$\alpha_{21}/\pi$ &0.353191  &0.353191  &$\alpha_{21}/\pi$ &1.31196  &1.14629  \\\hline
$\alpha_{31}/\pi$ &0.528182 &$\alpha_{31}/\pi$ &1.25878  &1.25878  &$\alpha_{31}/\pi$ &0.509414  &1.99551  \\\hline
$m_1/\meV$ &14.0612 &$m_1/\meV$ & 12.1830  & 12.1830  &$m_1/\meV$ &13.6298  &11.1822  \\\hline
$m_2/\meV$ &16.4808 &$m_2/\meV$ &14.9106  &14.9106  &$m_2/\meV$ &16.1143  &14.1047  \\\hline
$m_3/\meV$ &51.7056 &$m_3/\meV$ &51.5812  &51.5812  &$m_3/\meV$ &51.8358  &51.2977   \\\hline
$m_{\beta}/\meV$ &16.5872 &$m_{\beta}/\meV$ &15.0395  &15.0395  &$m_{\beta}/\meV$ &16.2311  &14.2324   \\\hline
$m_{\beta\beta}/\meV$ & 9.03230  &$m_{\beta\beta}/\meV$ &12.0458  &12.0458  &$m_{\beta\beta}/\meV$ &8.80743  &3.45226 \\\hline
$\chi^2_{\text{min}}$ &9.01 & $\chi^2_{\text{min}}$ &18.97  &18.97  & $\chi^2_{\text{min}}$ & 12.31  & 16.68 \\\hline
\end{tabular}}
\resizebox{\textwidth}{!}
{
\begin{tabular}{|c|c||c|c||c|c|c|}
\hline
Models &L6 &Models &L7 &Models &L8 &L9 \\\hline
$\mathrm{Re}\langle\tau\rangle$ &0.0546503  &$\mathrm{Re}\langle\tau\rangle$ &0.101532  &$\mathrm{Re}\langle\tau\rangle$ & $-0.482373$  & 0.193695   \\\hline
$\mathrm{Im}\langle\tau\rangle$ &1.17782  &$\mathrm{Im}\langle\tau\rangle$ &1.01583  &$\mathrm{Im}\langle\tau\rangle$ &1.27223  &0.991159   \\\hline
$\beta^e/\alpha^e_1$ &0.0248284  & $\beta^e/\alpha^e$ & 37.5920  & $\beta^e/\alpha^e$ &2834.28  &2833.91  \\\hline
$\alpha^e_2/\alpha^e_1$ & $-0.901642$  & $\alpha'{}^{e}/\alpha^e$ & $-1.01173$  & $\gamma^e/\alpha^e$ & 160.150  &160.149  \\\hline
$\alpha'{}^{e}/\alpha^e_1$ & $-1.17279$  & $\gamma^D/\beta^D$ &0.0139492  & $\beta^D/\alpha^D$ &1.07525  & $-1.07524$ \\\hline
$\alpha^e_1v_d(\MeV)$ &202.486  & $\alpha^e v_d(\MeV)$ &12.5174  & $\alpha^e v_d(\MeV)$ & 0.459000  &0.394985  \\\hline
$\frac{(\alpha^D v_u)^2}{\beta^N\Lambda}(\meV)$ &22.5089  & $\frac{(\beta^D v_u)^2}{\alpha^N\Lambda}(\meV)$ &9.40482  & $\frac{(\alpha^Dv_u)^2}{\beta^N\Lambda}(\meV)$ &12.2127  &  6.90400  \\\hline\hline
$\sin^2\theta^{l}_{12}$ &0.303992 &$\sin^2\theta_{12}$ &0.305479  &$\sin^2\theta^{l}_{12}$ &0.301955  &0.301963  \\\hline
$\sin^2\theta^{l}_{13}$ &0.0221904 &$\sin^2\theta_{13}$ &0.0221666  &$\sin^2\theta^{l}_{13}$ &0.022121  &0.0221229 \\\hline
$\sin^2\theta^{l}_{23}$ &0.573536 &$\sin^2\theta_{23}$ &0.485966  &$\sin^2\theta^{l}_{23}$ &0.613488  &0.613493  \\\hline
$\delta^l_{CP}/\pi$ &1.35611 &$\delta^l_{CP}/\pi$ &1.64081  &$\delta^l_{CP}/\pi$ & 1.27970  &1.27971  \\\hline
$\alpha_{21}/\pi$ &1.11691 &$\alpha_{21}/\pi$ &0.353732  &$\alpha_{21}/\pi$ &1.26955  &1.26956  \\\hline
$\alpha_{31}/\pi$ &1.99179 &$\alpha_{31}/\pi$ &1.25913  &$\alpha_{31}/\pi$ &0.411336  &0.411352  \\\hline
$m_1/\meV$ & 11.1010 &$m_1/\meV$ &12.1879  &$m_1/\meV$ &30.6674  & 30.6690    \\\hline
$m_2/\meV$ &14.0403 &$m_2/\meV$ &14.9146  &$m_2/\meV$ &31.8495  &31.8511   \\\hline
$m_3/\meV$ &51.2817 &$m_3/\meV$ &51.5824   &$m_3/\meV$ &58.6517  &58.6547 \\\hline
$m_{\beta}/\meV$ &14.1711 &$m_{\beta}/\meV$ &15.0435   &$m_{\beta}/\meV$ &31.8999  &31.9016 \\\hline
$m_{\beta\beta}/\meV$ &3.77388 &$m_{\beta\beta}/\meV$ &12.0465   &$m_{\beta\beta}/\meV$ &17.6025  &17.6035 \\\hline
$\chi^2_{\text{min}}$ &0.00134 & $\chi^2_{\text{min}}$ &18.96  &$\chi^2_{\text{min}}$ &6.45  &6.45 \\\hline \hline
\end{tabular}}
\caption{\label{tab:lepton bset-fit}The best fit values of the input parameters for the minimal lepton models listed in table~\ref{tab:lepton models}, where neutrinos are assumed to be Majorana particles. We give the values of the neutrino mixing angles, Dirac and Majorana CP violating phases
and the neutrino masses at the best fitting points. The notations $m_{\beta}$
and $m_{\beta\beta}$ denote the effective neutrino masses measured in beta decay and neutrinoless double decay respectively.
Note that the transformation $\tau\rightarrow-\tau^{*}$ leaves all observables unchanged except shifting the signs of the CP phases $\delta^{l}_{CP}$, $\alpha_{21}$ and $\alpha_{31}$. }
\end{table}

\begin{table}[]
\resizebox{1.0\textwidth}{!}{
\begin{tabular}{|c|c|c|c||c|c|c|}
\hline\hline
Models &$\rho_{L}$ &$\rho_{E^c}$ &$\rho_{N^c}$ &$k_{L}$ &$k_{E^c}$ &$k_{N^c}$  \\ \hline
D1 &$\mathbf{2}\oplus\mathbf{1}$ &$\mathbf{3}$ &$\mathbf{3}$ &$k,k$ &$4-k$ &$2-k \,(k>2)$  \\ \hline
D2 &$\mathbf{3}$ &$\mathbf{2}\oplus\mathbf{1}'$ &$\mathbf{3}'$ &$k$ &$4-k,4-k$ &$2-k \,(k>2)$  \\ \hline
D3 &$\mathbf{3}$ &$\mathbf{1}\oplus\mathbf{1}'\oplus\mathbf{1}$ &$\mathbf{3}'$ &$k$ &$2-k,4-k,4-k$ &$2-k \,(k>2)$ \\ \hline
D4 &$\mathbf{3}$ &$\mathbf{1}\oplus\mathbf{1}'\oplus\mathbf{1'}$ &$\mathbf{3}'$ &$k$ &$2-k,4-k,6-k$ &$2-k \,(k>2)$  \\ \hline
\end{tabular}}
\resizebox{1.0\textwidth}{!}{
\begin{tabular}{|c|c||c|c||c|c|c|}\hline
Model &D1 &Model &D2 &Models &D3 &D4  \\ \hline
$\mathrm{Re}\langle\tau\rangle$ &0.341595 &$\mathrm{Re}\langle\tau\rangle$ & 0.106064&$\mathrm{Re}\langle\tau\rangle$ &0.105759 &0.105759  \\\hline
$\mathrm{Im}\langle\tau\rangle$ &1.36933 &$\mathrm{Im}\langle\tau\rangle$ &1.00322 &$\mathrm{Im}\langle\tau\rangle$ &1.00325 &1.00325  \\\hline
$\beta^{\nu}/\alpha^{\nu}$ &1.02513 &$\beta^{e}/\alpha^{e}_1$ &37.5323 &$\beta^{e}/\alpha^{e}$ &11.0459 &11.0458  \\ \hline
$\alpha^{e}_2/\alpha^{e}_1$ & $-0.984449$ & $\gamma^\nu/\beta^{\nu}$ &0.00502027 &$\gamma^{e}/\alpha^{e}$ &0.00258625 &0.00203595  \\ \hline
$\beta^{e}/\alpha^{e}_1$ &21.8849 &$\alpha^{e}_2/\alpha^{e}_1$ & $-1.01168$ & $\gamma^\nu/\beta^{\nu}$ &0.00502515 &0.00502516  \\ \hline
$\alpha^{e}_1v_d(\MeV)$ &22.8857 &$\alpha^{e}_1v_d(\MeV)$ &12.2227 &$\alpha^{e}v_d(\MeV)$ &41.5209 &41.5205  \\ \hline
$\alpha^{\nu}v_u(\meV)$ & 18.5350 & $\beta^{\nu}v_u(\meV)$ &24.5106 & $\beta^{\nu}v_u(\meV)$ &24.5128 &24.5128  \\\hline \hline
$\sin^2\theta^{l}_{12}$ &0.302259 &$\sin^2\theta_{12}$ &0.307374 &$\sin^2\theta^{l}_{12}$ &0.307295 &0.307294  \\ \hline
$\sin^2\theta^{l}_{13}$ &0.0221697 &$\sin^2\theta_{13}$ &0.0221678 &$\sin^2\theta^{l}_{13}$ &0.0221678 &0.221678  \\ \hline
$\sin^2\theta^{l}_{23}$ &0.596554 &$\sin^2\theta_{23}$ &0.477773 &$\sin^2\theta^{l}_{23}$ &0.477732 &0.477732  \\ \hline
$\delta^l_{CP}/\pi$ &1.36268 &$\delta^l_{CP}/\pi$ &1.56812 &$\delta^l_{CP}/\pi$ &1.56839 &1.56839  \\ \hline
$m_1/\meV$ &25.4036 &$m_1/\meV$ &29.1666 &$m_1/\meV$ &29.1665 &29.1665  \\ \hline
$m_2/\meV$ &26.8187 &$m_2/\meV$ &30.4071 &$m_2/\meV$ & 30.4070 & 30.4070  \\ \hline
$m_3/\meV$ &56.1239 &$m_3/\meV$ &58.0277 &$m_3/\meV$ &58.0364 &58.0364  \\ \hline
$m_{\beta}/\meV$ &26.8833 &$m_{\beta}/\meV$ &30.4744 &$m_{\beta}/\meV$ &30.4746 &30.4745  \\ \hline
$\chi^2_{\text{min}}$ & 2.29 & $\chi^2_{\text{min}}$ & 22.77 & $\chi^2_{\text{min}}$ & 27.79 & 22.79  \\ \hline\hline
\end{tabular}}
\caption{\label{tab:Dirac lepton models}Summary of the lepton models based on $S_4$ modular symmetry and gCP symmetry, where neutrinos are Dirac particles. The integer $k$ should be greater than 2 so that the modular weight of $N^c$ is negative and modular invariance forbids the Majorana mass term of right-handed neutrinos. Notice that the Higgs fields are invariant under $S_4$ with zero modular weight. The best fit values of the input parameters are also included. We give the predictions for neutrino mixing angles, and Dirac CP violating phase, and the neutrino masses, and the effective neutrino masses $m_{\beta}$ probed by direct kinematic search in beta decay.
Note that the transformation $\tau\rightarrow-\tau^{*}$ leaves all observables unchanged except shifting the sign of the CP phase $\delta^{l}_{CP}$. }
\end{table}

\subsection{Leptogenesis}

Since we impose gCP as symmetry on the model, all couplings in the superpotential are constrained to be real in our working basis. As a consequence, all CP violations uniquely arise from the modulus vacuum expectation value.
Thus the CP violation in leptogenesis is naturally related to the CP violation phases $\delta^{l}_{CP}$, $\alpha_{21}$ and $\alpha_{31}$ in the lepton mixing matrix. Early studies of leptogenesis in the context of modular symmetry models without gCP symmetry can be found~\cite{Asaka:2019vev,Wang:2019ovr,Behera:2020sfe}. During the final preparations of this paper, a preprint discussing leptogenesis in a $A_4$ modular model with gCP appeared on the arXiv~\cite{Okada:2021qdf}. In this section,
we shall study whether the measured value of the baryon asymmetry of the universe $Y_{B0}=(0.870300 \pm 0.011288)\times10^{-10}$~\cite{Aghanim:2018eyx} can be correctly generated through leptogenesis in the minimal $S_4$ modular invariant models found in previous section, where the subscript 0 implies ``at present time''.

The right-handed neutrino masses depend on the overall scale $\Lambda$ in our model. In the present work, we assume that the right-handed neutrinos are heavy with masses above $10^{12}$ GeV, thus we work in the framework of unflavored thermal leptogenesis. The modular invariance is formulated in the supersymmetric context, as shown in section~\ref{subsec:framework}, therefore we should consider supersymmetric leptogenesis.
The out-of-equilibrium decays of the lightest right-handed neutrinos or sneutrinos in the early universe produce lepton asymmetries.
We denote the decay asymmetries for the decay of heavy neutrino into Higgs and lepton, neutrino into Higgsino and slepton, sneutrino into Higgsino and lepton, and sneutrino into Higgs and slepton as $\varepsilon_{1\ell}$, $\varepsilon_{1\tilde{\ell}}$, $\varepsilon_{\tilde{1}\ell}$ and $\varepsilon_{\tilde{1}{\ell}}$ respectively which are defined
by~\cite{Covi:1996wh,Antusch:2006cw}
\begin{align}
&\varepsilon_{1\ell}\equiv
\frac{\Gamma_{N_{1}\ell}-\Gamma_{N_{1}\bar{\ell}}}
{\Gamma_{N_{1}\ell}+\Gamma_{N_{1}\bar{\ell}}},~~~\quad
\varepsilon_{1\tilde\ell}\equiv
\frac{\Gamma_{N_{1}\widetilde\ell}-\Gamma_{N_{1}\widetilde{\ell}^{*}}}
{\Gamma_{N_{1}\widetilde\ell}+\Gamma_{N_{1}\widetilde{\ell}^{*}}},\\
&\varepsilon_{\tilde{1}\ell}\equiv
\frac{\Gamma_{\widetilde{N}_{1}^{*}\ell}-\Gamma_{\widetilde{N}_{1}\bar{\ell}}}
{\Gamma_{\widetilde{N}_{1}^{*}\ell}+\Gamma_{\widetilde{N}_{1}\bar{\ell}}},~~~\quad
\varepsilon_{\tilde1\tilde\ell}\equiv
\frac{\Gamma_{\widetilde{N}_{1}\tilde\ell}-\Gamma_{\widetilde{N}_{1}^{*}\tilde{\ell^{*}}}}
{\Gamma_{\widetilde{N}_{1}\tilde\ell}+\Gamma_{\widetilde{N}_{1}^{*}\tilde{\ell^{*}}}}.
\end{align}
In the minimal supersymmetric standard model, all the above decay asymmetries are equal $\varepsilon_{1\ell}=\varepsilon_{1\widetilde \ell}=\varepsilon_{\widetilde 1\ell}=\varepsilon_{\widetilde 1\widetilde \ell}$~\cite{Covi:1996wh,Antusch:2006cw}. In the basis where the Majorana mass matrix of the right-handed neutrinos is diagonal and real, the lepton asymmetry
parameter $\varepsilon_{1\ell}$ is given by~\cite{Covi:1996wh,Giudice:2003jh}
\begin{equation}
\varepsilon_{1\ell}=\frac{1}{8\pi{\left(\lambda_{\nu}\lambda_{\nu}^{\dagger}\right)}_{11}}\sum_{i=2,3}
\texttt{Im}\left\{\left[{\left(\lambda_{\nu}\lambda_{\nu}^{\dagger}\right)}_{1i}\right]^{2}\right\}~g\left(\frac{M_{i}^{2}}{M_{1}^{2}}\right)\,,
\end{equation}
where $\lambda_{\nu}$ is the neutrino Yukawa coupling matrix in the convention $(\lambda_{\nu})_{ij}N^c_i(L_j\cdot H_u)$
and the loop function
$g$ reads
\begin{align}
g(x)=\sqrt{x}\left[\frac{2}{1-x}-\mathrm{ln}\left(\frac{1+x}{x}\right)\right]\stackrel{x \gg 1}{\longrightarrow}-\frac{3}{\sqrt{x}}\,.
\end{align}
Non-vanishing asymmetry parameter $\varepsilon_{1\ell}$  requires that the off-diagonal entries of the product $\lambda_{\nu}\lambda_{\nu}^{\dagger}$ are complex and different from zero. For the models L1, L4, L5, L6, the product $\lambda_{\nu}\lambda_{\nu}^{\dagger}$ is proportional to the unity matrix. Consequently the lepton asymmetry $\varepsilon_{1\ell}$ is vanishing at LO and a net baryon asymmetry can not be generated. The masses of the three right-handed neutrinos are degenerate in the models L2, L3 and L7,
the baryon asymmetry is generated in the regime of resonant leptogenesis. Hence we take the model L9 as an example in the following.

The lepton number asymmetry is partially converted into a non-zero baryon number asymmetry by the fast sphaleron interactions in the thermal bath in the Early Universe. For all temperature ranges, the produced baryon asymmetry normalized to the entropy density can be computed from the $B-L$ asymmetry $\hat{Y}_{\Delta}$ as follows
\begin{align}
Y_{B}=\frac{10}{31}\hat{Y}_{\Delta}\,,
\end{align}
The $B-L$ asymmetry $\hat{Y}_{\Delta}$ can be computed by solving the following Boltzmann equations~\cite{Antusch:2006cw,Bjorkeroth:2016lzs}
\begin{eqnarray}
\nonumber
\frac{\mathrm{d} Y_{N_1}}{\mathrm{d} z} &=&- \, 2 K\,z \,\frac{K_1 (z)}{K_2 (z)}\, f_1 (z) \,(Y_{N_1} - Y^\mathrm{eq}_{N_1}) \; ,\\
\nonumber\frac{\mathrm{d} Y_{\widetilde N_1}}{\mathrm{d} z} &=&-\,2 K\,z\,\frac{K_1 (z)}{K_2 (z)}\, f_1 (z) \,(Y_{\widetilde N_1} - Y^\mathrm{eq}_{\widetilde N_1}) \; ,\\
\nonumber\frac{\mathrm{d} \hat Y_{\Delta}}{\mathrm{d} z} &=&
-\;(\varepsilon_{1\ell}+ \varepsilon_{1\widetilde \ell}) \,K\, z \,
\frac{K_1 (z)}{K_2 (z)}\,  f_1 (z) \,(Y_{N_1} - Y^\mathrm{eq}_{N_1}) \\
\nonumber &&-\;(\varepsilon_{\widetilde1\ell}+\varepsilon_{\widetilde1\widetilde \ell}) \,K\, z \,\frac{K_1 (z)}{K_2 (z)}\,  f_1 (z) \,(Y_{\widetilde N_1} - Y^\mathrm{eq}_{\widetilde N_1})\nonumber \\
\label{eq:MSSM_be}&&-\;K \, z\, \frac{K_1 (z)}{K_2 (z)} \, f_2 (z) \,\frac{\hat Y_{\Delta}}{\hat Y^\mathrm{eq}_{\ell}} \,(Y^\mathrm{eq}_{N_1}+Y^\mathrm{eq}_{\widetilde N_1})\;,
\end{eqnarray}
where $z=M_{1}/T$ with $T$ being the temperature. $K_{1}(z)$ and $K_{2}(z)$ are the modified Bessel functions of the second kind. $Y_{N_{1}}$ and $Y_{\widetilde N_1}$ denotes the density of the lightest right-handed neutrino $N_{1}$ with mass $M_1$ and its supersymmetric partners $\widetilde N_1$. The notations $Y^\mathrm{eq}_{N_1}$, $Y^\mathrm{eq}_{\widetilde N_1}$ and $\hat Y^\mathrm{eq}_{\ell}$ are corresponding equilibrium number densities and they take the following form
\begin{equation}\label{eq:MSSM-Yleq-Ynieq}
\hat{Y}_{\ell}^{\mathrm{eq}}=Y^{\mathrm{eq}}_{\tilde{\ell}}+Y^{eq}_{\ell},\quad
	Y^{\mathrm{eq}}_{\tilde{\ell}}\simeq Y^{eq}_{\ell}\simeq\frac{45}{\pi^4g_{*}} ,\quad Y_{N_1}^{\mathrm{eq}}(z)=Y_{\widetilde N_1}^{\mathrm{eq}}(z)\simeq\frac{45}{2\pi^4g_{*}}z^{2}K_{2}(z)\,,
\end{equation}
with $g_{*}=228.75$ being the number of degrees of freedom. Moreover, the washout parameters $K_{\alpha}$ and $K$ are defined as
\begin{align}\label{eq:MSSM_washout}
K_{\alpha}=\frac{\tilde{m}_{1\alpha}}{m^{*}},\qquad\tilde{m}_{1\alpha}\equiv\frac{|\lambda_{1\alpha}|^2v_{u}^2}{M_{1}}, \qquad\,K=\sum_{\alpha}K_{\alpha}  \,,
\end{align}
where
\begin{align}
v_{u}=v\sin\beta, \qquad \quad m^{*}\simeq\sin^2\beta\times1.58\times10^{-3}\,\text{eV}\,.
\end{align}
At the best fit point of model L9, we find the value of $K=1.293\times10^6\gg1$ which implies a strong washout. In the strong washout regime, the functions $f_{1}(z)$ and $f_{2}(z)$ can be approximated as~\cite{Buchmuller:2004nz}
\begin{equation}
\label{eq:f1f2}
f_{1}(z)=2f_{2}(z)=\left[\frac{K_{s}}{zK }+\frac{z}{t} \ln \left(1+\frac{t}{z}\right)\right] \frac{K_{2}(z)}{K_{1}(z)}\,,
\end{equation}
with
\begin{equation}\label{eq:a-Ks-over-K}
t=\frac{K}{K_{s}\ln(M_{1}/M_{h})}, \qquad \quad
\frac{K_{s}}{K}=\frac{9}{8\pi^2}\,.
\end{equation}
where $M_{h}=125$ GeV is the mass of Higgs boson. The free parameters are fixed at their best fit values shown in table~\ref{tab:lepton bset-fit}, notice that only the combination $(\alpha^D v_u)^2/(\beta^N\Lambda)$ can be determined by the data of lepton masses and mixing. Numerically solving the Boltzmann equations, we find that the observed baryon asymmetry can be produced for the following values of the flavor scale:
\begin{align}
\label{eq:Lambda}\Lambda=3.36\times 10^{15}\mathrm{GeV}.
\end{align}
Accordingly the right-handed neutrino masses are determined to be $M_1\simeq1.985\times10^{15}$ GeV, $M_2\simeq6.723\times10^{15}$ GeV and $M_3\simeq6.834\times10^{15}$ GeV. The VEV of $\tau$ is the unique source of modular symmetry and gCP symmetry breaking in this model. All CP violations at both low energy and high energy should significantly depend on $\langle\tau\rangle$. We plot the contour region of $Y_{B}$ in the plane Im$\langle\tau\rangle$ versus Re$\langle\tau\rangle$, where we fix all the coupling constants at their best fit values and $\Lambda$ at the value in Eq.~\eqref{eq:Lambda}. The green area indicates the $3\sigma$ allowed region by the experimental data of lepton masses and mixing angles in the same plane. We see that there exists a small region of $\tau$ where both the flavor structure of the lepton and the baryon asymmetry of the universe can be explained. At the boundary of the fundamental domain $\mathcal{D}$ and the imaginary axis, certain residual gCP symmetry is preserved such that the lepton asymmetry $\varepsilon_{1\ell}$ vanishes and no matter-antimatter asymmetry can be generated. Hence the VEV $\langle\tau\rangle$ should deviate from the CP conserved points in order to obtain non-trivial CP violation in neutrino oscillation as well as a net baryon asymmetry.

\begin{figure}[h]
\centering
\includegraphics[width=0.6\textwidth]{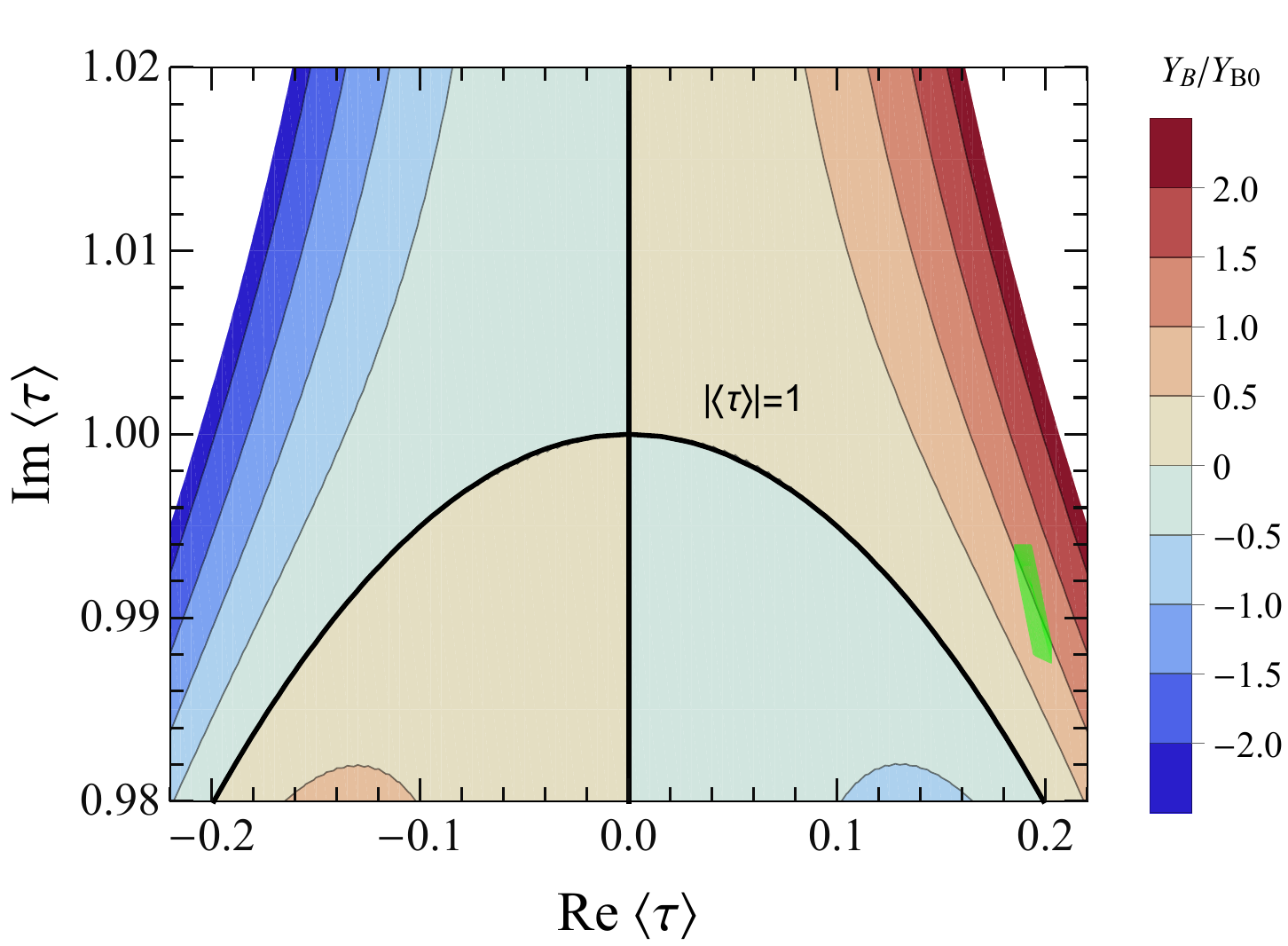}
\caption{The contour plot $Y_{B}$
in the Re$\langle\tau\rangle$-Im$\langle\tau\rangle$ plane.
The green region represent the region for which both lepton masses and lepton mixing angles are compatible with experimental data at the $3\sigma$ level or better.
\label{fig:Lep}}
\end{figure}

Furthermore, we plot the correlation between the $Y_{B}$ and the Dirac CP phase $\delta^{l}_{CP}$ in figure~\ref{fig:Lep}. Here the modulus vacuum expectation value $\langle\tau\rangle$ is treated as a random
complex number in the fundamental domain, the charged lepton masses and the neutrino mass squared differences and all three mixing angles are required to be within their experimentally preferred $3\sigma$ ranges~\cite{Esteban:2020cvm}. Imposing the observed baryon asymmetry $Y_B$, the allowed region of $\delta^{l}_{CP}$ would shrink considerably and it is around $1.28\pi$.

\begin{figure}[h]
\centering
\includegraphics[width=0.6\textwidth]{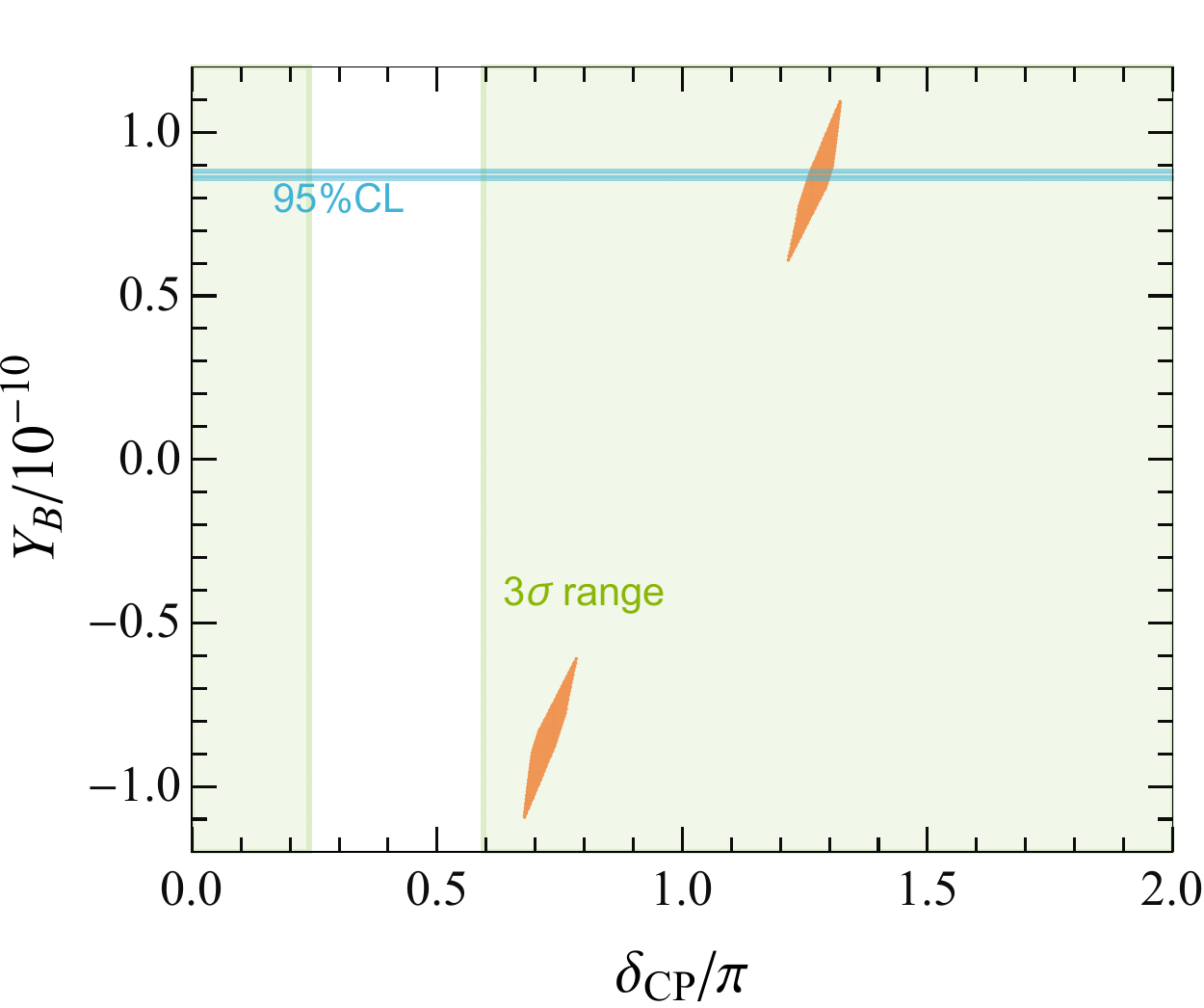}
\caption{The correlation between $Y_{B}$ and $\delta_{CP}$. The horizonal light blue band denotes the $95\%$CL region of $Y_{B}$, and the vertical light green band represents the 3$\sigma$ range of $\delta^l_{CP}$.}
\label{fig:LepDCP}
\end{figure}

\subsection{Quark models}

In the same fashion as we have done in the lepton sector, we can easily find out all possible quark models from the general results of section~\ref{sec:Dirac mass}, and subsequently we perform a $\chi^2$ analysis for each model to determine whether it can accommodate the precisely measured values of six quark masses $m_{u}$, $m_c$, $m_t$, $m_d$, $m_s$, $m_b$, and three quark mixing angles $\theta^{q}_{12}$, $\theta^q_{13}$, $\theta^q_{23}$, and a CP violation phase $\delta^{q}_{CP}$ in the quark sector.
We are concerned with the models which can reproduce the data with the smallest number of free parameters. It turns out that the minimal viable models labelled as Q1$\sim$Q10 make use of seven real coupling constants in addition to the modulus $\tau$, thus one prediction can be reached.
The transformation of quark fields under $S_4$ and their modular weights are listed in table~\ref{tab:quark models}. It can be seen that the left-handed quark fields $Q$ are assigned to a triplet or doublet plus singlet of $S_4$ modular group, while the right-handed quark fields $u^c$ and $d^c$ are singlets or the direct sum of doublet and singlet. We present the best fit values of the input parameters and the predictions for quark masses and CKM mixing parameters in table~\ref{tab:quark bset-fit}. Here we have omitted these models with a high $\chi^2_{\text{min}}$ or a large number of parameters. It is known that the quark masses and mixing parameters have been precisely measured and their errors are quite small. The hierarchical patterns of quark masses and CKM matrix generally require more free parameters in a concrete model and it is quite difficult to explain the quark data with few (less than nine) parameters.

\begin{table}[]
\resizebox{1.0\textwidth}{!}{
\begin{tabular}{|c|c|c|c|c|c|c|c|c|}\hline\hline
Models &Q1 &Q2 &Q3 &Q4 &Q5 \\
\hline
$\rho_Q$ & $\mathbf{3}$ & $\mathbf{3}$ & $\mathbf{3}$ & $\mathbf{3}$ & $\mathbf{3}$  \\ \hline
$\rho_{u^c}$ & $\mathbf{2}\oplus\mathbf{1}$ & $\mathbf{2}\oplus\mathbf{1}'$ & $\mathbf{2}\oplus\mathbf{1}'$ & $\mathbf{2}\oplus\mathbf{1}$ & $\mathbf{2}\oplus\mathbf{1}'$   \\ \hline
$\rho_{d^c}$ & $\mathbf{2}\oplus\mathbf{1}$ & $\mathbf{2}\oplus\mathbf{1}$ & $\mathbf{2}\oplus\mathbf{1}$ & $\mathbf{1}\oplus\mathbf{1}'\oplus\mathbf{1}$ &  $\mathbf{1}\oplus\mathbf{1}\oplus\mathbf{1}$  \\ \hline\hline
$k_{Q}$ & $k$ & $k$ & $k$ & $k$ & $k$   \\ \hline
$k_{u^c}$ & $6-k,4-k$ & $6-k,4-k$ & $6-k,6-k$ & $2-k,6-k$ & $2-k,8-k$   \\ \hline
$k_{d^c}$ & $2-k,6-k$ & $2-k,6-k$ & $2-k,6-k$ & $2-k,4-k,8-k$ & $2-k,4-k,6-k$   \\ \hline\hline
Models &Q6 &Q7 &Q8 &Q9 &Q10 \\ \hline
$\rho_Q$ & $\mathbf{2}\oplus\mathbf{1}$ & $\mathbf{2}\oplus\mathbf{1}$ & $\mathbf{2}\oplus\mathbf{1}$ & $\mathbf{2}\oplus\mathbf{1}$ & $\mathbf{2}\oplus\mathbf{1}$   \\ \hline
$\rho_{u^c}$ & $\mathbf{2}\oplus\mathbf{1}'$ & $\mathbf{2}\oplus\mathbf{1}'$ & $\mathbf{2}\oplus\mathbf{1}$ & $\mathbf{2}\oplus\mathbf{1}$ & $\mathbf{2}\oplus\mathbf{1}$   \\ \hline
$\rho_{d^c}$ & $\mathbf{1}\oplus\mathbf{1}\oplus\mathbf{1}$ & $\mathbf{1}'\oplus\mathbf{1}\oplus\mathbf{1}$ & $\mathbf{1}'\oplus\mathbf{1}'\oplus\mathbf{1}$ & $\mathbf{1}'\oplus\mathbf{1}\oplus\mathbf{1}$ & $\mathbf{1}'\oplus\mathbf{1}'\oplus\mathbf{1}$   \\ \hline\hline
$k_{Q}$ & $k,k$ & $k,k$ & $k,k$ & $k,k$ & $k,k$   \\ \hline
$k_{u^c}$ & $-k,6-k$ & $-k,6-k$ & $-k,4-k$ & $-k,4-k$ &  $-k,4-k$  \\ \hline
$k_{d^c}$ & $4-k,2-k,-k$ & $4-k,4-k,2-k$ & $6-k,4-k,2-k$ & $6-k,2-k,-k$ & $6-k,2-k,-k$   \\ \hline\hline
\end{tabular}}
\caption{\label{tab:quark models}
Classification of quark fields in the minimal quark models with $S_4$ modular symmetry and gCP symmetry, where $k$ can be any integer. Notice that the Higgs fields are invariant under $S_4$ with zero modular weight. }
\end{table}

\begin{table}[]
\resizebox{\textwidth}{100mm}{
\begin{tabular}{|c|c|c|c||c|c|c|}\hline
\hline
Models &Q1 &Q2 &Q3 &Models &Q4 &Q5   \\\hline
$\mathrm{Re}\langle\tau\rangle$  &$-0.436841$ & $-0.437014$ & $-0.436195$ &$\mathrm{Re}\langle\tau\rangle$ & $-0.00687942$ &0.493588\\\hline
$\mathrm{Im}\langle\tau\rangle$  &1.81494 &1.81474 &1.81557 &$\mathrm{Im}\langle\tau\rangle$ &1.00472 &0.874580\\\hline
$\beta^u/\alpha_1^u$  &0.00343099 &0.000237927 &0.000235881 &$\beta_1^u/\alpha^u$ &0.479893 &1505.91\\\hline
$\beta_1^d/\alpha^d$  &0.304628 &0.304084 &0.304619 & $\beta^d/\alpha^d$ &680.772 &612.613\\\hline
$\alpha_2^u/\alpha_1^u$  &1.03883 &1.03878 &1.03904 & $\gamma^d_1/\alpha^d$ & 3.32280 &34.3723\\\hline
$\alpha'{}^u/\alpha_1^u$  &1.00004 &1.00006 &1.00000 & $\beta_2^u/\alpha_1^u$ & $-226.241$ & $-0.186995$\\\hline
$\beta_2^d/\alpha^d$  &10.9113 &10.8935 &10.9002 & $\gamma^d_2/\alpha^d$ &39.7179 & $-5.96481$\\\hline
$\alpha_1^u v_u(\GeV)$ &8.84642 &8.84655 &8.84579 & $\alpha^u v_u(\GeV)$ &0.0808452 &0.0865103\\\hline
$\alpha^d v_d(\GeV)$ &0.0236747 & 0.0237130 &0.0236988 & $\alpha^d v_d(\GeV)$ &0.000336231 &0.000279067\\\hline\hline
$\theta^q_{12}$ &0.227433 &0.227325 &0.227439 &$\theta^q_{12}$ &0.227402 &0.227351\\ \hline
$\theta^q_{13}$ &0.00338504 &0.00337795 &0.00340799 &$\theta^q_{13}$ & 0.00318930 &0.00310537\\ \hline
$\theta^q_{23}$ &0.0388938 &0.0389309 &0.0387763 &$\theta^q_{23}$ &0.0386561 &0.0399389\\\hline
$\delta^q_{CP}/{}^\circ$ &69.4363 &69.5128 & 69.4400 &$\delta^q_{CP}/{}^\circ$ & 69.4280 & 70.1150\\ \hline
$m_u/m_c$ &0.00192985 &0.00192161 &0.00192901 &$m_u/m_c$ &0.00260041 &0.00310718\\ \hline
$m_c/m_t$ &0.00273544 &0.00273426 &0.00274461 &$m_c/m_t$ &0.00265548 &0.00297164\\ \hline
$m_d/m_s$ &0.0458926 &0.0458843 & 0.0459400 &$m_d/m_s$ &0.0504604 &0.0507499\\ \hline
$m_s/m_b$ &0.0176515 &0.0176858 &0.0176518 &$m_s/m_b$ &0.0177025 &0.0176849\\\hline
$\chi^2_{\text{min}}$ &0.58521 &0.588937 &0.663495 &$\chi^2_{\text{min}}$ &4.26505 &16.2615\\\hline\hline\hline
Models &Q7 &Q8 &Q10 &Models &Q6 &Q9   \\\hline
$\mathrm{Re}\langle\tau\rangle$ & $-0.495895$ &0.307358 &0.307719 &$\mathrm{Re}\langle\tau\rangle$  & $-0.495886$ & $-0.191783$ \\ \hline
$\mathrm{Im}\langle\tau\rangle$  &0.875601 &2.21966 &2.21896 &$\mathrm{Im}\langle\tau\rangle$  &0.875587 &2.21983\\\hline
$\delta^u/\alpha^u$ &0.218871 &0.607105 &0.606631 &$\delta^u/\alpha^u$  &0.219074 &0.612005 \\\hline
$\gamma^u/\alpha^u$  &1687.47 &258.331 &257.759 & $\gamma^u/\alpha^u$  &1692.88 &258.949 \\\hline
$\beta^d/\alpha^d$ &4.33152 &0.461192 &0.455737 & $\beta^d/\alpha^d$  &102.878 &116.077 \\\hline
$\gamma^d/\alpha^d$ &457.852 & 119.140 &119.036 & $\delta^d_3/\alpha^d$  &0.167218 &0.150554 \\\hline
$\delta^d/\alpha^d$ & 27.0580 &8.82734 &8.79643 &$\delta^d_1/\alpha^d$  &5.93625 &-8.38832 \\\hline
$\alpha^u v_u(\GeV)$ &0.244549 &0.244977 & 0.245520 & $\alpha^u v_u(\GeV)$ &0.243972 &0.244414 \\\hline
$\alpha^d v_d(\GeV)$ &0.00114978 &0.00554085 &0.00554569 & $\alpha^d v_d(\GeV)$ &0.00511694 &0.00568697 \\\hline\hline
$\theta^q_{12}$ &0.227357 &0.227366 & 0.227330 & $\theta^q_{12}$ &0.227368 &0.227365 \\\hline
$\theta^q_{13}$ &0.00333108 &0.00332068 &0.00332544 &$\theta^q_{13}$ &0.00332646 &0.00333852 \\\hline
$\theta^q_{23}$ & 0.0389070 &0.0389306 &0.0390165 &$\theta^q_{23}$ &0.0388726 &0.0389099 \\\hline
$\delta^q_{CP}/{}^\circ$ &69.2159 &69.3956 &69.2651 &$\delta^q_{CP}/{}^\circ$ &69.4425 &69.0265 \\\hline
$m_u/m_c$ &0.00333311 &0.00332201 &0.00332677 &$m_u/m_c$ &0.00332849 &0.00334163 \\ \hline
$m_c/m_t$ &0.00273137 &0.00273614 &0.00274221 &$m_c/m_t$ &0.00272492 &0.00272986 \\ \hline
$m_d/m_s$ &0.0505789 &0.0505368 &0.0499235 &$m_d/m_s$ &0.0504933 &0.0505272 \\ \hline
$m_s/m_b$ &0.0176638 &0.0176765 &0.0176811 &$m_s/m_b$ &0.0176839 &0.0176835 \\ \hline
$\chi^2_{\text{min}}$ &5.64026 &5.64895 &5.70768 &$\chi^2_{\text{min}}$ &5.63916 &5.65984 \\\hline\hline
\end{tabular}}
\caption{\label{tab:quark bset-fit}
Results of fit for the quark models listed in table~\ref{tab:quark models}. }
\end{table}

\subsection{Toward unified description of quarks and leptons }

The flavor structures of quarks and leptons are drastically different from each other, and it is not known at present whether the quark and lepton sectors are dictated by the same fundamental principle or not. In the previous two sections we have discussed individually the possible lepton and quark models with the smallest number of free parameters . In the following, we shall investigate whether quarks and leptons can be simultaneously described by the $S_4$ modular symmetry and gCP. In this scenario, both lepton and quark mass matrices would depend on a common complex modulus $\tau$, and all the CP violation phases in lepton and quark sectors arise from the modulus VEV $\langle \tau \rangle$. The quark-lepton unification has been studied in the context of $A_4$~\cite{Okada:2018yrn,Okada:2019uoy,Okada:2020rjb,Yao:2020qyy}, $T'$~\cite{Lu:2019vgm}, $S'_4$~\cite{Liu:2020akv} and $A'_5$~\cite{Yao:2020zml} modular symmetries. The most predictive model contains fifteen parameters including the real and imaginary part of the modulus $\tau$~\cite{Liu:2020akv}, as far as we know. The unification description of quark and lepton mixing can also be achieved in the paradigm of traditional flavor symmetry combined with gCP, the resulting lepton and quark mixing matrices can be predicted in terms of only four rotation angles if the flavor group and gCP are spontaneously broken down to $Z_2\times CP$ by certain flavons in the neutrino, charged lepton, up quark and down quark  sectors\cite{Lu:2016jit,Li:2017abz,Lu:2018oxc,Lu:2019gqp}. However, the fermion masses are not constrained in this approach, additional symmetries and fields are necessary to realize the required residual symmetry.

\begin{figure}[t!]
\centering
\includegraphics{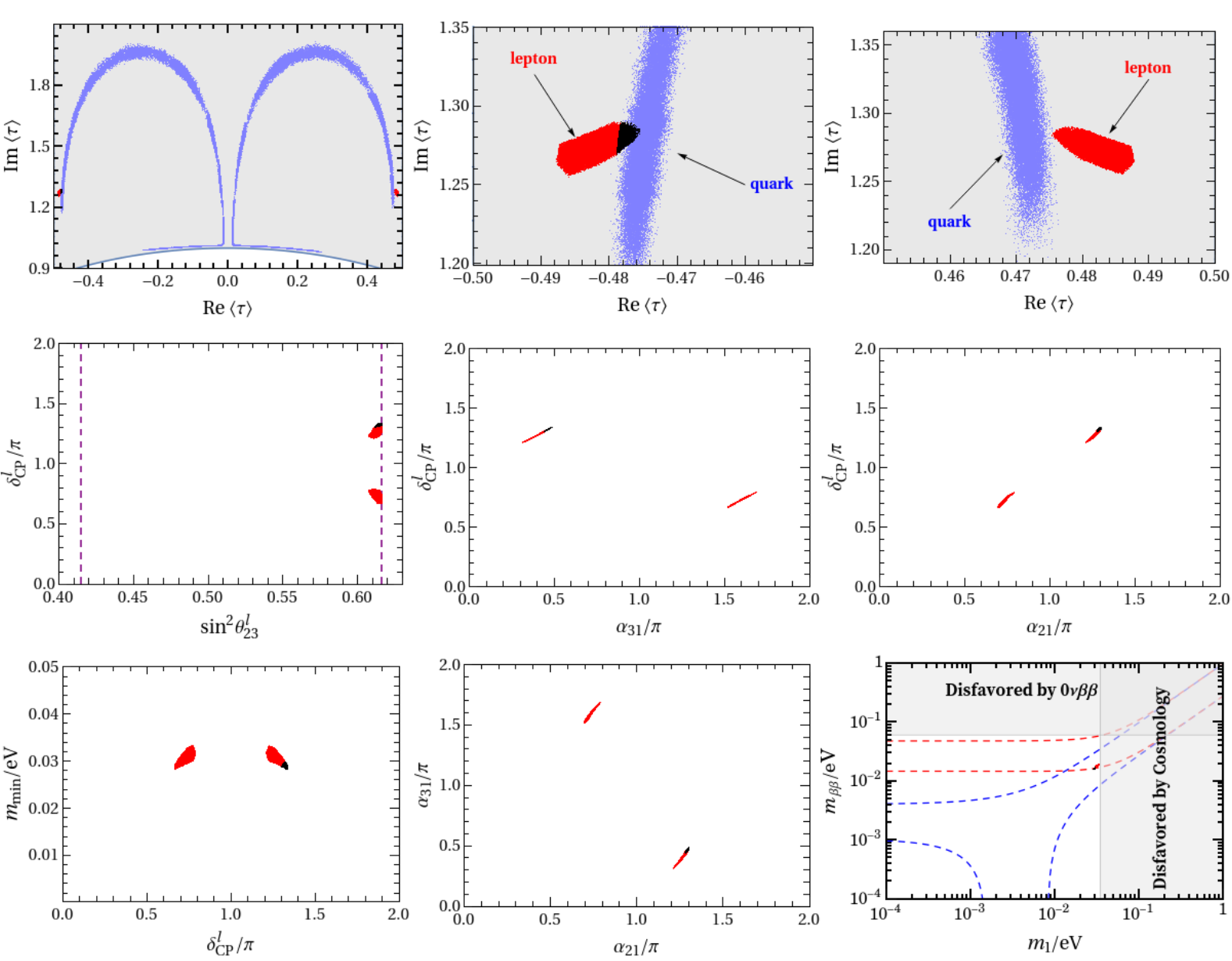}
\caption{The experimentally favored values of $\tau$ are displayed in the first row in the quark-lepton unification model. The quark (lepton) masses and mixing parameters are compatible with experimental data at $3\sigma$ level or better in the light blue (red) area, and the common values of $\tau$ are indicated with black. The second and the third rows are for the correlation among the neutrino masses and mixing parameters.
\label{fig:unified model}}
\end{figure}

By comprehensively scanning the possible quark-lepton models based on $S_4$ modular symmetry and gCP, we find that the minimal models use fifteen independent parameters including $\text{Re}\tau$ and $\text{Im}\tau$ to explain twenty-two observables: six quark masses $m_{u,c,t,d,s,b}$, three quark mixing angles $\theta^{q}_{12,13, 23}$, a quark CP violation phase $\delta^q_{CP}$, three charged lepton masses $m_{e,\mu,\tau}$, three neutrino
masses $m_{1,2,3}$, three lepton mixing angles $\theta^{l}_{12,13,23}$ and three leptonic CP violation phases $\delta^{l}_{CP}$, $\alpha_{21, 31}$.
In the following, we will present a benchmark model which contains five real couplings in the lepton sector and eight free couplings in the quark sector.
The right-handed neutrinos $N^c$ are $S_4$ triplet $\mathbf{3}$, while all the other lepton fields $L$, $E^c$ and quark fields $Q$, $u^c$, $d^c$ are assigned to $\mathbf{2}\oplus\mathbf{1}$. The transformation properties of the matter fields are given by
\begin{eqnarray}
\nonumber&&\rho_{u^c}=\rho_{d^c}=\rho_Q=\rho_{E^c}=\rho_L=\mathbf{2}\oplus\mathbf{1}\,,\,\rho_N^c=\mathbf{3}\,,\\
\nonumber&&k_{u^c_D}=k_{d^c_D}=2-k\,,\,k_{u^c_3}=-k\,,\,k_{d^c_3}=8-k\,,\,k_{Q_D}=k_{Q_3}=k\,,\,\\
&&k_{E^c_D}=k_{E^c_3}=k_{L_D}=k_{L_3}=k_{N^c}=1\,,
\end{eqnarray}
where $k$ is an arbitrary integer. We see that the lepton sector is exactly the aforementioned lepton model L8. From the general results in sections~\ref{sec:Dirac mass} and~\ref{sec:Majorana-mass}, we can read out the fermion mass matrices as follows,
\begin{eqnarray}
\nonumber M_e&=&\left(\begin{matrix}
\alpha^e Y_{\mathbf{2},1}^{(2)} &0 &\beta^e Y_{\mathbf{2},2}^{(2)} \\
0 & \alpha^e Y_{\mathbf{2},2}^{(2)} & \beta^e Y_{\mathbf{2},1}^{(2)} \\
\gamma^e Y_{\mathbf{2},2}^{(2)} & \gamma^e Y_{\mathbf{2},1}^{(2)} &0
\end{matrix}\right)v_d\,,~~~~
M_D=\left(\begin{matrix}
\alpha^DY_{\mathbf{3},2}^{(2)} &\alpha^DY_{\mathbf{3},3}^{(2)} &\beta^DY_{\mathbf{3},1}^{(2)} \\
\alpha^DY_{\mathbf{3},1}^{(2)} &\alpha^DY_{\mathbf{3},2}^{(2)} &\beta^DY_{\mathbf{3},3}^{(2)} \\
\alpha^DY_{\mathbf{3},3}^{(2)} &\alpha^DY_{\mathbf{3},1}^{(2)} &\beta^DY_{\mathbf{3},2}^{(2)}
\end{matrix}\right)v_u\,,\\
\nonumber M_{N^c}&=&\beta^N\left(\begin{matrix}
0&Y_{\mathbf{2},1}^{(2)}&Y_{\mathbf{2},2}^{(2)}\\
Y_{\mathbf{2},1}^{(2)}&Y_{\mathbf{2},2}^{(2)}&0\\
Y_{\mathbf{2},2}^{(2)}&0&Y_{\mathbf{2},1}^{(2)}
\end{matrix}\right)\Lambda\,,\, ~~~~
M_u=\left(\begin{matrix}
\alpha^u Y_{\mathbf{2},1}^{(2)} &0 & \beta^u Y_{\mathbf{2},2}^{(2)} \\
0 &\alpha^u Y_{\mathbf{2},2}^{(2)} & \beta^u Y_{\mathbf{2},1}^{(2)} \\
0 &0 &\delta^u
\end{matrix}\right)v_u\,,\,\\
M_d&=&\left(\begin{matrix}
\alpha^d Y_{\mathbf{2},1}^{(2)} &0 &\beta^d Y_{\mathbf{2},2}^{(2)} \\
0 &\alpha^d Y_{\mathbf{2},2}^{(2)} &\beta^d Y_{\mathbf{2},1}^{(2)} \\
\gamma^d_1 Y_{\mathbf{2}I,2}^{(8)} + \gamma^d_2 Y_{\mathbf{2}II,2}^{(8)}  &\gamma^d_1 Y_{\mathbf{2}I,1}^{(8)} + \gamma^d_2 Y_{\mathbf{2}II,1}^{(8)} &\delta^d Y_{\mathbf{1}}^{(8)}
\end{matrix}\right)v_d\,.
\end{eqnarray}
The best fit values of the input parameters for this unified model is determined to be
\begin{eqnarray}
\nonumber&& \mathrm{Re}\langle\tau\rangle=-0.477059\,,\,~\mathrm{Im}\langle\tau\rangle=1.28145\,,\,~\beta^u/\alpha^u=285.975\,,\,~\delta^u/\alpha^u=1.09959\,,\\
\nonumber& &\gamma^d_1/\alpha^d=2.52377\,,\,~\beta^d/\alpha^d=-94.0867\,,\,~\gamma^d_2/\alpha^d=6.99468\,,\,\delta^d/\alpha^d=-83.5316\,,\\
\nonumber&&\beta^e/\alpha^e=2820.85\,,\,~\gamma^e/\alpha^e=159.393\,,\,\beta^D/\alpha^D=1.07257\,,\,~\alpha^uv_u=0.216464\,\GeV\,,\\
&&
\alpha^dv_d=0.00294862\,\GeV\,,~\alpha^e v_d=0.461746\,\MeV\,,\,~\dfrac{(\alpha^Dv_u)^2}{\beta^N\Lambda}=11.9256\,\meV\,.
\end{eqnarray}
The masses and mixing parameters of leptons and quarks are predicted to be
\begin{eqnarray}
\nonumber&&\sin^2\theta^l_{12}=0.339580\,,\,~~\sin^2\theta^l_{13}=0.0215912\,,\,~~\sin^2\theta^l_{23}=0.615854\,,\\
\nonumber&&\delta^l_{CP}/\pi=1.33145\,,\,~~\alpha_{21}/\pi=1.29710\,,\,~~\alpha_{31}/\pi=0.476112\,,\\
\nonumber&&m_1=29.1179\,\meV\,,~~\,m_2=30.3604\,\meV\,,\,~~m_3=57.1162\,\meV\,,\\
\nonumber&&m_e/m_{\mu}=0.00480040\,,\,~~m_{\mu}/m_{\tau}=0.0565053\,,\,~~m_{\beta\beta}=16.3936\,\meV\,, \\
\nonumber&&m_{\beta}=30.4062\,\meV\,, \,~~ m_u/m_c=0.00192854\,,\,~~m_c/m_t=0.00272947\,,\\
\nonumber&& m_d/m_s=0.0459963\,,\,~~m_s/m_b=0.0178060\,,\,~~\theta^q_{12}=0.227380\,,\\
&&\theta^q_{13}=0.00311374\,,\,~~\theta^q_{23}=0.0394220\,,\,~~\delta^q_{CP}=68.6949^\circ\,,
\end{eqnarray}
which are compatible with experimental data at $3\sigma$ level. All the coupling constants as well as complex modulus $\tau$ are treated as random numbers, and the $3\sigma$ bounds of the mass ratios and mixing angles of both quarks and leptons are imposed. The values of $\tau$ compatible with experimental data are shown in figure~\ref{fig:unified model},  the light blue and red areas represent the regions favored by the quark and lepton data respectively. We see that there indeed exists a small overlap region of $\tau$ indicated by black in which the flavor structure of quarks and leptons can be accommodated simultaneously. Moreover, we display the correlations among neutrino masses and mixing parameters in figure~\ref{fig:unified model}. Since the common $\tau$ region of quark and lepton sectors is very small, the allowed values of all observables scatter in quite narrow ranges around their best fit values.

\section{\label{sec:conclusion}Conclusions and discussions }

Modular invariance is an promising framework to address the flavor puzzle of the standard model. In recent years, much effort has gone into the study of lepton models based on inhomogeneous and homogeneous finite modular groups. In the present work, we have performed a systematical analysis of the possible lepton and quark models with $S_4$ modular symmetry. Aiming at the minimal and predictive models, we impose the gCP symmetry so that all coupling constants are constrained to be real in our working basis and the vacuum expectation of the modulus is the unique source of modular and CP symmetry breaking. In the known $S_4$ modular symmetry models~\cite{Penedo:2018nmg,Novichkov:2018ovf,deMedeirosVarzielas:2019cyj,Kobayashi:2019mna,King:2019vhv,Criado:2019tzk,Wang:2019ovr,Wang:2020dbp,King:2021fhl,Ding:2021zbg},  usually the three generations of left-handed lepton fields and right-handed charged lepton are assumed to transform as triplet and singlet under $S_4$. Besides the singlet representations $\mathbf{1}$, $\mathbf{1}'$ and triplet representations $\mathbf{3}$ and $\mathbf{3}'$, the $S_4$ group has a doublet irreducible representation $\mathbf{2}$. The presence of doublet representation not only introduces new features in the modular invariant lepton models, but also provides a new expedient to describe the quark sector. We give the most general analytical expressions of the modular invariant Yukawa superpotential of charged fermions and the Majorana mass terms of right-handed neutrinos. We have analyzed both scenarios that neutrinos are Majorana particles and Dirac particles. Under the assumption of Majorana neutrinos, the light neutrino masses are generated by the type I seesaw mechanism, and the conventional seesaw models with three right-handed neutrinos and the minimal seesaw models with two right-handed neutrinos are analyzed.

We have comprehensively searched for the $S_4$ modular invariant lepton and quark models with the lowest possible number of free parameters. After heavy numerical analysis, we find that the minimal lepton models make use of five real couplings together with the modulus $\tau$ to describe the charged lepton masses, neutrino masses, lepton mixing angles and CP violation phases. Thirteen minimal lepton models are obtained, including nine Majorana neutrino models and four Dirac neutrino models, the classification of the matter fields under modular symmetry is summarized in table~\ref{tab:lepton models} and table~\ref{tab:Dirac lepton models}. Notice that the models L2 and D3 were already presented in Ref.~\cite{Novichkov:2019sqv} and Ref.~\cite{Wang:2020dbp} respectively while all others are new. The experimental data from neutrino oscillation, neutrinoless double decay, tritium beta decay and cosmology on neutrino mass sum can be well accommodated, as shown in table~\ref{tab:lepton bset-fit} and table~\ref{tab:Dirac lepton models}. Moreover, the predictions of these models are expected to be tested at forthcoming experiments with higher sensitivities. In most modular symmetry models, the right-handed are assumed to be singlets of modular group so that at least one parameter is introduced for each charged lepton and the hierarchies among electron, muon and tau masses can be reproduced by tuning the coupling constants. From table~\ref{tab:lepton models}, we see that other assignments such as the triplet and double plus singlet can also be in agreement with the experimental data. The model L1 is particularly interesting, all the lepton fields $L$, $E^c$ and $N^c$ transform as triplet $\mathbf{3}$ under $S_4$, the light neutrino mass matrix only depends on the modulus $\tau$ and an overall scale, the four coupling constants in the charged lepton mass matrices are of the same order of magnitude and the charged lepton mass hierarchies arise from the deviation from the fixed point $\tau=i$.

Because gCP symmetry enforces all coupling constant to be real and the complex phases in the mass matrices originate from the modular forms in our models, the CP violation in leptogenesis is strongly correlated with the Dirac and Majorana CP violation phases.
As an example, we have studied the leptogenesis in the model L9. By numerically solving the Boltzmann equations, we find that the baryon asymmetry of the universe, lepton masses and mixing angles can be correctly obtained in a small region of $\tau$. The allowed range of the Dirac CP phase $\delta^{l}_{CP}$ would shrink significantly if the measured value of the baryon asymmetry $Y_B$ is taken into account.

As regards the quark models with $S_4$ modular symmetry, at least seven real coupling constants are necessary to describe the hierarchies patterns of quark masses and mixing angles, and ten minimal quark models are found, as listed in table~\ref{tab:quark models}.
Furthermore, we investigate whether the $S_4$ modular symmetry can address both the lepton and quark flavor problems, and then the single complex modulus $\tau$ would be shared by the quark and lepton sectors. A typical quark-lepton unification model is presented, the lepton sector is the model L8 which contains five couplings, and the quark sector uses eight parameters. The value of $\tau$ is dominantly fixed by the experimental data of lepton masses and mixing angles. The allowed range of $\tau$ and the allowed values of quark and lepton masses and mixing parameters are very narrow, and this model can be tested in future neutrino experiments.

\subsection*{Acknowledgements}
This work is supported by the National Natural Science Foundation of China under Grant Nos 11975224, 11835013 and 11947301.

\section*{Appendix}

\setcounter{equation}{0}
\renewcommand{\theequation}{\thesection.\arabic{equation}}

\begin{appendix}

\section{\label{sec:S4_group}$S_{4}$ group }

The finite modular group $\Gamma_4$ is isomorphic to the permutation group $S_4$ which is the group of all permutations of four elements. Geometrically $S_4$ is the group of orientation-preserving symmetries of the cube or equivalently the octahedron. As shown in Eq.~\eqref{eq:mul-rules-GammaN}, the inhomogeneous finite modular group $\Gamma_4\cong S_4$ can be generated by the modular generators $S$ and $T$ satisfying the relations
\begin{equation}
S^2=(ST)^3=T^4=\mathbb{1}\,.
\end{equation}
In the paradigm of traditional flavor symmetry, it is convenient to express the $S_4$ group in terms of three generators $\hat{S}$, $\hat{T}$ and $\hat{U}$ obeying the multiplication rules~\cite{Ding:2013hpa}:
\begin{eqnarray}
\hat{S}^2=\hat{T}^3=\hat{U}^2=(\hat{S}\hat{T})^3=(\hat{S}\hat{U})^2=(\hat{T}\hat{U})^2=(\hat{S}\hat{T}\hat{U})^4=1\,.
\end{eqnarray}
The generators $\hat{S}$ and $\hat{T}$ alone generate the group $A_4$, and the generators $\hat{T}$ and $\hat{U}$ alone generate the group $S_3$. The two different choices of generators are related as follows
\begin{equation}
S = \hat{S} \hat{U},\, T = \hat{S} \hat{T}^2 \hat{U},\,ST = \hat{T}\,,~~\text{or}~~\hat{S} = (ST^2) ^2,\, \hat{T} = ST,\, \hat{U} = T^2ST^2\,.
\end{equation}
The $S_4$ group has 24 elements and 5 irreducible representations including two singlet representations $\mathbf{1}$, $\mathbf{1}'$, a doublet representation $\mathbf{2}$ and two triplet representations $\mathbf{3}$ and $\mathbf{3}'$. In this work, we choose the same representation basis as that of~\cite{Ding:2013hpa}, i.e. the representation matrix of the generator $\hat{T}$ is diagonal. We summarize the representation matrices of the generators in table~\ref{tab:S4_rep}.

\begin{table}[h!]
\renewcommand{\tabcolsep}{0.58mm}
\centering
\resizebox{1.0\textwidth}{!}{
\begin{tabular}{|c||c|c|c||c|c|}\hline\hline
 ~~  &  $\rho_\mathbf{r}(\hat{S}$)  &   $\rho_\mathbf{r}(\hat{T})$    &  $\rho_\mathbf{r}(\hat{U})$ &  $\rho_\mathbf{r}(S)$  &   $\rho_\mathbf{r}(T)$  \\ \hline
~~~${\bf 1}$, ${\bf 1^\prime}$ ~~~ & 1   &  1  & $\pm1$ & $\pm1$   &  $\pm1$ \\ \hline
   &   &    &   &   &  \\ [-0.16in]
${\bf 2}$ &  $\left( \begin{array}{cc}
    1&~0 \\
    0&~1
    \end{array} \right) $
    & $\left( \begin{array}{cc}
    \omega&~0 \\
    0&~\omega^2
    \end{array} \right) $
    & $\left( \begin{array}{cc}
    0&~1 \\
    1&~0
    \end{array} \right)$ &  $\left( \begin{array}{cc}
    0&~1 \\
    1&~0
\end{array} \right) $
& $\left( \begin{array}{cc}
0 ~&~ \omega^2 \\
\omega ~&~ 0
\end{array} \right) $ \\ [0.12in]\hline
   &   &    &   &  &  \\ [-0.16in]
${\bf 3}$, ${\bf 3^\prime}$ & $\dfrac{1}{3} \left(\begin{array}{ccc}
    -1&~ 2  ~& 2  \\
    2  &~ -1  ~& 2 \\
    2 &~ 2 ~& -1
    \end{array}\right)$
    & $\left( \begin{array}{ccc}
    1 &~ 0 ~& 0 \\
    0 &~ \omega^{2} ~& 0 \\
    0 &~ 0 ~& \omega
    \end{array}\right) $
    & $\mp\left( \begin{array}{ccc}
    1 &~ 0 ~& 0 \\
    0 &~ 0 ~& 1 \\
    0 &~ 1 ~& 0
    \end{array}\right)$ & $\pm\dfrac{1}{3} \left(\begin{array}{ccc}
    1&~ -2  ~& -2  \\
    -2  &~ -2  ~& 1 \\
    -2 &~ 1 ~& -2
\end{array}\right)$
& $\pm\dfrac{1}{3}\left( \begin{array}{ccc}
1 &~ -2\omega^2 ~& -2\omega \\
-2 &~ -2\omega^2 ~& \omega \\
-2 &~ \omega^2 ~& -2\omega
\end{array}\right) $
\\[0.22in] \hline \hline
\end{tabular}}
\caption{\label{tab:S4_rep}The representation matrices of the generators $\hat{S}$,$\hat{T}$,$\hat{U}$ as well as $S,T$ in the five irreducible representations of $S_4$ in the chosen basis, where $\omega=e^{2\pi i/3}$ is the cube root of unit. }
\end{table}

We now list the Kronecker products and the corresponding Clebsch-Gordan coefficients which are quite useful when constructing $S_4$ modular invariant models. For convenience, we denote $\mathbf{1}\equiv\mathbf{1}^0$, $\mathbf{1}'\equiv\mathbf{1}^1$, $\mathbf{3}\equiv\mathbf{3}^0$, $\mathbf{3}'\equiv\mathbf{3}^1$ for the singlet or triplet representations. We shall use $\alpha_i$ to denote the elements of first representation and $\beta_i$ stands for the elements of the second representation of the tensor product $\mathbf{R_1}\otimes \mathbf{R_2}$, where $\mathbf{R_1}$ and $\mathbf{R_2}$ are two irreducible representations of $S_4$.

\begin{equation}
\begin{aligned}
&{}\mathbf{1}^i\otimes\mathbf{1}^j=\mathbf{1}^{<i+j>}\sim\alpha\beta \,,\\
&\mathbf{1}^i\otimes\mathbf{2}=\mathbf{2}\sim\left(\begin{matrix}
\alpha\beta_1\\
(-1)^{i}\alpha\beta_2
\end{matrix}\right) \,,\\
&\mathbf{1}^i\otimes\mathbf{3}^j=\mathbf{3}^{<i+j>}\sim\left(\begin{matrix}
\alpha\beta_1\\
\alpha\beta_2\\
\alpha\beta_3
\end{matrix}\right) \,,\\
&\mathbf{2}\otimes\mathbf{2}=\mathbf{1}^0\oplus\mathbf{1}^1\oplus\mathbf{2}\,,\quad
\begin{cases}
\mathbf{1}^l\sim\alpha_1\beta_2+(-1)^l\alpha_2\beta_1 \\
\mathbf{2}\sim\left(\begin{matrix}
\alpha_2\beta_2\\
\alpha_1\beta_1
\end{matrix}\right)
\end{cases} \,,\\
&\mathbf{2}\otimes\mathbf{3}^i=\mathbf{3}^0\oplus\mathbf{3}^1\,,\quad
\mathbf{3}^l\sim\left(\begin{matrix}
\alpha_1\beta_2\\
\alpha_1\beta_3\\
\alpha_1\beta_1
\end{matrix}\right)+(-1)^{i+l}
\left(\begin{matrix}
\alpha_2\beta_3\\
\alpha_2\beta_1\\
\alpha_2\beta_2
\end{matrix}\right) \,,\\
&\mathbf{3}^i\otimes\mathbf{3}^j=\mathbf{1}^{<i+j>}\oplus\mathbf{2}\oplus\mathbf{3}^0\oplus\mathbf{3}^1\,,\quad
\begin{cases}
\mathbf{1}^{<i+j>}\sim\alpha_1\beta_1+\alpha_2\beta_3+\alpha_3\beta_2\\
\mathbf{2}\sim\left(\begin{matrix}
\alpha_1\beta_3+\alpha_2\beta_2+\alpha_3\beta_1\\
(-1)^{i+j}(\alpha_1\beta_2+\alpha_2\beta_1+\alpha_3\beta_3)
\end{matrix}\right)\\
\mathbf{3}^{l}\sim\left(\begin{matrix}
\alpha_1\beta_1-\alpha_3\beta_2\\
\alpha_3\beta_3-\alpha_2\beta_1\\
\alpha_2\beta_2-\alpha_1\beta_3
\end{matrix}\right)-(-1)^{i+j+l}\left(\begin{matrix}
\alpha_1\beta_1-\alpha_2\beta_3\\
\alpha_3\beta_3-\alpha_1\beta_2\\
\alpha_2\beta_2-\alpha_3\beta_1
\end{matrix}\right)
\end{cases}\,,
\end{aligned}
\end{equation}
where $i, j, l=0, 1$ and we define the notation $<i>\equiv i~(\mathrm{mod}~2)$.

\section{\label{sec:modularform_of_N=4}Modular multiplets of weight $4, 6, 8$ at level $N=4$}


From the tensor products of lower weight modular forms with the help of the Clebsch Gordan coefficients of $S_4$ in Appendix~\ref{sec:S4_group}, we can get the higher weight modular forms. In the following we construct the weight 4, weight 6 and weight 8 modular multiplets. By expressing the modular forms $Y_{1,2,3,4,5}$ in terms of $\vartheta_1$ and $\vartheta_2$ as shown in Eq.~\eqref{eq:wt2MF_express}, we can easily identify the linearly independent modular multiplets of higher weights without examining the cumbersome non-linear constraints which relate redundant multiplets coming from tensor products. The linear space of modular forms of level 4 and weight $k$ has dimension $2k+1$. In the following, we shall present the explicit form of the modular forms of weight $4, 6, 8$.

The weight 4 modular forms can be generated from the tensor products of $Y^{(2)}_{\mathbf{2, 3}}$. Using the Kronecker products $\mathbf{2}\otimes\mathbf{2}=\mathbf{1}\oplus\mathbf{1'}\oplus\mathbf{2}$, $\mathbf{2}\otimes\mathbf{3}=\mathbf{3}\oplus\mathbf{3'}$ and $\mathbf{3}\otimes\mathbf{3}=\mathbf{1}\oplus\mathbf{2}\oplus\mathbf{3}\oplus\mathbf{3'}$, the weight 4 modular forms can be arranged into different $S_4$ irreducible representations $\mathbf{1}$, $\mathbf{2}$, $\mathbf{3}$ and $\mathbf{3'}$.
To be more specific, we have
\begin{eqnarray}
\nonumber&& Y^{(4)}_{\mathbf{1}}=\left(Y^{(2)}_{\mathbf{2}}Y^{(2)}_{\mathbf{2}}\right)_{\mathbf{1}}= 2Y_1Y_2 \,, \\
\nonumber&& Y^{(4)}_{\mathbf{2}}=\left(Y^{(2)}_{\mathbf{2}}Y^{(2)}_{\mathbf{2}}\right)_{\mathbf{2}}=\Big( Y^2_2,~ Y^2_1 \Big)^T \,, \\
\nonumber&& Y^{(4)}_{\mathbf{3}}=\left(Y^{(2)}_{\mathbf{2}}Y^{(2)}_{\mathbf{3}}\right)_{\mathbf{3}}=\Big(Y_1Y_4+Y_2Y_5, ~ Y_1Y_5+Y_2Y_3,~ Y_1Y_3+Y_2Y_4 \Big)^T, \\
&& Y^{(4)}_{\mathbf{3'}}=\left(Y^{(2)}_{\mathbf{2}}Y^{(2)}_{\mathbf{3}}\right)_{\mathbf{3'}}=\Big(Y_1Y_4-Y_2Y_5, ~ Y_1Y_5-Y_2Y_3, ~ Y_1Y_3-Y_2Y_4 \Big)^T\,.
\end{eqnarray}
Then we proceed to construct weight 6 modular multiplets from the tensor products of $Y^{(2)}_{\mathbf{2, 3}}$ and $Y^{(4)}_{\mathbf{1,2,3,3'}}$.
We find thirteen linearly independent weight 6 modular forms which can be decomposed into two singlets $\mathbf{1}$, $\mathbf{1}'$, a doublet $\mathbf{2}$ and three triplets $\mathbf{3}$ and $\mathbf{3}'$ of $S_4$ as follows
\begin{eqnarray}
\nonumber&&Y^{(6)}_{\mathbf{1}}=\left(Y^{(2)}_{\mathbf{2}}Y^{(4)}_{\mathbf{2}}\right)_{\mathbf{1}}= Y^3_1+Y^3_2 \,, \\
\nonumber&&Y^{(6)}_{\mathbf{1'}}=\left(Y^{(2)}_{\mathbf{2}}Y^{(4)}_{\mathbf{2}}\right)_{\mathbf{1'}}= Y^3_1-Y^3_2\,, \\
\nonumber&&Y^{(6)}_{\mathbf{2}}=\left(Y^{(2)}_{\mathbf{2}}Y^{(4)}_{\mathbf{1}}\right)_{\mathbf{2}}=2Y_1Y_2\Big(Y_1,~ Y_2\Big)^T\,, \\
\nonumber&&Y^{(6)}_{\mathbf{3}I}=\left(Y^{(2)}_{\mathbf{3}}Y^{(4)}_{\mathbf{1}}\right)_{\mathbf{3}}=2Y_1Y_2\Big(Y_3,~ Y_4, ~ Y_5\Big)^T\,, \\
\nonumber&&Y^{(6)}_{\mathbf{3}II}=\left(Y^{(2)}_{\mathbf{3}}Y^{(4)}_{\mathbf{2}}\right)_{\mathbf{3}}=\Big(Y_2^2Y_4 + Y_1^2Y_5,~Y_2^2Y_5  + Y_1^2Y_3,~ Y_2^2Y_3 + Y_1^2 Y_4 \Big)^T\,,\\
&&Y^{(6)}_{\mathbf{3'}}=\left(Y^{(2)}_{\mathbf{3}}Y^{(4)}_{\mathbf{2}}\right)_{\mathbf{3'}}=\Big(Y^2_2Y_4-Y^2_1Y_5,~Y^2_2Y_5-Y^2_1Y_3,~ Y^2_2Y_3-Y^2_1Y_4 \Big)^T\,.
\end{eqnarray}
Finally, the weight 8 modular multiplets can be obtained from the tensor products of $Y^{(2)}_{\mathbf{2,3}}$ and $Y^{(6)}_{\mathbf{1,1',2},\mathbf{3}I,\mathbf{3}II,\mathbf{3'}}$,
and they decompose as $\mathbf{1}\oplus\mathbf{2}\oplus\mathbf{2}\oplus\mathbf{3}\oplus\mathbf{3}\oplus\mathbf{3}'\oplus\mathbf{3}'$ under $S_4$:
\begin{eqnarray}
\nonumber&&Y^{(8)}_{\mathbf{1}}=\left(Y^{(2)}_{\mathbf{2}}Y^{(6)}_{\mathbf{2}}\right)_{\mathbf{1}}= 4Y_1^2Y_2^2 , \\
\nonumber&&Y^{(8)}_{\mathbf{2}I}=\left(Y^{(2)}_{\mathbf{2}}Y^{(6)}_{\mathbf{1}}\right)_{\mathbf{2}}=(Y_1^3+Y_2^3)\Big(Y_1,~ Y_2\Big)^T, \\
\nonumber&&Y^{(8)}_{\mathbf{2}II}=\left(Y^{(2)}_{\mathbf{2}}Y^{(6)}_{\mathbf{1'}}\right)_{\mathbf{2}}=(Y_1^3-Y_2^3)\Big(Y_1,~ -Y_2\Big)^T, \\
\nonumber&&Y^{(8)}_{\mathbf{3}I}=\left(Y^{(2)}_{\mathbf{2}}Y^{(6)}_{\mathbf{3}I}\right)_{\mathbf{3}}=2Y_1Y_2\Big(Y_1Y_4+Y_2Y_5,~ Y_1Y_5+Y_2Y_3, ~ Y_1Y_3+Y_2Y_4\Big)^T, \\
\nonumber&&Y^{(8)}_{\mathbf{3}II}=\left(Y^{(2)}_{\mathbf{3}}Y^{(6)}_{\mathbf{1}I}\right)_{\mathbf{3}}=(Y_1^3+Y_2^3)\Big(Y_3,~ Y_4,~ Y_5 \Big)^T, \\
\nonumber&&Y^{(8)}_{\mathbf{3'}I}=\left(Y^{(2)}_{\mathbf{2}}Y^{(6)}_{\mathbf{3}I}\right)_{\mathbf{3'}}=2Y_1Y_2\Big(Y_1Y_4-Y_2Y_5,~ Y_1Y_5-Y_2Y_3, ~ Y_1Y_3-Y_2Y_4\Big)^T, \\
&&Y^{(8)}_{\mathbf{3'}II}=\left(Y^{(2)}_{\mathbf{3}}Y^{(6)}_{\mathbf{1'}I}\right)_{\mathbf{3'}}=(Y_1^3-Y_2^3)\Big(Y_3,~ Y_4,~ Y_5 \Big)^T.
\end{eqnarray}
The dimension of the modular forms space ${\cal M}_8(\Gamma(4))$ is equal to $17$ .
\end{appendix}

\section{\label{sec:two right-handed neutrino}Classifying the minimal seesaw models with $S_4$ modular symmetry }

If the light neutrino masses originate from the type-I seesaw mechanism, the non-zero solar and atmospheric neutrino mass squared differences requires at least two right-handed neutrinos. The two right-handed neutrino models are the so-called minimal seesaw models.
In the following, we shall systematically classify the neutrino superpotential for both doublet and singlet assignments of right-handed neutrinos.

\begin{itemize}[labelindent=-0.8em, leftmargin=1.5em]

\item{$N^c_D=(N^c_1, N^c_2)\sim\mathbf{2}$}

The modular weight of $N^c$ is denoted as $k_{N^c}$. Since the $S_4$ contraction $\bf{2}\otimes\bf{2}\rightarrow\bf{1}'$ is antisymmetric with respect to the two components of the doublet, the modular forms in the $\mathbf{1}'$ representation can not appear in the Majorana mass terms of the right-handed neutrinos. The most general form of the superpotential for the heavy neutrino masses is given by
\begin{equation}
\begin{aligned}
\mathcal{W}_{N^c}=&\Lambda(N^c_DN^c_Df_M(Y))_{\bf{1}}\\
=&\sum_{a=1}^2\sum_{b=1}^2N^c_aN^c_b\left(\alpha Y_{\mathbf{1},\prec 1-a-b\succ}^{(2k_{N^c_D})}+\sum_A\beta_{A}Y_{\mathbf{2}A,\prec -a-b\succ}^{(2k_{N^c_D})}\right)\Lambda\,,
\end{aligned}
\end{equation}
which leads to
\begin{equation}
M_{N^c}=\left(\begin{matrix}
\beta_{A}Y_{\mathbf{2}A,1}^{(2k_{N^c_D})}&\alpha Y_{\mathbf{1}}^{(2k_{N^c_D})}\\
\alpha Y_{\mathbf{1}}^{(2k_{N^c_D})}&\beta_{A}Y_{\mathbf{2}A,2}^{(2k_{N^c_D})}
\end{matrix}\right)\Lambda\,.
\end{equation}
We proceed to consider the neutrino Dirac Yukawa couplings. If the  left-handed leptons transform as a triplet $\mathbf{3}^{j}$ under $S_4$ with modular weight $k_L$, the superpotential for the Dirac neutrino masses is of the form
\begin{equation}
\begin{aligned}
\mathcal{W}_D=&{}\alpha(N^c_DLf_{D}(Y))_{\bf{1}}H_u\\
=&\sum_{a=1}^2\sum_{b=1}^3N^c_aL_b\sum_{l=0}^1\sum_A (-1)^{(a+1)(j+l)}\alpha^l_{A}Y_{\mathbf{3}^lA,\prec 2+a-b\succ}^{(k_{N^c_D}+k_L)}H_u\,,
\end{aligned}
\end{equation}
The Dirac neutrino mass matrix can be read off as follows
\begin{equation}
	M_D=\left(\begin{matrix}
		\alpha^l_{A}Y_{\mathbf{3}^lA,2}^{(k_{N^c_D}+k_L)}&\alpha^l_{A}Y_{\mathbf{3}^lA,1}^{(k_{N^c_D}+k_L)} &\alpha^l_{A}Y_{\mathbf{3}^lA,3}^{(k_{N^c_D}+k_L)}\\
		(-1)^{j+l}\alpha^l_{A}Y_{\mathbf{3}^lA,3}^{(k_{N^c_D}+k_L)} &(-1)^{j+l}\alpha^l_{A}Y_{\mathbf{3}^lA,2}^{(k_{N^c_D}+k_L)} &(-1)^{j+l}\alpha^l_{A}Y_{\mathbf{3}^lA,1}^{(k_{N^c_D}+k_L)}
	\end{matrix}\right)v_u\,.
\end{equation}
Under the representation transformation $L:\mathbf{3}^j\rightarrow\mathbf{3}^{j+1}$, the mass matrix $M_D$ changes into $M_D\rightarrow\mathrm{diag}\{1,-1\}M_D$. As a consequence, the light  neutrino mass matrix given by seesaw formula is left invariant if the free coupling $\alpha$ is changed into $-\alpha$. The transformation of the charged lepton mass matrix under $L:\mathbf{3}^j\rightarrow\mathbf{3}^{j+1}$ can be read from table~\ref{tab:mass-trans}, then we know that the predictions for lepton masses and mixing matrix are preserved. Therefore it is sufficient to only consider the case of $L\sim\mathbf{3}$ for the triplet assignment.

For the doublet plus singlet assignment of the left-handed leptons: $L_D\equiv(L_1,L_2)\sim\mathbf{2}$ and $L_3\sim\mathbf{1}^{j}$ whose modular weights are denoted as $k_{L_D}$ and $k_{L_3}$ respectively, the superpotential of the neutrino Yukawa coupling is
\begin{equation}
\begin{aligned}
\mathcal{W}_D=&{}\alpha(N^c_DL_Df_{DD}(Y))_{\bf{1}}H_u+\beta(N^c_DL_3f_{D3}(Y))_{\bf{1}}H_u\\
=&\sum_{a=1}^2\sum_{b=1}^2N^c_aL_b\left[\alpha^l_{1}\sum_{l=0}^1(-1)^{l(a+1)}Y_{\mathbf{1}^l,\prec 1-a-b\succ}^{(k_{N^c_D}+k_{L_D})}+\sum_A \alpha_{2A}Y_{\mathbf{2}A,\prec -a-b\succ}^{(k_{N^c_D}+k_{L_D})}\right]H_u\\
&+\sum_{a=1}^2N^c_aL_3\sum_B \beta_{B}(-1)^{j(a+1)}Y_{\mathbf{2}B,\prec -a\succ}^{(k_{N^c_D}+k_{L_3})}H_u \,.
\end{aligned}
\end{equation}
Then Dirac neutrino mass matrix is given by
\begin{equation}
M_D=\left(\begin{matrix}
\alpha_{2A}Y_{\mathbf{2}A,1}^{(k_{N^c_D}+k_{L_D})} &\alpha^l_{1}Y_{\mathbf{1}^l}^{(k_{N^c_D}+k_{L_D})} &\beta_{B}Y_{\mathbf{2}B,2}^{(k_{N^c_D}+k_{L_3})} \\
\alpha^l_{1}(-1)^{l}Y_{\mathbf{1}^l}^{(k_{N^c_D}+k_{L_D})} &\alpha_{2A}Y_{\mathbf{2}A,2}^{(k_{N^c_D}+k_{L_D})} &\beta_{B}(-1)^{j}Y_{\mathbf{2}B,1}^{(k_{N^c_D}+k_{L_3})}
\end{matrix}\right)v_u \,.
\end{equation}
If we change the representation $L_3:\mathbf{1}^j\rightarrow\mathbf{1}^{j+1}$ as well as the couplings $\alpha^l_1\rightarrow-\alpha^l_{1}$, the mass matrix $M_D$ becomes $M_D\rightarrow \mathrm{diag}\{1,-1\}M_D\mathrm{diag}\{1,-1,1\}$ and the light neutrino mass matrix transforms as $M_\nu\rightarrow\mathrm{diag}\{1,-1,1\}M_\nu\mathrm{diag}\{1,-1,1\}$.
Taking into account the charged lepton sector, the phase $\mathrm{diag}\{1,-1,1\}$ can be absorbed by field redefinition. Thus the field $L_3$ can be assigned to the trivial singlet $\mathbf{1}$ of $S_4$ without loss of generality.

\item{$N^c_1\sim\mathbf{1}^{i_1}$, $N^c_2\sim \mathbf{1}^{i_2}$}

In this case, the superpotential of the right-handed neutrino mass terms is
\begin{equation}
\mathcal{W}_{N^c}=\sum_{a=1}^2\sum_{b=1}^2 \Lambda \alpha_{ab}N^c_aN^c_bf_{ab}(Y)=\sum_{a=1}^2\Lambda \alpha_{aa}N^c_aN^c_aY_{\mathbf{1}}^{(2k_{N^c_a})}+2\Lambda \alpha_{12}N^c_1N^c_2Y_{\mathbf{1}^{<i_1+i_2>}}^{(k_{N^c_a}+k_{N^c_b})}\,.
\end{equation}
The mass matrix $M_{N^c}$ reads as
\begin{equation}
M_{N^c}=\left(\begin{matrix}
\alpha_{11}Y_{\mathbf{1}}^{(2k_{N^c_1})} ~&\alpha_{12}Y_{\mathbf{1}^{<i_1+i_2>}}^{(k_{N^c_1}+k_{N^c_2})} \\
\alpha_{12}Y_{\mathbf{1}^{<i_1+i_2>}}^{(k_{N^c_1}+k_{N^c_2})} ~& \alpha_{22}Y_{\mathbf{1}}^{(2k_{N^c_2})}
\end{matrix}\right)\Lambda\,.
\end{equation}
For the triplet assignment of left-handed lepton fields $L\sim\mathbf{3}^j$, the superpotential of the neutrino Dirac coupling takes the following form,
\begin{equation}
\begin{aligned}
\mathcal{W}_D=&{}[\alpha(N^c_1Lf_{1}(Y))_{\bf{1}}+\beta(N^c_2Lf_{2}(Y))_{\bf{1}}]H_{u}\\
=&\sum_{b=1}^3\sum_A\alpha_{A}N^c_1L_bY_{\mathbf{3}^{< i_1+j >}A,\prec 2-b\succ}^{(k_{N^c_1}+k_{L})}H_{u}
+\sum_{b=1}^3\sum_B\beta_{B}N^c_2L_bY_{\mathbf{3}^{< i_2+j >}A,\prec 2-b\succ}^{(k_{N^c_2}+k_{L})}H_{u}\,,
\end{aligned}
\end{equation}
which gives the Dirac mass matrix
\begin{equation}
M_D=\left(\begin{matrix}
\alpha_AY_{\mathbf{3}^{< i_1+j >}A,1}^{(k_{N^c_1}+k_{L})}&\alpha_AY_{\mathbf{3}^{< i_1+j >}A,3}^{(k_{N^c_1}+k_{L})}&\alpha_AY_{\mathbf{3}^{< i_1+j >}A,2}^{(k_{N^c_1}+k_{L})}\\
\beta_BY_{\mathbf{3}^{< i_2+j >}B,1}^{(k_{N^c_2}+k_{L})}&\beta_BY_{\mathbf{3}^{< i_2+j >}B,3}^{(k_{N^c_2}+k_{L})}&\beta_BY_{\mathbf{3}^{< i_2+j >}B,2}^{(k_{N^c_2}+k_{L})}
\end{matrix}\right)v_{u}\,.
\end{equation}
{If the left-handed lepton fields transform as doublet and singlet under $S_4$: $L_D\equiv(L_1,L_2)\sim\mathbf{2}$ and $L_3\sim\mathbf{1}^{j}$,}
For the doublet and singlet assignment: $L_D\equiv(L_1,L_2)\sim\mathbf{2}$ and $L_3\sim\mathbf{1}^{j}$, the Dirac neutrino mass terms are
\begin{eqnarray}
\nonumber\mathcal{W}_D&=&[\alpha(N^c_1L_Df_{1D}(Y))_{\bf{1}}+\beta(N^c_2L_Df_{2D}(Y))_{\bf{1}}+\delta_1 (N^c_1L_3f_{13}(Y))_{\mathbf{1}}+\delta_2 (N^c_2L_3f_{23}(Y))_{\mathbf{1}}]H_{u}\\
\nonumber&=&\Big[\alpha_A N^c_1(L_1Y_{\mathbf{2}A,2}^{(k_{N^c_1}+k_{L_D})}+(-1)^{i_1}L_2Y_{\mathbf{2}A,1}^{(k_{N^c_1}+k_{L_D})})+\delta_1 N^c_1L_3Y_{\mathbf{1}^{<i_1+j>}}^{(k_{N^c_1}+k_{L_3})}\\
&&+\beta_B N^c_2(L_1Y_{\mathbf{2}B,2}^{(k_{N^c_2}+k_{L_D})}+(-1)^{i_2}L_2Y_{\mathbf{2}B,1}^{(k_{N^c_2}+k_{L_D})})+\delta_2 N^c_2L_3Y_{\mathbf{1}^{<i_2+j>}}^{(k_{N^c_2}+k_{L_3})}\Big]H_{u}\,.
\end{eqnarray}
The mass matrix $M_D$ is found to be
\begin{equation}
M_D=\left(\begin{matrix}
\alpha_A Y_{\mathbf{2}A,2}^{(k_{N^c_1}+k_{L_D})}&(-1)^{i_1}\alpha_A Y_{\mathbf{2}A,1}^{(k_{N^c_1}+k_{L_D})} &\delta_1 Y_{\mathbf{1}^{<i_1+j>}}^{(k_{N^c_1}+k_{L_3})}\\
\beta_B Y_{\mathbf{2}B,2}^{(k_{N^c_2}+k_{L_D})}&(-1)^{i_2}\beta_B Y_{\mathbf{2}B,1}^{(k_{N^c_2}+k_{L_D})} &\delta_2 Y_{\mathbf{1}^{<i_2+j>}}^{(k_{N^c_2}+k_{L_3})}
\end{matrix}\right)v_{u}\,.
\end{equation}

\end{itemize}

In the same fashion as section~\ref{sec:numerical results}, we have numerically analyzed the possible minimal seesaw models with $S_4$ modular symmetry, we find that at least eight parameters including $\mathrm{Re}\langle\tau\rangle$ and $\mathrm{Im}\langle\tau\rangle$ should be used to accommodate the experimental data of leptons. Since the resulting models contain one more free parameter than the minimal models listed in table~\ref{tab:lepton bset-fit} and  table~\ref{tab:Dirac lepton models}, we don't give concrete examples here.


\providecommand{\href}[2]{#2}\begingroup\raggedright\endgroup

\end{document}